\documentclass[titlepage,12pt]{utarticle}

\usepackage{latexsym}
\usepackage{amsmath,amssymb,amsxtra,amscd,amsthm}
\usepackage[mathscr]{eucal}
\usepackage{bbm,mathrsfs}
\usepackage[dvips]{graphics,graphicx,psfrag}
\usepackage{epsfig}
\usepackage{color}
\usepackage{diagrams}
 
\def\cA{\mathcal{A}}

\def\cF{\mathcal{F}}

\def\cK{\mathcal{K}}
\def\cL{\mathcal{L}}

\def\cR{\mathcal{R}}

\def\cT{\mathcal{T}}

\def\cW{\mathcal{W}}

\def\mC{\mathbb{C}}

\def\mP{\mathbb{P}}

\def\mR{\mathbb{R}}

\def\mZ{\mathbb{Z}}

\def\tC{\mathrm{C}}
\def\tD{\mathrm{D}}

\def\tG{\mathrm{G}}
\def\tH{\mathrm{H}}

\def\tL{\mathrm{L}}

\def\tS{\mathrm{S}}

\DeclareMathOperator{\chern}{ch}
\DeclareMathOperator{\ch}{c}

\DeclareMathOperator{\Hom}{Hom}

\renewcommand{\mod}{\mathop{\mathrm{mod}}}
\def\bb1{\textup{\small{1}} \kern-3.8pt \textup{1}}
\newcommand{\diff}[2]{\textrm{d}^{#1}{#2}}
\newcommand{\e}[1]{\mathrm{e}^{#1}}

\newcommand{\BS}[1]{|\; #1 \;\rangle\!\rangle_B}

\newcommand{\pd}[2]{\frac{\partial #1}{\partial #2}}
\renewcommand{\d}{\partial}

\newcommand{\ibar}{\bar{\imath}}

\newcommand{\n}[2]{n^{(#1)}_{#2}}

\def\SL2Z{\tS\tL(2,\mZ)}

\numberwithin{equation}{section}
\psfrag{v01}{$(0,1)$}
\psfrag{v10}{$(1,0)$}
\psfrag{v11}{$(1,1)$}
\psfrag{v21}{$(2,0)$}
\psfrag{v31}{$(3,1)$}
\psfrag{v2-1}{$(2,-1)$}
\psfrag{v012}{$(0,1,2)$}
\psfrag{v010}{$(0,1,0)$}
\psfrag{v810}{$(8,1,0)$}
\psfrag{v021}{$(0,2,1)$}
\psfrag{v001}{$(0,0,1)$}
\psfrag{v501}{$(5,0,1)$}
\psfrag{v1001}{$(10,0,1)$}
\psfrag{v013}{$(0,1,3)$}
\psfrag{v610}{$(6,1,0)$}

\begin{document}
\preprint{
 MPP-2008-42\\
}

\title{\vskip-1cm  Towards Open String Mirror Symmetry\medskip\\ for One--Parameter Calabi--Yau Hypersurfaces}

\author{Johanna Knapp$^1$\footnote{knapp@mppmu.mpg.de}$\quad$ and $\:\:$ Emanuel Scheidegger$^2$\footnote{emanuel.scheidegger@math.uni-augsburg.de}}
\oneaddress{$^1$  Max--Planck--Institut f\"ur Physik \\   
F\"ohringer Ring 6\\
D--80805 Munich\\ Germany\\
{~}\\
$^2$        
Institut f\"ur Mathematik\\ Universit\"at Augsburg\\
D--86135 Augsburg\\ Germany}

 \nobreak
\Abstract{
This work is concerned with branes and differential equations for one--parameter Calabi--Yau hypersurfaces in weighted projective spaces. For a certain class of B--branes we derive the inhomogeneous Picard--Fuchs equations satisfied by the brane superpotential. 
In this way we arrive at a prediction for the real BPS invariants for holomorphic maps of worldsheets with low Euler characteristics, ending on the mirror A--branes.
}
\date{May 2008}
\maketitle
\tableofcontents
\section{Introduction}
Mirror symmetry for the closed string is the oldest, best understood and most thoroughly checked string duality. It is mathematically well--defined and there are elaborate techniques to calculate for example BPS invariants which are of interest to both physicists and mathematicians. The first cornerstone has been laid in~\cite{Candelas:1990rm} where the then abstract concept of mirror symmetry was put into a computational scheme to produce genus zero Gromov--Witten invariants in the example of a compact Calabi--Yau threefold, the quintic in $\mP^4$. Soon after that, it was realized in~\cite{Witten:1991zz} that the natural setting for mirror symmetry is topological string theory, and this led to the formulation of the A-- and B--model. The second cornerstone consisted of a thorough analysis of topological string theory in~\cite{Bershadsky:1993ta,Bershadsky:1993cx} leading to the holomorphic anomaly equations which govern its structure. They provide a powerful formalism to calculate BPS invariants at higher genus.\\

As compared to the closed string case, open string mirror symmetry is in many respects unexplored territory. For non--compact Calabi--Yau manifolds the subject is fairly well understood. The breakthrough was the formulation of the open string BPS invariants in~\cite{Ooguri:1999bv} which were then first computed in~\cite{Aganagic:2000gs}. By now, there exists a considerable amount of literature dealing with the open mirror symmetry on non--compact Calabi--Yau threefolds. In particular, recently in~\cite{Bouchard:2007ys} a set of recursion relations was found that allows to completely describe the topological B--model on non--compact Calabi--Yau threefolds and to compute the various BPS invariants. \\

However, for compact Calabi--Yau threefolds mirror symmetry with D--branes is much less understood. The reason for this is that it is in general much harder to deal with compact Calabi--Yaus because one has to take into account many new features. One complication related also to phenomenology is that, if one is interested in computing consistent models which have resemblance to the real world, one has to take into account the effects of fluxes and orientifold planes. These aspects may not directly enter in certain calculations but on the long run one cannot neglect these issues. Another difficulty when dealing with compact Calabi--Yau manifolds are the D--branes themselves. In non--compact models the branes typically sit at singularities and/or stretch into infinity. In compact Calabi--Yau manifolds branes obviously wrap compact cycles, leading to additional interesting, and phenomenologically relevant, structure. Upon the study of boundary conditions in topological string theory in~\cite{Witten:1992fb}, Kontsevich conjectured in~\cite{Kontsevich} that the mathematical framework to deal with open string mirror symmetry are categories. Whether one can make use of this abstract concept very much depends on whether one is interested in A--branes or B--branes. B--branes are quite well understood and can be approached in various ways which are also accessible to physicists. The relevant categories are the category of coherent sheaves and the category of matrix factorizations. A--branes are captured by the Fukaya category, which is hardly understood, even by mathematicians, and not many non--trivial examples for A--branes on compact Calabi--Yau threefolds are known. This is one reason why phenomenologically inclined physics papers mostly deal with models based on torus orbifolds where the A--branes are quite simple. For general Calabi--Yau threefolds Kontsevich's homological mirror symmetry conjecture states that the two categories are equivalent. However, this has not yet been useful for computing open BPS invariants.\\

Recently, in a pioneering series of articles~\cite{Walcher:2006rs,Pandharipande:2006ab,Walcher:2007tp,Neitzke:2007yw,Morrison:2007bm,Walcher:2007qp} Walcher and various collaborators took the first steps towards understanding open string mirror symmetry for compact Calabi--Yau threefolds. Further work includes~\cite{Cook:2007dj,Bonelli:2007gv,Alim:2007qj,Konishi:2007qx,Cook:2008eu}\\

In the first paper,~\cite{Walcher:2006rs}, disk instantons have been computed for the quintic using mostly A--model techniques. A particular Lagrangian A--brane, defined by the real quintic, was identified. It admits two vacua separated by a domain wall. The instantons are then maps from the disk into this Lagrangian. The generating function of these instantons is the BPS domain wall tension~\cite{Ooguri:1999bv}. It was shown that this object is determined by an inhomogeneous Picard--Fuchs differential equation. A particular differential equation was proposed in~\cite{Walcher:2006rs} and it was verified by A--model localization techniques in~\cite{Pandharipande:2006ab} that its solution produces the correct instanton numbers. In the second paper,~\cite{Morrison:2007bm}, the authors focused on the B--model. The D--brane which is the mirror of the real quintic was identified. From the associated geometric boundary conditions the inhomogeneous Picard--Fuchs equation could be derived, thus completing for the first time an explicit open string mirror symmetry computation on a compact Calabi--Yau threefold. In~\cite{Walcher:2007tp} the holomorphic anomaly equations were extended to include a particular set of D--branes. As mentioned above, the presence of D--branes on compact spaces generally requires the introduction of orientifolds leading to tadpoles that must be cancelled. In~\cite{Walcher:2007qp} it has been argued  that also in topological string theory the effects of orientifolds play a crucial role, and the holomorphic anomaly equations were further extended to include also unoriented worldsheets.\\

The aim of the present article is to deepen the understanding of the concepts introduced in \cite{Walcher:2006rs,Morrison:2007bm}, focusing on the B--model. The models we will consider are the one--parameter hypersurfaces in weighted $\mP^4$. These models are slightly more complicated than the quintic but also exhibit enough similarities to provide a testing ground for the ideas of \cite{Walcher:2006rs,Morrison:2007bm}.\\\\
The paper is organized as follows. In Section~\ref{sec:general-remarks} we give an overview on the subject and introduce the notation. Section~\ref{sec:d-branes-effective} is concerned with a certain class of D--branes and their (obstructed) moduli on the one--parameter hypersurfaces. In this discussion we use techniques of matrix factorizations and boundary conformal field theory. In Section~\ref{sec:geom-bound-cond} we establish the relation to geometric boundary conditions. In Section~\ref{sec:toric} we discuss how to resolve the singularities at the points on the boundary which are fixed by an action of the orbifold group. This is necessary preparatory work for the derivation of the inhomogeneous Picard--Fuchs equations, which we do in Section~\ref{sec:picard-fuchs} for a particular choice of boundary conditions. In Section~\ref{sec:instantons} we discuss the properties of the BPS domain wall tension and compute the certain real BPS invariants. We close the main part of the paper with some conclusions and open questions in Section~\ref{sec:conclusions}. In the Appendix we provide some information on orientifolds on the one--parameter hypersurfaces.\\\\
Recently we have been informed by Johannes Walcher that he and Daniel Krefl also work on open string mirror symmetry on one--parameter hypersurfaces \cite{walcher1par}.


\section{General Remarks}
\label{sec:general-remarks}

In this section we start with a short review of the preceding work on open string mirror symmetry for compact Calabi--Yau threefolds with the aim of making the reader familiar with some new concepts and of defining the central objects and setting the notation.

\subsection{Review of open and closed mirror symmetry}
\label{sec:review-open-closed}

We begin with a mirror pair of families of Calabi--Yau threefolds $(X,Y)$ realized as hypersurfaces in a toric variety. We will refer to $X$ as the target space for the A--model, and $Y$ as the target space of the B--model, although at some point the roles will be interchanged. 
We extend this mirror pair by including families of D--branes to an open string mirror pair $\left( (X,L_{\alpha}) ,  (Y,E_\alpha) \right)$. This is a down--to--earth way of formulating the homological mirror symmetry conjecture~\cite{Kontsevich}. Let $\text{Fuk($X$)}$ denote the category of A--branes on $X$, and $\tD^b(\text{Coh}(Y))$ denote the category of B--branes on $Y$, then this conjecture states an equivalence between these two categories\footnote{Since we are ultimately working with orientifolds, one should endow these categories with a parity functor~\cite{Hori:2006ic},~\cite{Diaconescu:2006id}.}
\begin{equation}
  \label{eq:HMS}
  \begin{array}{ccc}
    \text{Fuk($X$)} &\cong &\tD^b(\text{Coh}(Y)),\\
    L_\alpha & \leftrightarrow & E_\alpha.
  \end{array}
\end{equation}
Here $L_\alpha$ is a choice of A--branes consisting of a family of special Lagrangian submanifolds $L$ of $X$ together with a local system on $L$, i.e. a choice of a flat connection. The flat connections are solutions to the equations of motion of the Chern--Simons functional on the world--volume of the A--brane~\cite{Witten:1992fb}. Recall that flat connections are equivalent to representations $\rho: \pi_1(L) \to \tG\tL(1,\mR)$, and hence are classified by the cohomology group $\Gamma_L = \Hom(\pi_1(L),\mR)$. $\alpha$ will denote an element of this group. On the other side, $E_\alpha$ is a choice of B--branes consisting of a family of complexes $E$ of holomorphic vector bundles on $Y$ together with a choice of a complex structure on $E$. Remember that a holomorphic vector bundle admits a unique Hermitian connection $a$ such that $\bar\partial_a = \bar\partial$, and vice versa. Hermitian connections are solutions to the equations of motion of the holomorphic Chern--Simons functional on the world--volume of the B--brane~\cite{Witten:1992fb,Donaldson:1998ab}. We will label the choice of the Hermitian connection on $E$ by $\alpha$. If open string mirror symmetry holds, it follows that generically there have to be as many local systems on $L$ as there are Hermitian connections on $E$. \\

In the physical realization of closed string mirror symmetry, the pivotal quantity is the holomorphic prepotential $\cF$. In the A--model on $X$, it is defined as the generating function of holomorphic maps of spheres into $X$
\begin{equation}
  \label{eq:F_A}
    \cF_{A}(t) = c(t) + \sum_{\beta\in \tH_2(X,\mZ)} \tilde n_{\beta}\; q^{\text{Area}(\beta)}, 
\end{equation}
where $q=\e{2\pi i t}$ and $c(t)$ is a cubic polynomial in the complexified K\"ahler moduli $t$ containing the classical part, i.e. topological information on $X$. $\beta$ is the class of the image of the map in $X$. \\
In the B--model, the prepotential can be written in terms of an integral symplectic basis $(w_3(z),w^{(1)}_2(z),\dots,w^{(n)}_2(z),w_0(z),w^{(1)}_1(z),\dots,w^{(n)}_1(z))$, $n=h^{2,1}(Y)$ of periods on $Y$ as follows:
\begin{equation}
  \label{eq:F_B}
  \cF_B(z) = \frac{1}{2}\left(w_3(z)w_0(z) - \sum_{i=1}^{n} w_1^{(i)}(z)w_2^{(i)}(z) \right),
\end{equation}
where $z$ denotes the complex structure moduli of the mirror manifold $Y$. The prepotential $\cF_B(z)$ is a solution to the Picard--Fuchs equations which comes from the fact that the periods are solutions to the Picard--Fuchs equations
\begin{equation}
  \label{eq:PF_for_periods}
  \cL_{\text{PF}} \varpi = 0, 
\end{equation}
for $\varpi \in \{w_3,w^{(1)}_2,\dots,w^{(n)}_2,w_0,w^{(1)}_1,\dots,w^{(n)}_1\}$. The most important property of the prepotentials is that closed string mirror symmetry relates the two in the following way
\begin{equation}
  \label{eq:prepotential}
  \cF_A(t) = \varpi_0(z(t))^{-2}\cF_B(z(t)),
\end{equation}
where the map $z(t)$ is the (inverse of the) mirror map, and $\varpi_0(z)=w_0(z)$ is the fundamental period, i.e. the period which is holomorphic near $z=0$. \\

Given the holomorphic prepotential $\cF_A=\cF^{(0)}_A$ as well as the topological data of $X$, one can then proceed to determine the generating function $\cF^{(1)}$ of holomorphic maps of genus 1 in the A--model,
\begin{equation}
  \label{eq:10}
   \cF^{(1)}_A(t) = l(t) + \sum_{\beta\in \tH_2(X,\mZ)} \tilde n_{\beta}\; q^{\text{Area}(\beta)},
\end{equation}
from the integration of its holomorphic anomaly equation~\cite{Bershadsky:1993ta} in the B--model
\begin{equation}
  \label{eq:HAE_genus1}
  {\bar \partial}_{\ibar} \partial_j \cF^{(1)}_B(z)=\frac{1}{2}
C_{\ibar}^{kl} C_{jkl} + \left( \frac{\chi}{24} -1\right )
G_{\ibar j},
\end{equation}
where $C_{ijk} = \partial_i\partial_j\partial_k\cF$ is the 3--point function, $G_{\ibar j}=\partial_{\ibar}\partial_j K$ is the Zamolodchikov metric on the moduli space $M$ of complex structures on $Y$, and $\chi$ is the Euler number of $Y$. In~(\ref{eq:10}), $l(t)$ is a linear polynomial in $t$ depending on the topological data of $X$. Thereby we pick up a holomorphic ambiguity $f^{(1,0)}$ which turns out to have a universal behavior $(1-z_c)^{-\frac{1}{6}}$ near the conifold point $z_c$. This allows one to fix it completely in the case of one--parameter Calabi--Yau hypersurfaces without having to rely on the absence of certain curves. Closed string mirror symmetry then states that
\begin{equation}
  \label{eq:14}
  \cF^{(1)}_A(t) = \cF^{(1)}_B(z(t)).
\end{equation}\\
Following the program initiated by Walcher in~\cite{Walcher:2006rs,Walcher:2007tp,Morrison:2007bm}, we now consider the open string analog of the prepotential $\cF$ which is the BPS domain wall tension $\cT$ and point out some of its properties. \\

In the A--model on $X$, $\cT_{A,\alpha}(t)$ is defined as the generating functional counting holomorphic maps of discs ending on $L_\alpha$. It has the form
\begin{equation}
  \label{eq:T_A}
  \cT_{A,\alpha}(t) = \frac{t}{2} + \cT_{\text{classical},\alpha} + \sum_{D\in \tH_2(X,L_{\alpha},\mZ)} \tilde n_{D}\; q^{\text{Area}(D)},
\end{equation}
where $\tH_2(X,L_{\alpha},\mZ)$ is the relative cohomology group labeling the classes $D$ of the image of the holomorphic discs. $ \cT_{\text{classical},\alpha}$ contains ``classical'' contributions, i.e. topological invariants such as the analytic or Ray--Singer torsion of $L_{\alpha}$, and is therefore independent of $t$. The goal in this work is to compute the BPS invariants $\tilde n_{D}$ for some choice of the pair $(X,L_\alpha)$. We will explain the way of choosing this pair below. Before, however, we need to introduce the B--model version of $\cT$.\\

In the B--model, $\cT_B(z)$ is defined as the difference of the holomorphic Chern--Simons functionals $\cW_\alpha(z)$ for two distinct Hermitian connections:
\begin{equation}
  \label{eq:T_B}
  \cT_{B,\alpha}(z) = \cW_\alpha(z) - \cW_0(z) = S_{\text{holCS}}(\alpha).
\end{equation}
Here $0$ denotes a reference connection, i.e. a reference complex structure. In order to make sense of~\eqref{eq:T_B} for an arbitrary complex $E$ of holomorphic vector bundles, we proceed as in~\cite{Morrison:2007bm}. Let us for the moment exhibit the complex structure dependence of a member of the family $(Y,E)$ by writing $Y_z$ and $E_z$ for that member. Then, to such a complex $E_z$ we can associate its algebraic second Chern class $\ch_2^{\text{alg}}(E_z)$. This is an element of the Chow group $\tC\tH^2(Y_z)$ of codimension 2 algebraic cycles in $Y_z$ modulo rational equivalence. There is a natural map from $\tC\tH^2(Y_z)$ into the cohomology group $\tH^4(Y_z,\mZ)$. The image of $\ch_2^{\text{alg}}(E_z)$ under this map is the topological second Chern class $\ch_2^{\text{top}}(E_z)$ characterizing the charges of the D--brane described by the complex $E$. We want to emphasize that $\ch_2^{\text{alg}}(E_z)$ contains more information than just the charges. An element of the Chow group $\tC\tH^2(Y_z)$ can be represented by a collection of curves $C_z$ given by a set of algebraic equations. In order to relate $C_z$ to $\cT_B$ we have to require that it is homologically trivial, i.e. that its image in $\tH^4(Y_z,\mZ)$ vanishes. In this case, there exists a so--called normal function $\nu_{C_z}=\nu_{\ch_2^{\text{top}}(E_z)}$ that has been introduced by Griffiths~\cite{griffiths1,griffiths2,Griffiths:1979ab}. If we pick any 3--chain $\Gamma_z$ such that $\partial \Gamma_z = C_z$ and integrate the holomorphic 3--form $\Omega(z)$ of $Y_z$ over this chain, we obtain the truncated normal function
\begin{equation}
  \label{eq:normal_function}
  \nu_{C_z}(\Omega) = \int_{\Gamma_z} \Omega(z).
\end{equation}
This is then the familiar expression for the holomorphic Chern--Simons functional for the special case where the B--brane is described by a holomorphic vector bundle on a holomorphic curve~\cite{Witten:1997ep,Kachru:2000ih,Aganagic:2000gs,Dijkgraaf:2002fc}. Hence, the formula for the domain wall tension in the B--model is~\cite{Morrison:2007bm}:
\begin{equation}
  \label{eq:T_B-normal}
  \cT_{B,\alpha}(z) = \nu_{C_{\alpha,z}}\left(\Omega\right) . 
\end{equation}
Note that the normal function is only well--defined only up to periods, i.e. up to integrals $\int_\gamma \Omega$ for some 3--cycle $\gamma\in \tH_3(Y_z,\mZ)$. In fact, the normal function should be viewed as a holomorphic section of the Griffiths intermediate Jacobian fibration over the moduli space of complex structures of the family $Y$. For more details about the mathematical properties of $\nu_C$, see~\cite{GreenM,Voisin1,Voisin2}. The main property of interest to us is that the normal function satisfies an inhomogeneous version of the Picard--Fuchs equations~\cite{Griffiths:1979ab}:
\begin{equation}
  \label{eq:PF_for_TB}
  \cL_{\text{PF}} \cT_{B,\alpha}(z)  =  f_\alpha(z),
\end{equation}
where $f_\alpha(z)$ is some function in $z$ and $\cL_{\text{PF}}$ is the differential operator from~\eqref{eq:PF_for_periods}. The function $f_\alpha(z)$ contains information about the B--brane realized by the complex $E_\alpha$ beyond its charges.\\ 

Having defined the BPS domain wall tension in both the A-- and the B--model, we can now state the analog of~\eqref{eq:prepotential} for open string mirror symmetry~\cite{Morrison:2007bm}:
\begin{equation}
  \label{eq:open-string-mirror}
  \cT_{A,\alpha}(t) = \varpi_0(z(t))^{-1} \cT_{B,\alpha}(z(t)).
\end{equation}
This conjecture has been proven for a particular choice of $(X,L_\alpha)$, namely for the quintic $X$ in $\mP^4$ and $L$ its real locus~\cite{Walcher:2006rs,Pandharipande:2006ab,Morrison:2007bm}.\\

As in the closed string situation, given the BPS domain wall tension $\cT_A = \cF_A^{(0,1)}$, one can then proceed to study holomorphic maps of Riemann surfaces with larger Euler number. In the A--model on $X$, $\cF^{(0,2)}_{A,\alpha}(t)$ is defined as the generating functional counting holomorphic maps of annuli ending on $L_\alpha$. It has the form
\begin{equation}
  \label{eq:A_A}
  \cF^{(0,2)}_{A,\alpha}(t) = \sum_{A\in \tH_2(X,L_{\alpha},\mZ)} \tilde n_{A}\; q^{\text{Area}(A)}.
\end{equation}
On the other hand, it has been recently shown in~\cite{Walcher:2007tp,Walcher:2007qp}, that in the B--model there is an extension of the holomorphic anomaly equation to Riemann surfaces with boundaries which for the annulus reads
\begin{equation}
  \label{eq:HAE_Annuli}
  \partial_{\ibar} \partial_j\cF^{(0,2)}_{B,\alpha} = -\Delta_{jk,\alpha} \Delta^{k}_{\ibar,\alpha} + \frac{N_\alpha}{2}G_{\ibar j} ,
\end{equation}
where $N_\alpha$, roughly speaking, is the number of generators of the unbroken gauge group on the B--brane $E_\alpha$. Similar to $\cT_{B,\alpha}$, $\Delta_{ij,\alpha}$ is a quantity from Hodge theory, the Griffiths infinitesimal invariant~\cite{Griffiths:1983ab} of the normal function $\nu_{C_\alpha}$ (see also \cite{GreenM,Voisin1,Voisin2}). In the holomorphic limit, they are related by
\begin{equation}
  \label{eq:37}
  \Delta_{ij,\alpha} = \lim_{\bar z\to 0} D_iD_j\cT_{B,\alpha},
\end{equation}
where $D_i$ is the covariant derivative of special geometry. For more details we refer to~\cite{Walcher:2007tp}. Integrating~(\ref{eq:HAE_Annuli}) again introduces a holomorphic ambiguity $f^{(0,2)}$. One would hope that it has a universal behavior near the conifold point such that the ambiguity can be fixed completely for sufficiently simple B--branes. Then the analog of~(\ref{eq:14}) becomes
\begin{equation}
  \label{eq:47}
  \cF^{(0,2)}_A(t) = \cF^{(0,2)}_B(z(t)).
\end{equation}
It turns out, however, that the invariants $\tilde n_A$ (after taking into account multiple cover contributions~\cite{Ooguri:1999bv}) need not be integral. The reason for this is the topological string version of the tadpole cancellation~\cite{Walcher:2007qp}. In the presence of branes, the A-- and B--model only decouple if the tadpoles are cancelled. This requires the presence of orientifold planes and hence unoriented worldsheets. It was argued that, upon inclusion of the contributions of the worldsheets, the real BPS invariants become integers. In particular, we will need the generating function $\cK_A$ for holomorphic maps of Klein bottles in the A--model. 
\begin{equation}
  \label{eq:K_A}
  \cK_{A,\alpha}(t) = \sum_{K\in \tH_2(X,L_{\alpha},\mZ)} n_{K}\; q^{\text{Area}(K)}.
\end{equation}
If the orientifold projection is trivial, the corresponding quantity $\cK_B$ in the B--model satisfies the holomorphic anomaly equation of $\cF^{(1)}$ in~(\ref{eq:HAE_genus1}) with $\chi=0$~\cite{Walcher:2007qp}
\begin{equation}
  \label{eq:HAE_Klein}
  {\bar \partial}_{\ibar} \partial_j \cK_B=\frac{1}{2}
C_{\ibar}^{kl} C_{jkl} -G_{\ibar j}.
\end{equation}
Here, it is conjectured that the holomorphic ambiguity has a universal behavior similar to $f^{(1)}$, but with a different exponent: $f^{(1,0)_k} = (1-z_c)^{-\frac{1}{4}}$.

\subsection{The program}
\label{sec:program}

After having introduced all the objects we need and having stated the (conjectured) relations among them, we now proceed to explain how the BPS invariants $n_{D}$ can be determined. Given an A--brane $L_{\alpha}$, there is, according to~(\ref{eq:HMS}), a mirror B--brane $E_{\alpha}$. We pick such a B--brane and compute its algebraic second Chern class $\ch_2^{\text{alg}}(E_{\alpha})$ to get the curve $C_{\alpha}$. These curves are homologically equivalent for distinct values of $\alpha$, i.e. choosing 0 as a reference value we have
\begin{equation}
  \label{eq:hom-equiv}
  C_\alpha \stackrel{\textrm{hom}}{\cong} C_0,
\end{equation}
hence the difference of two of them is homologically trivial: $[C_\alpha - C_0] = 0 \in \tH^4(Y,\mZ)$. We then construct a 3--chain $\Gamma_\alpha$ such that $\partial \Gamma_\alpha = C_\alpha - C_0$. Performing the integration of $\Omega$ over $\Gamma_\alpha$ yields the function $f_\alpha(z)$ and consequently $\cT_{B,\alpha}(z)$. Finally, we substitute the mirror map, use~\eqref{eq:open-string-mirror}, and expand $\cT_A(t)$ as in~\eqref{eq:T_A} in order to read off the BPS invariants $n_D^{0,\text{real}}=n_D$. From $\cT_B$ we can also determine the Griffiths infinitesimal invariant $\Delta_{ij}$, and together with the closed string quantities $C_{ijk}$, $G_{\ibar j}$ we can integrate the equations~\eqref{eq:HAE_Annuli} and~\eqref{eq:HAE_Klein}. From~\eqref{eq:A_A} and~\eqref{eq:K_A} we can then read off the BPS invariants $n_{A=K}^{1,\text{real}}=4n_A+n_K$. We will now give the details of each of these steps as well as the references to the various sections where the  corresponding computations are carried out. \\

The first step consists of the choice of the mirror pair $(X,Y)$. For this work, we will choose $X$ to be one of the four possible hypersurfaces in toric varieties $X$ with a one--parameter K\"ahler moduli space, i.e. with $h^{1,1}(X)=1$. They are all realized as degree $d$ hypersurfaces in weighted projective spaces $\mP(w)$ with $d=\sum_{i=1}^5 w_i$. So $X$ is any of the following families:
\begin{equation}
  \label{eq:the_Xs}
  \mP(1,1,1,1,1)[5], \qquad \mP(1,1,1,1,2)[6], \qquad \mP(1,1,1,1,4)[8], \qquad \mP(1,1,1,2,5)[10].
\end{equation}
The first one of these, the quintic in $\mP^4$, has been the central example in~\cite{Walcher:2006rs,Morrison:2007bm}. Therefore we focus here on the other three families. The mirror families $Y$ can be obtained through the Greene--Plesser orbifold construction~\cite{Greene:1990ud} and yields for the $Y$
\begin{equation}
  \label{eq:the_Ys}
  \begin{aligned}
    &\mP(1,1,1,1,1)[5]/(\mZ_5)^3, 
    &&\mP(1,1,1,1,2)[6]/(\mZ_6)^2\times \mZ_3,\\
    &\mP(1,1,1,1,4)[8]/(\mZ_8)^2\times \mZ_2, \qquad
    &&\mP(1,1,1,2,5)[10]/(\mZ_{10})^2,
  \end{aligned}
\end{equation}
respectively. These spaces are singular and have to be resolved. For this one can invoke the standard techniques of toric geometry. The equations for the latter three mirror families are
\begin{align}
  \label{eq:hypersurfaces}
  W^{(6)}(\psi)&= x_1^6+x_2^6+x_3^6+x_4^6+x_5^3-6\psi x_1x_2x_3x_4x_5,\\
  W^{(8)}(\psi)&= x_1^8+x_2^8+x_3^8+x_4^8+x_5^2-4\psi x_1^2x_2^2x_3^2x_4^2,\\
  W^{(10)}(\psi)&= x_1^{10}+x_2^{10}+x_3^{10}+x_4^5+x_5^2-5\psi x_1^2x_2^2x_3^2x_4^2,
\end{align}
where $\psi$ is the complex structure modulus and we denote by $W$ the superpotential of the associated Landau--Ginzburg model. Note that for $W^{(8)}$ and $W^{(10)}$ we do not use the standard deformation which would be $\psi x_1x_2x_3x_4x_5$. The deformations we use can be obtained from the standard one via the equations of motion for $x_5$. We will justify our choice of deformations in Section \ref{sec:moduli}. Closed string mirror symmetry for these families has been studied in detail in~\cite{Font:1992uk,Klemm:1992tx}, in particular the Picard--Fuchs system~\eqref{eq:PF_for_periods} and the prepotentials~\eqref{eq:F_A},~(\ref{eq:F_B}) were determined there. \\

When we now want to specify families of special Lagrangian submanifolds $L$ of $X$ we run into trouble because there is no general construction known. The only special Lagrangian submanifolds that are known in general are the so--called real Calabi--Yau manifolds. See e.~g.~\cite{Roiban:2002iv} for the case $W^{(8)}$. We circumnavigate this problem by directly specifying the mirror family $E$ of B--branes on $Y$ and assume that there exists a submanifold $L \subset X$ that is mirror to $E$. From the properties of $E$ we can infer some of the properties of $L$, in particular the number of flat connections on $L$.  \\

Let us discuss this last point in more detail and explain what we mean by a family of A-- or B--branes. One of the four fundamental facts about open strings on compact Calabi--Yau threefolds that were argued for in~\cite{Walcher:2007tp} states that at a generic point in the closed string moduli space, there are no continuous open string moduli. Hence, at a generic point $z_0$ of the complex structure moduli space, a D--brane can only depend on the complex structure modulus $z$ as well as on discrete open string moduli $\alpha$. The discreteness of the open string moduli means that there are potentially continuous moduli for which there is however a superpotential which forces them to be fixed at its critical locus. This means that the moduli are obstructed. This will be discussed in great detail in Section~\ref{sec:moduli}. Assuming that the critical locus is a finite set, we identify the points with choices $\alpha$ of a Hermitian structure on $E$.\\ 

Hence, our task is to specify the B-brane $E$ and study its deformations and obstructions. Definitely, the handiest way to describe B--branes is through the concept of matrix factorizations~\cite{Kapustin:2002bi,Brunner:2003dc} (for recent reviews see~\cite{Jockers:2007ng,Knapp:2007vc}). In particular, as Orlov has shown in~\cite{orlov1}, we can associate to every complex of holomorphic vector bundles $E$ on $Y$ a matrix factorization $Q$ of $W$ with $Q^2 = W\cdot\mathbbm{1}$ and vice versa. This correspondence is not unique but an explicit construction has recently been given in~\cite{Herbst:2008jq} where this correspondence was physically realized as the open string version of the Calabi--Yau/Landau--Ginzburg correspondence~\cite{Witten:1993yc}. We will apply the results of~\cite{Herbst:2008jq} in Section~\ref{sec:geom-bound-cond}. So instead of specifying a complex $E$ of holomorphic vector bundles on $Y$, we will instead give the corresponding matrix factorization $Q$ of $W$. At the Gepner point, a subset of the matrix factorizations can be identified~\cite{Brunner:2005fv} with the Recknagel--Schomerus boundary states $\BS{L,M,S}$~\cite{Recknagel:1997sb,Brunner:1999jq} in the corresponding Gepner model. The relation between D--branes on the last three families $X$ in~\eqref{eq:the_Xs} and boundary states in the corresponding Gepner model has been studied in~\cite{Scheidegger:1999ed} which will be useful along the way (for related work see~\cite{Naka:2000he,Govindarajan:2000vi}). This will explained in more detail in Section~\ref{sec:matr-fact-bound}, where we will also specify the matrix factorizations for the various $W$ in~\eqref{eq:hypersurfaces}. At this point, we have to take into account that we need the matrix factorizations on the mirror $Y$ and not on $X$. The reason why we focus on matrix factorizations corresponding to Recknagel--Schomerus boundary states is the following: Since the mirror construction only involves taking a quotient with respect to the Greene--Plesser group $G_{\text{GP}}$, we can simply take a $G_{\text{GP}}$--equivariant version of the matrix factorization of the $W$ in~\eqref{eq:hypersurfaces}. As mentioned above, given these matrix factorizations, we have to work out their deformations and obstructions in order to find the possible vacua $\alpha$. In particular, we have to make sure that the deformations are also $G_{\text{GP}}$--equivariant. This will be the content of Sections~\ref{sec:obstr-effect-superp} and~\ref{sec:moduli-mathc-all}, and the resulting object will be denoted by $Q_\alpha$. \\

Let us summarize the first step. Instead of specifying the A--brane $L_\alpha$ on $X$, we decide to start on the mirror side and to specify the mirror B--brane $E_\alpha$ on $Y$. For various technical reasons it is, however, simpler to first start with the corresponding $G_{\text{GP}}$--equivariant matrix factorization $Q_\alpha$ of $W$ and then construct the complex $E_\alpha$ from $Q_\alpha$ by using the open string version of the Calabi--Yau/Landau--Ginzburg correspondence~\cite{Herbst:2008jq}. This is done in great detail in Section~\ref{sec:LG-CY} for the examples we have selected in Section~\ref{sec:d-branes-effective}.  \\

Equipped with an explicit complex of holomorphic vector bundles $E_\alpha$ on $Y$ we can proceed to the second step and compute its topological Chern character $\chern^{\text{top}}(E_\alpha)$ as well as its algebraic second Chern class $\ch^{\text{alg}}(E_\alpha)$. The topological Chern character allows us to verify that the complexes we have constructed indeed come from Recknagel--Schomerus boundary states by comparing it to the Chern characters obtained in~\cite{Scheidegger:1999ed}. This is done in Section~\ref{sec:LG-CY}. Furthermore, it allows us to check whether it satisfies the tadpole cancellation condition along the lines of~\cite{Brunner:2004zd}. In fact, in the case of the quintic, the complex chosen in~\cite{Morrison:2007bm} is precisely the one for which the tadpole cancellation condition is satisfied~\cite{Brunner:2004zd}. This seems to be important for the following reason: Another of the four fundamental facts pointed out in~\cite{Walcher:2007tp} is that the topological charge of the D--brane configuration under consideration has to vanish. This is the topological string version of the tadpole cancellation condition. This was subsequently made more precise in~\cite{Walcher:2007qp} where it was argued that the decoupling of the B--brane from the A--type moduli only happens under this condition. This condition then requires the inclusion of unoriented worldsheets and therefore orientifolds. It was shown that only upon their inclusion the open string BPS invariants at higher order in perturbation theory become integral. We will come back to this issue at the end of this section. The algebraic second Chern class allows us to determine the curves $C_\alpha$. For this purpose one chooses generic sections of $\ker Q_\alpha$ and looks for the locus where they fail to be linearly independent. This locus is a representative of the algebraic Chern class. This is explained in Section~\ref{sec:algebr-second-chern}. \\

In the third step we have to select a 3--chain $\Gamma_\alpha$ on $Y$ such that $\partial \Gamma_\alpha = C_\alpha - C_0$ and integrate the holomorphic 3--form $\Omega$ over this 3--chain. This is typically done by putting an infinitesimal tube $T(\Gamma_\alpha)$ around $\Gamma_\alpha$ in the ambient weighted projective space and integrating over this 4--chain instead~\cite{griffiths1}. Furthermore, one expects on general grounds~\cite{Griffiths:1979ab} that this integral satisfies an inhomogeneous Picard--Fuchs equation of the form~\eqref{eq:PF_for_TB}. As explained in~\cite{Morrison:1991cd} there is a standard algorithm of reduction of the pole order due to Griffiths and Dwork~\cite{Dwork,griffiths1} which yields the following differential equation for the holomorphic 3--form $\Omega$ :
\begin{equation}
  \label{eq:PF_for_Omega}
  \cL_{\text{PF}}\Omega(z) = \diff{}{\varphi(z)}.
\end{equation}
We briefly review this in Section~\ref{sec:griffiths-dwork} and apply it to the various families $(Y,C_\alpha)$ in Sections~\ref{sec:d=8-hypersurface} and~\ref{sec:d=10-hypersurface}. This provides us with both $\cL_{\text{PF}}$ and $\diff{}{\varphi(z)}$. The integral of the term on the right--hand side of~\eqref{eq:PF_for_Omega} over the tube $T(\Gamma_\alpha)$ gives the inhomogeneous term $f(z)$ in~\eqref{eq:PF_for_TB} due to the fact that the 4--chain $T(\Gamma_\alpha)$ has a non--trivial boundary $T(C_\alpha-C_0)$
\begin{equation}
  \label{eq:inhomog}
  f_\alpha(z) = \int_{T(C_\alpha-C_0)} \varphi(z).
\end{equation}
As was pointed out in~\cite{Morrison:2007bm}, here one runs into a further technical problem. The tubes $T(C_\alpha)$ and $T(C_0)$ will intersect in general in some number of points $p_i \in Y$. Moreover, these points can coincide with the singular points from the action of the orbifold group $G_{\text{GP}}$ on $W=0$. The induced singularities have to be resolved. Since the ambient spaces are weighted projective spaces, this can be done straightforwardly in the framework of toric geometry. This is the subject of Section~\ref{sec:toric}. Once we are equipped with the resolution we can proceed to compute the integral in~(\ref{eq:inhomog}) along the lines of~\cite{Morrison:2007bm}. This is worked out for the various families $(Y,C_\alpha)$ also in the Sections~\ref{sec:d=8-hypersurface} and~\ref{sec:d=10-hypersurface}. In the end, we obtain from~(\ref{eq:normal_function}) and~(\ref{eq:T_B-normal}) the normal function $\cT_{B,\alpha}(z)$. \\

The last step then involves plugging the inverse mirror map $z(t)$ into $\cT_B(z)$, using the open string mirror formula~\eqref{eq:open-string-mirror}, and to expand the so obtained BPS domain wall tension $\cT_A(t)$ according to~\eqref{eq:T_A}. This is standard and will be carried out in Section~\ref{sec:instanton-expansion}. Before that, however, we study the solutions to~(\ref{eq:PF_for_Omega}) and their monodromy behavior along paths in the complex structure moduli space of $Y$. For this purpose, we analyze these solutions in Section~\ref{sec:solut-pf-equat} and their analytic continuation to large values of $z$ in Section~\ref{sec:analyt-cont}. The monodromy behavior is discussed in Section~\ref{sec:monodromies}. This will provide a consistency check on the results we have found in Section~\ref{sec:picard-fuchs}. In Section~\ref{sec:instanton-expansion} we try in addition to make a prediction for real BPS invariants with Euler number 1 by solving the holomorphic anomaly equations for the annulus~(\ref{eq:HAE_Annuli}) and the Klein bottle~(\ref{eq:HAE_Klein}). It is important at this point that we have carefully chosen our B--brane in Section~\ref{sec:d-branes-effective} such that the orientifold projection becomes trivial. \\

Finally, we will study the normal function $\cT_B(z)$ and the differential equation it satisfies in more detail in Section~\ref{sec:semiperiods}. It turns out that the $\cT_B(z)$ we found also satisfies a homogeneous differential equation
\begin{equation}
  \label{eq:homog}
  \cL_B \cT_B(z) = 0
\end{equation}
in a similar way as the one found for the quintic in~\cite{Walcher:2006rs}. We will argue that the solutions to the differential operator $\cL_B$ are so--called semi--periods. These are solutions to the GKZ hypergeometric system of differential equations (see~\cite{Stienstra:2005nr} for a nice review). This system arises naturally in the extension of the Greene--Plesser mirror construction to arbitrary Calabi--Yau hypersurfaces in toric varieties found by Batyrev~\cite{Batyrev:1994hm,Hosono:1993qy}. The GKZ system in particular contains the Picard--Fuchs system. Furthermore there is a construction of 3--chains $S$ such that the integral of the holomorphic 3--form $\Omega$ over $S$ is a semi--period. We speculate on the relation between the 3--chains $S$ and the 3--chain $\Gamma$ used in the construction of the normal function.

\section{D--branes and Effective Superpotentials}
\label{sec:d-branes-effective}
In this section we will discuss D--branes on the one--parameter hypersurfaces. We will make use of the description of Landau--Ginzburg branes in terms of matrix factorizations and of the boundary state formalism, available at the Gepner point. In all cases we will restrict ourselves to tensor product boundary states. We will discuss which branes have moduli and how they are obstructed by computing the effective superpotential. This will give a hint which branes admit two vacua separated by a domain wall.
\subsection{Matrix Factorizations and Boundary States}
\label{sec:matr-fact-bound}
Let us now discuss the class of matrix factorizations which characterize the D--branes we are interested in. At the Gepner point we can make an identification with the Recknagel--Schomerus boundary states. Given the $A_{d-2}$ minimal model with superpotential $W=x^{d}$ we can identify certain matrix factorizations with boundary states \cite{Brunner:2003dc,Brunner:2005fv}. In particular we have:
\begin{equation}
Q^{(k)}=\left(\begin{array}{cc}
0&x^k\\
x^{d-k}&0
\end{array}
\right)\qquad\Longleftrightarrow\qquad \BS{L,S}=\BS{k-1,0}
\end{equation}
The additional label $M$ is non--zero whenever an orbifold action is taken into account. The branes we are looking at will be tensor products of such boundary states. We will often use the boundary state notation to label the matrix factorizations, even when the deformation is turned on. We consider the following factorizations for the three hypersurfaces:
\begin{eqnarray}
Q^{d=6}&=&\sum_{i=1}^4(x_i^{k_i}\eta_i+x_i^{6-k_i}\bar{\eta}_i)+x_5\eta_5+(x_5^2-6\psi x_1x_2x_3x_4)\bar{\eta}_5\label{d6-mf}\\
Q^{d=8}_{\pm}&=&\sum_{i=1}^4(x_i^{k_i}\eta_i+x_i^{8-k_i}\bar{\eta}_i)+(x_5\pm\sqrt{4\psi}x_1x_2x_3x_4)\eta_5+(x_5\mp\sqrt{4\psi}x_1x_2x_3x_4)\bar{\eta}_5\label{d8-mf}\\
Q^{d=10}_{\pm}&=&\sum_{i=1}^3(x_i^{k_i}\eta_i+x_i^{10-k_i}\bar{\eta}_i)+x_4^{k_4}\eta_4+x_4^{5-k_4}\bar{\eta_4}\nonumber\\
&&+(x_5\pm\sqrt{5\psi}x_1x_2x_3x_4)\eta_5+(x_5\mp\sqrt{5\psi}x_1x_2x_3x_4)\bar{\eta}_5\label{d10-mf}
\end{eqnarray}
The $\eta_i,\bar{\eta}_i$ are boundary fermions satisfying Clifford algebra relations:
\begin{equation}
\{\eta_i,\bar{\eta}_j\}=\delta_{ij}\qquad \{\eta_i,\eta_j\}=0
\end{equation}
The R--charges of the variables $x_i$ are $\frac{2w_i}{d}$ where $w_i$ are their homogeneous weights. The R--charges of the boundary fermions are chosen such that the matrix factorization $Q$ has charge $1$.\\
Note that (\ref{d6-mf})--(\ref{d10-mf}) do not present the only way to incorporate the bulk deformation into the matrix factorizations of this type. Throughout this paper we will use the above expressions whenever speaking of the bulk deformed matrix factorizations. Let us mention that none of the above matrix factorizations has the structure of the factorization for the quintic given in \cite{Morrison:2007bm}. This particular form of matrix factorization is actually quite special and we have only found it for $d=8$ and $k_i=3$:
\begin{equation}
\tilde{Q}^{d=8}_{\pm}=\sum_{i=1}^4(x_i^3\eta_i+x_i^{5}\bar{\eta}_i)+(x_5\eta_5+x_5\bar{\eta}_5)\pm2\sqrt{\psi}\prod_{i=1}^4(\eta_i-x_i^2\bar{\eta}_i)(\eta_5-\bar{\eta}_5)
\end{equation}
At the Gepner point this matrix factorization can be identified with the $L=(2,2,2,2,0)$ boundary state.
\subsection{Which Branes have Moduli?}
\label{sec:moduli}
Only matrix factorizations with obstructed brane moduli can lead to a discrete number of brane vacua which are separated by domain walls. In order to find brane moduli one starts at the Gepner point and looks for open string states which are valid boundary deformations. A simultaneous bulk deformation will in general obstruct these boundary deformations. The information about the obstructions is encoded in the critical locus of the effective superpotential $\mathcal{W}_{eff}$. This will be the subject of the next section. In this section we confine ourselves to some rather trivial technical observations on how to assemble open string moduli from minimal model open string states.\\
In the following we will focus on tensor product branes. These B--branes are tensor products of boundary conditions of the minimal model components, which, in addition, have to be invariant under an orbifold action of a finite group $G_{\text{GP}}$ which is determined by the Greene--Plesser construction of mirror symmetry \cite{Greene:1990ud}. Orbifold invariance greatly constrains the number of possible open string states. So, when making an ansatz for an allowed open string state it is essential that the constrains coming from the orbifold are included.\\
What we are interested in are (at least at first order) marginal deformations of a matrix factorization, i.e. open string states which have R--charge 1 and odd $\mZ_2$--degree. This leads to the following obvious criteria on the minimal model components:
\begin{itemize}
\item The R--charges of the minimal model components of the open string state have to add up to $1$.
\item In order for the $\mZ_2$--degree to be odd we must compose the open string state of an odd number of fermionic minimal model components.
\end{itemize} 
These restrictions are usually not strong enough for practical purposes -- for the models we discuss here the number of marginal boundary fermions may still be of order a hundred. What cuts down this number to a handful is the orbifold condition. If we are interested in obstructed deformations there are some additional constraints. See \cite{Hori:2004ja} for examples of obstructed and unobstructed boundary deformations. A boundary deformation is obstructed at second order when the (Massey) product of the associated open string state with all the other open string states gives a $\mZ_{2}$--even open string state. In particular these bosonic open string states may be bulk deformations $\Phi_i$ which are also in the boundary cohomology, i.e. $\Psi_i=\Phi_i\cdot\mathbbm{1}$. These are responsible for the fact that obstructed boundary parameters can be expressed in terms (unobstructed) bulk parameters via the relations defined by the critical locus of $\mathcal{W}_{eff}$. If one is specifically interested in boundary deformations leading to a cubic effective superpotential and therefore to a simple domain wall structure, we get additional constraints on the form of the marginal boundary deformations:
\begin{itemize}
\item In at least one open string state all the $x_i$ which appear in the bulk deformation have to appear. This is a necessary condition for the bulk moduli to enter the effective superpotential and for $\mathcal{W}_{eff}$ to be cubic\footnote{Only then it is possible that an open string state squares to the bulk deformation. If the equations for the critical locus do not contain bulk parameters the generic solution of the equation is that all boundary parameters are 0. In order for the effective superpotential to be cubic all Massey products must give obstructions or $0$ at order $2$ in deformation theory. So, in particular, one open string state must square to a bulk deformation.}.
\item In at least one open string state the powers of the $x_i$ must not be higher than the $x_i$--powers in the bulk deformations.
\end{itemize}
Let us now focus on the deformations we have in the one--parameter hypersurfaces in weighted $\mP^4$. There are only two possibilities: $\Phi_1=x_1x_2x_3x_4x_5$ or $\Phi_2=x_1^2x_2^2x_3^2x_4^2$. In order to find marginal deformations which are obstructed at order $2$ we have to search for minimal model open string states which are either linear or quadratic in the $x_i$. From these conditions we will also get constraints for the form of the matrix factorizations we have to use. Let us thus consider a minimal model with superpotential \begin{equation}
W=x^d
\end{equation}
and a generic matrix factorization
\begin{equation}
Q^{(k)}=\left(\begin{array}{cc}
0&x^k\\
x^{d-k}&0
\end{array}
\right)
\end{equation}
Without loss of generality we will assume that $k\leq d-k$.\\
Let us first discuss the fermionic open string states. These are:
\begin{equation}
\psi_l=\left(\begin{array}{cc}
0&x^{l}\\
-x^{d-2k+l}&0
\end{array}\right)
\qquad l=0,\ldots,k-1
\end{equation}
The $R$--charges of these fermions are $q_{\psi_l}=\frac{d-2k+2l}{d}$. Note that for our choice for $k$, we have $l\leq d-2k+l$ which means that the exponent of the lower left entry of the matrix is always greater or equal to the exponent of the upper right entry. \\
What are now the fermionic building blocks we can have when we have the bulk deformations $\Phi_{1,2}$? It is easy to see that we can only have $l=0,1,2$. It makes sense to treat the cases $d=even$ and $d=odd$ separately. \\
For odd $l$ the only interesting open string states are those for $l=0,1$:
\begin{equation}
\psi_{0}=\left(\begin{array}{cc}
0&1\\
-x^{d-2k}&0
\end{array}\right)\qquad
\psi_{1}=\left(\begin{array}{cc}
0&x\\
-x^{d-2k+1}&0
\end{array}\right)
\end{equation}
In order to satisfy the criteria above we must have $d-2k=1$, which shows that for every odd $d$ there is only one matrix factorization which leads to the desired open string state. In addition to that the state $\psi_2$ only exists if $k\geq 2$ which means that $d\geq 5$.\\
Let us now discuss the case $k=even$. There, the following fermionic open string states are of interest:
\begin{equation}
\psi_{0}=\left(\begin{array}{cc}
0&1\\
-x^{d-2k}&0
\end{array}\right)\qquad
\psi_{1}=\left(\begin{array}{cc}
0&x\\
-x^{d-2k+1}&0
\end{array}\right)\qquad
\psi_{2}=\left(\begin{array}{cc}
0&x^2\\
-x^{d-2k+2}&0
\end{array}\right)
\end{equation}
Since we do not allow $x$--powers in our open string states which are higher than those appearing in the bulk deformation, $\psi_1$ and $\psi_2$ only need to be considered if $k=\frac{d}{2}$. For our bulk deformations we must have $d-2k=2$ which only works if $d\geq 4$.\\
The bosonic open string states have a simpler structure:
\begin{equation}
\phi_l=\left(\begin{array}{cc}
x^l&0\\
0&x^l
\end{array}\right)\qquad l=0,\ldots,k-1
\end{equation}
The $R$--charges are $q_{\phi_r}=\frac{2l}{d}$.\\\\
Out of this reasoning we can make an interesting observation for the degree $8$ hypersurface. Since all the $d_i$ of the minimal model components are even and the $x_5$--variable only appears quadratic in the Landau--Ginzburg superpotential it is impossible to find a charge $1$ fermionic open string state\footnote{Actually there won't be any open string state where $x_5$ appears.} which squares to the deformation $\Phi_1=x_1x_2x_3x_4x_5$. The situation is entirely different in we choose the bulk deformation $\Phi_2=x_1^2x_2^2x_3^2x_4^2$. In the bulk theory these two deformations would be equivalent modulo the equations of motion, when we have a boundary this situation is different.\\
We can apply similar arguments to the degree $6$ hypersurface. In this model we can only have the bulk deformation $\Phi_1=x_1x_2x_3x_4x_5$. Since the minimal model superpotential for the $x_5$--component is cubic, we can only get an $x_5$ into an open string state through the charge $\frac{1}{3}$ fermion $\left(\begin{array}{cc}0&1\\-x_5&0\end{array}\right)$. Therefore the $R$--charges of the other minimal model components must add up to $\frac{2}{3}$. But the even and odd open string states with only linear $x_i$--entries have $R$--charge $\frac{2}{3}$ and $\frac{1}{3}$, respectively. From this we can conclude that the boundary deformations cannot be of the structure that one marginal open string state squares to the bulk deformation which tells us that the effective superpotential will not be a cubic polynomial. 
\subsection{Obstructions and the Effective superpotential}
\label{sec:obstr-effect-superp}
In this section we will make a systematic search for open string moduli on the (mirror) hypersurfaces and compute the effective superpotential by computing Massey products using an algorithm described in \cite{siqveland,Knapp:2006rd}. We will refrain from describing the algorithm here since the structure of the branes is so simple that we do not need the technical details. In the previous section we have discussed certain conditions on the minimal model open string states in order for the effective superpotential to be cubic. It actually turns out that if a brane has obstructed deformations, the effective superpotential encoding the obstructions is either cubic or bicubic. We will now give an example for each case. Similar discussions for the quintic can be found in \cite{Hori:2004ja,Hori:2004zd}. \\
\subsubsection{The $d=8$ brane $L=(3,3,2,1,0)$}
This is an example for a boundary state which yields a cubic effective superpotential. This boundary state and its associated matrix factorization will be our main example throughout the paper. The reason for this is that we can manage to cancel the tadpoles by adding an O9--plane, as we will show later. We consider the Gepner point $\psi=0$ of the mirror hypersurface where we have to take into account a $(\mZ_{8})^2\times\mZ_2$ orbifold. At the Gepner point the matrix factorizations can be decomposed as follows in terms of the minimal model components:
\begin{equation}
Q_0=\left(\begin{array}{cc}
0&x_1^4\\
x_1^4&0
\end{array}\right)\otimes
\left(\begin{array}{cc}
0&x_2^4\\
x_2^4&0
\end{array}\right)\otimes
\left(\begin{array}{cc}
0&x_3^3\\
x_3^5&0
\end{array}\right)\otimes
\left(\begin{array}{cc}
0&x_4^2\\
x_4^6&0
\end{array}\right)\otimes
\left(\begin{array}{cc}
0&x_5\\
x_5&0
\end{array}\right)
\end{equation}
Here the $\otimes$ is understood as a graded tensor product.
There is only one marginal orbifold invariant boundary operator given by the tensor product of four even $R$--charge $\frac{1}{4}$ open string states and one charge $0$ odd open string state of the five minimal models:
\begin{equation}
\Psi=\left(\begin{array}{cc}
x_1&0\\
0&x_1
\end{array}\right)\otimes
\left(\begin{array}{cc}
x_2&0\\
0&x_2
\end{array}\right)\otimes
\left(\begin{array}{cc}
x_3&0\\
0&x_3
\end{array}\right)\otimes
\left(\begin{array}{cc}
x_4&0\\
0&x_4
\end{array}\right)\otimes
\left(\begin{array}{cc}
0&1\\
-1&0
\end{array}\right)
\end{equation}
We observe that 
\begin{equation}
\Psi^2=-x_1^2x_2^2x_3^2x_4^2,
\end{equation}
which is precisely the bulk deformation. If we combine this deformation with a bulk deformation we get constraints on the bulk-- and boundary operators. If we deform $Q_0$ with the marginal operator,
\begin{equation}
Q=Q_0+u\Psi,
\end{equation}
$Q$ will square to the deformed Landau--Ginzburg superpotential if and only if 
\begin{equation}
u^2-4\psi=0
\end{equation}
The obstructions to the deformations of branes are encoded by the critical locus of the effective superpotential and the above relation can be integrated to give:
\begin{equation}
\mathcal{W}_{eff}=\frac{1}{3}u^3-4u\psi
\end{equation}
From this, we see that at $\psi\neq 0$ the boundary modulus $u$ can only have two values corresponding to two different vacua. This is evidence that this brane may support structure like normal functions and domain walls similar to the quintic.
\subsubsection{The $d=8$ brane $L=(2,2,2,2,0)$}
This brane has a bicubic superpotential as we will now show. At the Gepner point the associated matrix factorization looks as follows:
\begin{equation}
Q_0=\left(\begin{array}{cc}
0&x_1^3\\
x_1^5&0
\end{array}\right)\otimes
\left(\begin{array}{cc}
0&x_2^3\\
x_2^5&0
\end{array}\right)\otimes
\left(\begin{array}{cc}
0&x_3^3\\
x_3^5&0
\end{array}\right)\otimes
\left(\begin{array}{cc}
0&x_4^3\\
x_4^5&0
\end{array}\right)\otimes
\left(\begin{array}{cc}
0&x_5\\
x_5&0
\end{array}\right)
\end{equation}
There are, up to $Q_0$--exact pieces, only two orbifold invariant open string states.  The first one comes from the tensor product of four charge $\frac{1}{4}$ open string fermions from the four $A_6$--components of the Gepner model and the charge 0 fermion from the $A_0$--piece:
\begin{equation}
\Psi_1=\left(\begin{array}{cc}
0&1\\
-x_1^2&0
\end{array}\right)\otimes
\left(\begin{array}{cc}
0&1\\
-x_2^2&0
\end{array}\right)\otimes
\left(\begin{array}{cc}
0&1\\
-x_3^2&0
\end{array}\right)\otimes
\left(\begin{array}{cc}
0&1\\
-x_4^2&0
\end{array}\right)\otimes
\left(\begin{array}{cc}
0&1\\
-1&0
\end{array}\right)
\end{equation}
The second marginal fermionic open string state corresponds to a tensor product of four charge $\frac{1}{4}$ bosons:
\begin{equation}
\Psi_2=\left(\begin{array}{cc}
x_1&0\\
0&x_1
\end{array}\right)\otimes
\left(\begin{array}{cc}
x_2&0\\
0&x_2
\end{array}\right)\otimes
\left(\begin{array}{cc}
x_3&0\\
0&x_3
\end{array}\right)\otimes
\left(\begin{array}{cc}
x_4&0\\
0&x_4
\end{array}\right)\otimes
\left(\begin{array}{cc}
0&1\\
-1&0
\end{array}\right)
\end{equation}
We are now ready to calculate the superpotential from this data. The first step is to deform the matrix factorization:
\begin{equation}
Q=Q_0+u_1\Psi_1+u_2\Psi_2
\end{equation}
The matrix factorization condition is:
\begin{equation}
\label{mfcond}
Q^2\stackrel{!}{=}W_0-4 \psi x_1^2x_2^2x_3^2x_4^2 +\sum_if_i(u_i,\psi)\Phi_i,
\end{equation}
where $W_0$ is the Landau--Ginzburg superpotential at the Gepner point and $\Phi_i$ are bosonic open string states. We could actually also absorb the second summand into the last terms since we can write the bulk deformation as an open string state\footnote{It is easy to check that this state is actually in the boundary open string spectrum.} $\Phi_0=x_1^2x_2^2x_3^2x_4^2\mathbbm{1}$. For the matrix factorization condition to be satisfied we must impose   $f_i(u_i,\psi)=0$. These vanishing relations determine the critical locus of the effective superpotential \cite{siqveland,Knapp:2006rd}.\\
In order to determine the obstructions to the deformations with bulk and boundary operators we compute the 'matric Massey products':
\begin{eqnarray}
\Psi_1\Psi_1&=&-x_1^2x_2^2x_3^2x_4^2\nonumber\\
\Psi_2\Psi_2&=&-x_1^2x_2^2x_3^2x_4^2
\end{eqnarray}
Furthermore we have to compute $\Psi_1\Psi_2+\Psi_2\Psi_1$. The calculation gives a $\mZ_2$--even state with $R$--charge $2$. At the Gepner point, this state is not $Q_0$--exact but $Q_0$--closed and therefore must be a bosonic open string state. One easily checks that this state is also orbifold invariant. So this symmetric product of open string states will give a contribution to the vanishing relations in the third term of (\ref{mfcond}). The deformation theory algorithm then implies that there are no higher Massey products to compute.\\
We therefore conclude that the deformations are fully obstructed at order 2 and the critical locus of the effective superpotential is:
\begin{eqnarray}
\label{critloc}
u_1^2+u_2^2-4\psi&=&0\nonumber\\
u_1u_2&=&0
\end{eqnarray}
These conditions have four solutions:
\begin{eqnarray}
\label{vacua}
u_1=0&\qquad&u_2=\pm 2\sqrt{\psi}\nonumber\\
u_2=0&\qquad&u_1=\pm2\sqrt{\psi}
\end{eqnarray}
In order to be sure that the equations (\ref{critloc}) really describe the minima of an effective superpotential we should also check if we can actually get this potential by integrating them. The canonical approach to get the the effective superpotential is to integrate homogeneous linear combinations of the vanishing relations and to determine the free coefficients by requiring that the second order derivatives with respect to the boundary parameters $u_{1,2}$ match. Doing this, one gets a symmetric form for the effective superpotential:
\begin{equation}
\label{weffsym}
\mathcal{W}_{eff}^{sym}=\frac{u_1^3}{3}+\frac{u_2^3}{3}+u_1^2u_2+u_1u_2^2-4\psi(u_1+u_2)
\end{equation}
The critical locus of this effective superpotential gives the equation
\begin{equation}
(u_1+u_2)^2-4\psi\stackrel{!}{=}0,
\end{equation}
which not only has the solutions (\ref{vacua}) but also an additional pair of solutions $u_1=u_2=\pm\sqrt{\psi}$. There are two more choices of effective superpotentials whose critical locus is exactly (\ref{critloc}):
\begin{eqnarray}
\label{weff12}
\mathcal{W}_{eff}^1&=&\frac{u_1^3}{3}+u_1u_2^2-4\psi u_1\nonumber\\
\mathcal{W}_{eff}^2&=&\frac{u_2^3}{3}+u_1^2u_2-4\psi u_2
\end{eqnarray}
These three results are actually equivalent in the sense that they can be related via field redefinitions. We can get (\ref{weff12}) out of (\ref{weffsym}) by applying the transformations $\{u_1\rightarrow u_1-u_2,u_2\rightarrow u_2\}$ and $\{u_1\rightarrow u_1,u_2\rightarrow u_2-u_1\}$\footnote{In general, i.e. if we consider also massive deformations, which means that the parameters $u_i$ have different degrees of homogeneity, these transformations become non--linear.}.\\
These ambiguities are due to a gauge freedom which cannot be fixed in the topological sector \cite{Herbst:2004jp,Ashok:2004xq}. This reflects the presence of "$A_{\infty}$--morphisms" in the underlying $A_{\infty}$--category. In the language of matrix factorizations this can be traced back to an ambiguity in choosing open string states and higher order deformations. In this particular example we could have as well chosen some linear combinations of the boundary fermions $\Psi_{1/2}$ as our basis of open string states. But here we {\em have} chosen a particular set, so we should at least find out which of the above realizations of the effective superpotential fits to our choice of open string states. In this particular example we can check this explicitly. We use the fact that the effective superpotential can also be interpreted as the generating functional of disk amplitudes. Amplitudes which do not have any integrated insertions, i.e. amplitudes with one bulk and one boundary insertion or amplitudes with three boundary insertions, can be evaluated explicitly using the residue formula of Kapustin and Li \cite{Kapustin:2003ga}. In this case we can even determine the full superpotential by computing correlators. With our choice of boundary fermions, we find:
\begin{eqnarray}
\langle \Phi\Psi_1\rangle&=&-1\nonumber \\
\langle \Phi\Psi_2\rangle&=&0\nonumber\\
\langle \Psi_1\Psi_1\Psi_1\rangle&=&1 \nonumber \\
\langle \Psi_1\Psi_1\Psi_2\rangle&=&1\nonumber\\
\langle\Psi_1\Psi_2\Psi_2\rangle=\langle\Psi_2\Psi_2\Psi_2\rangle&=&0
\end{eqnarray}
This picks the second choice of effective superpotential in (\ref{weff12}) as the one compatible with our choice of open string states.
\subsection{Moduli and $\mathcal{W}_{eff}$ for all tensor product branes}
\label{sec:moduli-mathc-all}
We now list all tensor product branes which have moduli and compute the effective superpotential. The boundary states which do not appear in the tables below do not have any moduli.
\subsubsection{$d=6$}
Our systematic search shows that, after implementing the $(\mZ_6)^2\times\mZ_3$--orbifold, which puts us on the mirror there are no fermionic charge $1$ open string states left. Thus, none of the $d=6$ boundary states we have considered has open string moduli.
\subsubsection{$d=8$}
We collect the data about the boundary states with moduli in table \ref{tabd8mod}. We have taken into account the $\mZ_8^2\times\mZ_2$ orbifold action with generators
\begin{eqnarray}
g'_1:&&(6,1,1,0,0)\nonumber\\
g'_2:&&(3,1,0,0,4)\nonumber\\
g'_3:&&(4,0,0,0,4),
\end{eqnarray}
where $g_j': x_i\rightarrow e^{2\pi i g'_{j,i}/d}x_i$.\\
We give the structure of the boundary states by giving $R$--charge and $\mZ_2$--degree of each minimal model component formatted as $\mathrm{R}^{\mZ_2}$. 
\begin{table}
\begin{center}
\begin{tabular}{|c||c|c|c|}
\hline
Boundary state & Number of Moduli & Structure of Moduli & $\mathcal{W}_{eff}$\\
\hline\hline
$(1,1,1,1,0)$&$1$&$\frac{1}{4}^0\otimes\frac{1}{4}^0\otimes\frac{1}{4}^0\otimes\frac{1}{4}^0\otimes0^1$&cubic\\\hline
$(2,1,1,1,0)$&$1$&$\frac{1}{4}^0\otimes\frac{1}{4}^0\otimes\frac{1}{4}^0\otimes\frac{1}{4}^0\otimes0^1$&cubic\\\hline
$(3,1,1,1,0)$&$1$&$\frac{1}{4}^0\otimes\frac{1}{4}^0\otimes\frac{1}{4}^0\otimes\frac{1}{4}^0\otimes0^1$&cubic\\\hline
$(2,2,1,1,0)$&$1$&$\frac{1}{4}^0\otimes\frac{1}{4}^0\otimes\frac{1}{4}^0\otimes\frac{1}{4}^0\otimes0^1$&cubic\\\hline
$(3,2,1,1,0)$&$1$&$\frac{1}{4}^0\otimes\frac{1}{4}^0\otimes\frac{1}{4}^0\otimes\frac{1}{4}^0\otimes0^1$&cubic\\\hline
$(2,2,2,1,0)$&$1$&$\frac{1}{4}^0\otimes\frac{1}{4}^0\otimes\frac{1}{4}^0\otimes\frac{1}{4}^0\otimes0^1$&cubic\\\hline
$(3,2,2,1,0)$&$1$&$\frac{1}{4}^0\otimes\frac{1}{4}^0\otimes\frac{1}{4}^0\otimes\frac{1}{4}^0\otimes0^1$&cubic\\\hline
$(3,3,2,1,0)$&$1$&$\frac{1}{4}^0\otimes\frac{1}{4}^0\otimes\frac{1}{4}^0\otimes\frac{1}{4}^0\otimes0^1$&cubic\\\hline
$(3,3,3,1,0)$&$1$&$\frac{1}{4}^0\otimes\frac{1}{4}^0\otimes\frac{1}{4}^0\otimes\frac{1}{4}^0\otimes0^1$&cubic\\\hline
$(2,2,2,2,0)$&$2$&$\begin{array}{c}\frac{1}{4}^0\otimes\frac{1}{4}^0\otimes\frac{1}{4}^0\otimes\frac{1}{4}^0\otimes0^1\\ \frac{1}{4}^1\otimes\frac{1}{4}^1\otimes\frac{1}{4}^1\otimes\frac{1}{4}^1\otimes0^1\end{array}$&bicubic\\\hline
$(3,2,2,2,0)$&$2$&$\begin{array}{c}\frac{1}{4}^0\otimes\frac{1}{4}^0\otimes\frac{1}{4}^0\otimes\frac{1}{4}^0\otimes0^1\\ \frac{1}{4}^1\otimes\frac{1}{4}^1\otimes\frac{1}{4}^1\otimes\frac{1}{4}^1\otimes0^1\end{array}$&bicubic\\\hline
$(3,3,2,2,0)$&$2$&$\begin{array}{c}\frac{1}{4}^0\otimes\frac{1}{4}^0\otimes\frac{1}{4}^0\otimes\frac{1}{4}^0\otimes0^1\\ \frac{1}{4}^1\otimes\frac{1}{4}^1\otimes\frac{1}{4}^1\otimes\frac{1}{4}^1\otimes0^1\end{array}$&bicubic\\\hline
$(3,3,3,2,0)$&$2$&$\begin{array}{c}\frac{1}{4}^0\otimes\frac{1}{4}^0\otimes\frac{1}{4}^0\otimes\frac{1}{4}^0\otimes0^1\\ \frac{1}{4}^1\otimes\frac{1}{4}^1\otimes\frac{1}{4}^1\otimes\frac{1}{4}^1\otimes0^1\end{array}$&bicubic\\\hline
$(3,3,3,3,0)$&$2$&$\begin{array}{c}\frac{1}{4}^0\otimes\frac{1}{4}^0\otimes\frac{1}{4}^0\otimes\frac{1}{4}^0\otimes0^1\\ \frac{1}{4}^1\otimes\frac{1}{4}^1\otimes\frac{1}{4}^1\otimes\frac{1}{4}^1\otimes0^1\end{array}$&bicubic\\\hline
\end{tabular}
\end{center}\caption{Tensor product branes with moduli for $d=8$.}\label{tabd8mod}
\end{table}
\subsubsection{$d=10$}
The list of tensor product branes with moduli can be found in table \ref{tabd10mod}. We have taken into account the $(\mZ_{10})^2$ orbifold action with generators
\begin{eqnarray}
g_1:&&(1,9,0,0,0)\nonumber\\
g_2:&&(1,0,9,0,0).
\end{eqnarray}
\begin{table}
\begin{center}
\begin{tabular}{|c||c|c|c|}
\hline
Boundary state & Number of Moduli & Structure of Moduli & $\mathcal{W}_{eff}$\\
\hline\hline
$(1,1,1,1,0)$&$1$&$\frac{1}{5}^0\otimes\frac{1}{5}^0\otimes\frac{1}{5}^0\otimes\frac{2}{5}^0\otimes0^1$&cubic\\\hline
$(2,1,1,1,0)$&$1$&$\frac{1}{5}^0\otimes\frac{1}{5}^0\otimes\frac{1}{5}^0\otimes\frac{2}{5}^0\otimes0^1$&cubic\\\hline
$(3,1,1,1,0)$&$1$&$\frac{1}{5}^0\otimes\frac{1}{5}^0\otimes\frac{1}{5}^0\otimes\frac{2}{5}^0\otimes0^1$&cubic\\\hline
$(4,1,1,1,0)$&$1$&$\frac{1}{5}^0\otimes\frac{1}{5}^0\otimes\frac{1}{5}^0\otimes\frac{2}{5}^0\otimes0^1$&cubic\\\hline
$(3,2,1,1,0)$&$1$&$\frac{1}{5}^0\otimes\frac{1}{5}^0\otimes\frac{1}{5}^0\otimes\frac{2}{5}^0\otimes0^1$&cubic\\\hline
$(4,2,1,1,0)$&$1$&$\frac{1}{5}^0\otimes\frac{1}{5}^0\otimes\frac{1}{5}^0\otimes\frac{2}{5}^0\otimes0^1$&cubic\\\hline
$(4,3,1,1,0)$&$1$&$\frac{1}{5}^0\otimes\frac{1}{5}^0\otimes\frac{1}{5}^0\otimes\frac{2}{5}^0\otimes0^1$&cubic\\\hline
$(2,2,2,1,0)$&$1$&$\frac{1}{5}^0\otimes\frac{1}{5}^0\otimes\frac{1}{5}^0\otimes\frac{2}{5}^0\otimes0^1$&cubic\\\hline
$(3,2,2,1,0)$&$1$&$\frac{1}{5}^0\otimes\frac{1}{5}^0\otimes\frac{1}{5}^0\otimes\frac{2}{5}^0\otimes0^1$&cubic\\\hline
$(4,2,2,1,0)$&$1$&$\frac{1}{5}^0\otimes\frac{1}{5}^0\otimes\frac{1}{5}^0\otimes\frac{2}{5}^0\otimes0^1$&cubic\\\hline
$(3,3,2,1,0)$&$1$&$\frac{1}{5}^0\otimes\frac{1}{5}^0\otimes\frac{1}{5}^0\otimes\frac{2}{5}^0\otimes0^1$&cubic\\\hline
$(4,3,2,1,0)$&$1$&$\frac{1}{5}^0\otimes\frac{1}{5}^0\otimes\frac{1}{5}^0\otimes\frac{2}{5}^0\otimes0^1$&cubic\\\hline
$(4,4,2,1,0)$&$1$&$\frac{1}{5}^0\otimes\frac{1}{5}^0\otimes\frac{1}{5}^0\otimes\frac{2}{5}^0\otimes0^1$&cubic\\\hline
$(3,3,3,1,0)$&$2$&$\begin{array}{c}\frac{1}{5}^0\otimes\frac{1}{5}^0\otimes\frac{1}{5}^0\otimes\frac{2}{5}^0\otimes0^1\\\frac{1}{5}^1\otimes\frac{1}{5}^1\otimes\frac{1}{5}^1\otimes\frac{2}{5}^0\otimes 0^0 \end{array}$&bicubic\\\hline
$(4,3,3,1,0)$&$2$&$\begin{array}{c}\frac{1}{5}^0\otimes\frac{1}{5}^0\otimes\frac{1}{5}^0\otimes\frac{2}{5}^0\otimes0^1\\\frac{1}{5}^1\otimes\frac{1}{5}^1\otimes\frac{1}{5}^1\otimes\frac{2}{5}^0\otimes 0^0 \end{array}$&bicubic\\\hline
$(4,4,3,1,0)$&$2$&$\begin{array}{c}\frac{1}{5}^0\otimes\frac{1}{5}^0\otimes\frac{1}{5}^0\otimes\frac{2}{5}^0\otimes0^1\\\frac{1}{5}^1\otimes\frac{1}{5}^1\otimes\frac{1}{5}^1\otimes\frac{2}{5}^0\otimes 0^0 \end{array}$&bicubic\\\hline
$(4,4,4,1,0)$&$2$&$\begin{array}{c}\frac{1}{5}^0\otimes\frac{1}{5}^0\otimes\frac{1}{5}^0\otimes\frac{2}{5}^0\otimes0^1\\\frac{1}{5}^1\otimes\frac{1}{5}^1\otimes\frac{1}{5}^1\otimes\frac{2}{5}^0\otimes 0^0 \end{array}$&bicubic\\\hline
\end{tabular}
\end{center}\caption{Tensor product branes with moduli for $d=10$.}\label{tabd10mod}
\end{table}
\subsubsection{Comments}
Tables \ref{tabd8mod} and \ref{tabd10mod} show that the general structure of the open string moduli and the shape of the effective superpotential is always the same. Note however that the open string states on the various branes have different matrix entries since the fermionic minimal model state which has the required $R$--charge looks different for every degree and $L$--label. \\
The bosonic open string states have the same structure and charge for every $L$--label, only their number increases for increasing $L$. Bosonic open string states with linear entries in the $x_i$ appear as soon as $L\geq1$. It so happens that they also have the right $R$--charge in $d=8$ and $d=10$ such that four of them can be tensored to give, together with the charge $0$ fermionic state of the $x_5^2$ piece, a modulus. Whenever there is only one modulus this is made up of these bosonic minimal model components. \\
If the $L$--label is high enough there are also fermionic minimal model states which have the correct $R$--charge. These can then in principle be tensored in every possible combination with the bosonic minimal model states of correct charge. One would therefore naively expect much more moduli than just two. The reason that there are at most two boundary moduli on our branes is due to the orbifold actions which allow only for highly symmetric combinations of the minimal model components. 
\section{Geometric Boundary Conditions and Normal Functions}
\label{sec:geom-bound-cond}

In this section we discuss how to extract the complexes $E_\pm$ of holomorphic vector bundles on $X$ and the geometric boundary conditions $C_\pm$ from the matrix factorization $Q_\pm$. Although this is not strictly necessary to determine the algebraic second Chern class which tells us about the existence of a normal function, we can get valuable information from the bundle data. For instance, we can check whether the charges of the given brane can be cancelled by a suitable choice of orientifolds. We discuss a certain class of O--planes in the appendix. In this section we furthermore compute the algebraic second Chern class $C_{\pm}=\ch_2^{\text{alg}}(E_\pm)$ for two branes of the $d=8$ and $d=10$ hypersurfaces. The branes are chosen by the conditions that they have a cubic effective superpotential and a tadpole cancellation condition which is satisfied by the simplest O--plane configuration we could find.\\

Unlike in the other sections of this work, we will deal here with B--branes $E$ on $X$. The reason is that $Y$ can be obtained as an orbifold of $X$ by the Greene--Plesser group $G_{\text{GP}}$, see~(\ref{eq:the_Ys}). This enormously simplifies our lives since we will have to vary only one K\"ahler parameter. At the end of this procedure, we will view the complexes $E_\pm$ as complexes on the singular space $X/G_{\text{GP}}$, and hence $C_{\pm}$ as curves on this singular space. In order to make them into B--branes on $Y$ we will have to resolve the singularities of $X/G_{\text{GP}}$. This is then the topic of Section~\ref{sec:toric}. 

\subsection{Calabi--Yau/Landau--Ginzburg Correspondence with Branes}
\label{sec:LG-CY}
The authors of \cite{Herbst:2008jq} give an explicit algorithm how to extract geometric data out of a matrix factorization by making a detour through the linear sigma model. The algorithm can be implemented in the following steps:
\begin{itemize}
\item Determine the $R$--charges of the matrix factorization and take into account the twisted sectors of the $\mZ_d$ orbifold action. The representation of the orbifold group on the matrix factorizations is related to the $R$--charges in the following way \cite{Walcher:2004tx}:
\begin{equation}
\label{rorb}
\gamma^i=\sigma e^{i\pi R}e^{-i\pi\varphi^i},
\end{equation}
where $\sigma=\mathrm{diag}(\mathbbm{1}_r,-\mathbbm{1}_r)$ and the $\varphi^i$ are determined by the condition $(\gamma^i)^d=\mathbbm{1}$. This gives $d$ $\mZ_d$ equivariant matrix factorizations with $R$--charges shifted by the values of $\varphi^i$. The branes in the twisted sector are in one-to-one correspondence with  boundary states $|L,M,S\rangle$ with non--zero $M$--labels. This gives $R$--charges $R_n$, where $n$ labels the twisted sectors.
\item Going from the Landau--Ginzburg model to the Calabi--Yau manifold we have to pass through the conifold point. In order to safely get through the singularity we have to apply the 'grade restriction rule'. Defining $S:=\sum_{Q_i>0}Q_i$, where the $Q_i$ are the positive linear sigma model charges, we define a set of integers,
\begin{equation}
\label{grr}
\Lambda=
\left\{q\in\mZ| -\frac{S}{2}<\frac{\theta}{2}+q<\frac{S}{2}\right\},
\end{equation}
for any given $\theta$ in a 'window' of length $2\pi$. An appropriate choice of this window always allows us to set $\Lambda=\{0,\ldots,d-1\}$ with $d$ the degree of the hypersurface equation.
\item Determine the linear sigma model charges $(\tilde{R}_n,q_n)$ which are defined via the following relations:
\begin{equation}
R_n=\tilde{R}_n-\frac{2q_n}{d}
\end{equation}
 The $\tilde{R}_n$ are integers with values $\tilde{R}_n=s\:\:\mathrm{mod}\:\:2$  where $s$ is even or odd depending on the $\mZ_2$--degree of the composite of boundary fermions, and $q_n\in\Lambda$.
\item Construct a semi--infinite complex by placing $\mathcal{O}(q_n+dk)^{\oplus m}$, where $m$ is the multiplicity of the charge $\tilde{R}_n$, at the position (i.e. the homological degree) $\mathrm{deg}=\tilde{R_n}+2k$, where $k$ goes from $0$ to $\infty$.
\item From these complexes one can extract the bundle data using '$q$--isomorphisms' which relate the infinite complexes to finite ones. For tensor product branes this is easily done by subtracting the complex associated to a suitable trivial brane in the linear sigma model.
\end{itemize}
Note that for this procedure it does not matter whether the marginal bulk deformation $\psi$ is turned on or not since only the $R$--charges of the boundary fermions enter in the calculation.\\ 
We will now perform these steps for two branes on the $d=8$ and $d=10$ hypersurfaces.
\subsubsection{The $d=8$ brane $L=(3,3,2,1,0)$}
The $R$--charges of the boundary fermions $\eta_i$, $\bar{\eta}_i$ can be read off from the matrix factorization:
\begin{equation}
Q=\sum_{i=1}^2(x_i^4\eta_i+x_i^4\bar{\eta}_i)+x_3^3\eta_3+x_3^5\bar{\eta}_3+x_4^2\eta_4+x_4^6\bar{\eta}_4+x_5\eta_5+x_5\bar{\eta}_5
\end{equation}
We have listed the $R$--charges in table \ref{fercharge-tab}.
\begin{table}
\begin{center}
\begin{tabular}{|c|c|c|c|c|c|c|c|c|c|}
\hline
$\eta_1$&$\bar{\eta}_1$&$\eta_2$&$\bar{\eta}_2$&$\eta_1$&$\bar{\eta}_3$&$\eta_4$&$\bar{\eta}_4$&$\eta_5$&$\bar{\eta}_5$\\
\hline
$0$&$0$&$0$&$0$&$\frac{1}{4}$&$-\frac{1}{4}$&$\frac{1}{2}$&$-\frac{1}{2}$&$0$&$0$\\
\hline
\end{tabular}\caption{$R$--charges of the boundary fermions of the $L=(3,3,2,1,0)$ boundary state.}\label{fercharge-tab}
\end{center}
\end{table}
Using the procedure of \cite{Herbst:2008jq}, we get the following semi--infinite complexes describing the D--branes in the twisted sectors, labeled by $n$, in the geometric regime\footnote{The numbers in brackets denote the homological degree of the first term in the complex.}:\\\\
$n=0:\:[0]$\\
\begin{equation}
\label{33210a}
\begin{diagram}
\mathcal{O}(0)^{\oplus 4}&\rTo&\begin{array}{c}\mathcal{O}(4)^{\oplus 4}\\\oplus\\\mathcal{O}(3)^{\oplus 4}\\\oplus\\\mathcal{O}(2)^{\oplus 4}\\\oplus\\\mathcal{O}(1)^{\oplus 4}\end{array}&\rTo&\begin{array}{c}\mathcal{O}(8)^{\oplus 4}\\\oplus\\\mathcal{O}(7)^{\oplus 4}\\\oplus\\\mathcal{O}(6)^{\oplus 4}\\\oplus\\\mathcal{O}(5)^{\oplus 4}\end{array}&\rTo&\begin{array}{c}\mathcal{O}(12)^{\oplus 4}\\\oplus\\\mathcal{O}(11)^{\oplus 4}\\\oplus\\\mathcal{O}(10)^{\oplus 4}\\\oplus\\\mathcal{O}(9)^{\oplus 4}\end{array}&\rTo&\cdots
\end{diagram}
\end{equation}
$n=1:\:[1]$\\
\begin{equation}
\label{33210b}
\begin{diagram}
\begin{array}{c}\mathcal{O}(3)^{\oplus 4}\\\oplus\\\mathcal{O}(2)^{\oplus 4}\\\oplus\\\mathcal{O}(1)^{\oplus 4}\\\oplus\\\mathcal{O}(0)^{\oplus 4}\end{array}&\rTo&\begin{array}{c}\mathcal{O}(7)^{\oplus 4}\\\oplus\\\mathcal{O}(6)^{\oplus 4}\\\oplus\\\mathcal{O}(5)^{\oplus 4}\\\oplus\\\mathcal{O}(4)^{\oplus 4}\end{array}&\rTo&\begin{array}{c}\mathcal{O}(11)^{\oplus 4}\\\oplus\\\mathcal{O}(10)^{\oplus 4}\\\oplus\\\mathcal{O}(9)^{\oplus 4}\\\oplus\\\mathcal{O}(8)^{\oplus 4}\end{array}&\rTo&\cdots
\end{diagram}
\end{equation}
$n=2:\:[1]$\\
\begin{equation}
\label{33210c}
\begin{diagram}
\begin{array}{c}\mathcal{O}(2)^{\oplus 4}\\\oplus\\\mathcal{O}(1)^{\oplus 4}\\\oplus\\\mathcal{O}(0)^{\oplus 4}\end{array}&\rTo&\begin{array}{c}\mathcal{O}(6)^{\oplus 4}\\\oplus\\\mathcal{O}(5)^{\oplus 4}\\\oplus\\\mathcal{O}(4)^{\oplus 4}\\\oplus\\\mathcal{O}(3)^{\oplus 4}\end{array}&\rTo&\begin{array}{c}\mathcal{O}(10)^{\oplus 4}\\\oplus\\\mathcal{O}(9)^{\oplus 4}\\\oplus\\\mathcal{O}(8)^{\oplus 4}\\\oplus\\\mathcal{O}(7)^{\oplus 4}\end{array}&\rTo&\cdots
\end{diagram}
\end{equation}
$n=3:\:[1]$\\
\begin{equation}
\label{33210d}
\begin{diagram}
\begin{array}{c}\mathcal{O}(1)^{\oplus 4}\\\oplus\\\mathcal{O}(0)^{\oplus 4}\end{array}&\rTo&\begin{array}{c}\mathcal{O}(5)^{\oplus 4}\\\oplus\\\mathcal{O}(4)^{\oplus 4}\\\oplus\\\mathcal{O}(3)^{\oplus 4}\\\oplus\\\mathcal{O}(2)^{\oplus 4}\end{array}&\rTo&\begin{array}{c}\mathcal{O}(9)^{\oplus 4}\\\oplus\\\mathcal{O}(8)^{\oplus 4}\\\oplus\\\mathcal{O}(7)^{\oplus 4}\\\oplus\\\mathcal{O}(6)^{\oplus 4}\end{array}&\rTo&\cdots
\end{diagram}
\end{equation}
We do not write down the remaining four complexes since they are the same as the above ones, shifted one position to the right. This is a manifestation of the selfduality of the brane, i.e. the fact that this brane is its own antibrane. \\
We now have to extract the relevant information about the vector bundle. We proceed as described in \cite{Herbst:2008jq} and subtract the semi--infinite complex corresponding to a suitable trivial brane. This brane is given in terms of a matrix factorization in the linear sigma model. For our case, the right choice is the following:
\begin{equation}
Q^{LSM}_{triv}=\sum_{i=1}^2(x_i^4\eta_i+Px_i^4\bar{\eta}_i)+x_3^3\eta_3+Px_3^5\bar{\eta}_3+x_4^2\eta_4+Px_4^6\bar{\eta}_4+x_5\eta_5+Px_5\bar{\eta}_5
\end{equation} 
This matrix factorization contains the $P$--field of the linear sigma model. \\
In order to build up the semi--infinite complex associated to this matrix factorization we have to determine the linear sigma model charges $(\tilde{R}_i,q_i)$ of the boundary fermions. These are determined by the following conditions \cite{Herbst:2008jq}:
\begin{eqnarray}
\tilde{R}(\lambda)Q^{LSM}(\lambda^2P,x_i)\tilde{R}(\lambda)^{-1}&=&\lambda Q(P,x_i)\qquad \Rightarrow\qquad \tilde{R}_i\nonumber\\
\rho(g^{-1})Q(g^{-N}P,g x_i)\rho(g)&=&Q(P,x_i)\qquad \;\;\:\Rightarrow \qquad q_i
\end{eqnarray} 
This yields the results presented in table \ref{ferchargelsm-tab}.
\begin{table}
\begin{center}
\begin{tabular}{|c|c|c|c|c|c|c|c|c|c|c|}
\hline
&$\eta_1$&$\bar{\eta}_1$&$\eta_2$&$\bar{\eta}_2$&$\eta_1$&$\bar{\eta}_3$&$\eta_4$&$\bar{\eta}_4$&$\eta_5$&$\bar{\eta}_5$\\
\hline
$\tilde{R}$&$1$&$-1$&$1$&$-1$&$1$&$-1$&$1$&$-1$&$1$&$-1$\\
\hline
$q$&$-4$&$4$&$-4$&$4$&$-4$&$3$&$-2$&$2$&$-4$&$4$\\
\hline
\end{tabular}\caption{LSM--charges of the trivial brane.}\label{ferchargelsm-tab}
\end{center}
\end{table}
From these data we can extract the following complex:
\begin{equation}
\label{33210triv}
\begin{diagram}
\mathcal{O}(0)&\rTo&\begin{array}{c}\mathcal{O}(4)^{\oplus 3}\\\oplus\\\mathcal{O}(3)\\\oplus\\\mathcal{O}(2)\end{array}&\rTo&\begin{array}{c}\mathcal{O}(8)^{\oplus 4}\\\oplus\\\mathcal{O}(7)^{\oplus 3}\\\oplus\\\mathcal{O}(6)^{\oplus 3}\\\oplus\\\mathcal{O}(5)\end{array}&\rTo&\begin{array}{c}\mathcal{O}(12)^{\oplus 4}\\\oplus\\\mathcal{O}(11)^{\oplus 4}\\\oplus\\\mathcal{O}(10)^{\oplus 4}\\\oplus\\\mathcal{O}(9)^{\oplus 3}\end{array}&\rTo&\begin{array}{c}\mathcal{O}(16)^{\oplus 4}\\\oplus\\\mathcal{O}(15)^{\oplus 4}\\\oplus\\\mathcal{O}(14)^{\oplus 4}\\\oplus\\\mathcal{O}(13)^{\oplus 4}\end{array}&\rTo&\cdots
\end{diagram}
\end{equation}
We can now compare this complex to the complexes we have computed above. These have additional entries which make up the non--trivial information about the brane. Subtracting (\ref{33210triv}) from (\ref{33210a}) we get:\\
$n=0:\:[0]$\\
\begin{equation}
\begin{diagram}
\mathcal{O}(0)^{\oplus 3}&\rTo&\begin{array}{c}\mathcal{O}(4)\\\oplus\\\mathcal{O}(3)^{\oplus 3}\\\oplus\\\mathcal{O}(2)^{\oplus 3}\\\oplus\\\mathcal{O}(1)^{\oplus 4}\end{array}&\rTo&\begin{array}{c}\mathcal{O}(7)\\\oplus\\\mathcal{O}(6)\\\oplus\\\mathcal{O}(5)^{\oplus 3}\end{array}&\rTo&\mathcal{O}(9)
\end{diagram}
\end{equation}
In order to get the non--trivial piece of (\ref{33210b}) we have to tensor (\ref{33210triv}) with $\mathcal{O}(3)$ and shift it by one position to the right. Then we get:\\
$n=1:\:[1]$\\
\begin{equation}
\begin{diagram}
\begin{array}{c}\mathcal{O}(3)^{\oplus 3}\\\oplus\\\mathcal{O}(2)^{\oplus 4}\\\oplus\\\mathcal{O}(1)^{\oplus 4}\\\oplus\\\mathcal{O}(0)^{\oplus 4}\end{array}&\rTo&\begin{array}{c}\mathcal{O}(7)\\\oplus\\\mathcal{O}(6)^{\oplus 3}\\\oplus\\\mathcal{O}(5)^{\oplus 3}\\\oplus\\\mathcal{O}(4)^{\oplus 4}\end{array}&\rTo&\begin{array}{c}\mathcal{O}(10)\\\oplus\\\mathcal{O}(9)\\\oplus\\\mathcal{O}(8)^{\oplus 3}\end{array}&\rTo&\mathcal{O}(12)
\end{diagram}
\end{equation}
Tensoring (\ref{33210triv}) with $\mathcal{O}(2)$ and shifting one position to the right, we get the interesting information out of (\ref{33210c}):\\
$n=2:\:[1]$\\
\begin{equation}
\begin{diagram}
\begin{array}{c}\mathcal{O}(2)^{\oplus 3}\\\oplus\\\mathcal{O}(1)^{\oplus 4}\\\oplus\\\mathcal{O}(0)^{\oplus 4}\end{array}&\rTo&\begin{array}{c}\mathcal{O}(6)\\\oplus\\\mathcal{O}(5)^{\oplus 3}\\\oplus\\\mathcal{O}(4)^{\oplus 3}\\\oplus\\\mathcal{O}(3)^{\oplus 4}\end{array}&\rTo&\begin{array}{c}\mathcal{O}(9)\\\oplus\\\mathcal{O}(8)\\\oplus\\\mathcal{O}(7)^{\oplus 3}\end{array}&\rTo&\mathcal{O}(11)
\end{diagram}
\end{equation}
Finally, we tensor (\ref{33210triv}) with $\mathcal{O}(1)$ and shift by one to the right and subtract this from (\ref{33210d}) to get:\\
$n=3:\:[2]$\\
\begin{equation}
\begin{diagram}
\begin{array}{c}\mathcal{O}(1)^{\oplus 3}\\\oplus\\\mathcal{O}(0)^{\oplus 4}\end{array}&\rTo&\begin{array}{c}\mathcal{O}(5)\\\oplus\\\mathcal{O}(4)^{\oplus 3}\\\oplus\\\mathcal{O}(3)^{\oplus 3}\\\oplus\\\mathcal{O}(2)^{\oplus 4}\end{array}&\rTo&\begin{array}{c}\mathcal{O}(8)\\\oplus\\\mathcal{O}(7)\\\oplus\\\mathcal{O}(6)^{\oplus 3}\end{array}&\rTo&\mathcal{O}(10)
\end{diagram}
\end{equation}
The complexes for $n=4,\ldots,7$ are the same as the above ones, shifted to the right by one position. Computing the Chern characters, we get:
\begin{eqnarray}
n=0:&\qquad&-4-4H+10H^2+\frac{16}{3}H^3\nonumber\\
n=1:&\qquad&-8+4H+10H^2-\frac{16}{3}H^3\nonumber\\
n=2:&\qquad&-4+8H+4H^2-\frac{38}{3}H^3\nonumber\\
n=3:&\qquad&8H-4H^2-\frac{38}{3}H^3\nonumber
\end{eqnarray}
This is in agreement with results obtained from conformal field theory calculations \cite{Scheidegger:1999ed}. The other four Chern characters are the same as the ones given with an overall negative sign. The brane of interest is the one with $n=1$ and its antibrane with $n=5$. Comparing with table \ref{tab-d8orientifold} in the appendix, these branes have the correct charges to satisfy the tadpole cancellation condition with an O9--plane. Furthermore, we observe that, as in the quintic case, this brane is associated to a semi infinite complex which is periodic from the beginning.
\subsubsection{The $d=10$ brane $L=(4,3,2,1,0)$}
We have the following matrix factorization at the Gepner point:
\begin{equation}
Q=x_1^5\eta_1+x_1^5\bar{\eta}_1+x_2^4\eta_2+x_2^6\bar{\eta}_2+x_3^3\eta_3+x_3^7\bar{\eta}_3+x_4^2\eta_4+x_4^3\bar{\eta}_4+x_5\eta_5+x_5\bar{\eta}_5
\end{equation}
Taking into account the $\left(\mZ_{10}\right)^2$ orbifold action, we find a particular brane which is associated to the following semi--infinite complex via the algorithm of \cite{Herbst:2008jq}:\\
$n=1:\:[1]$\\
\begin{equation}
\label{43210semininf}
\begin{diagram}
\begin{array}{c}\mathcal{O}(4)^{\oplus 2}\\\oplus\\\mathcal{O}(3)^{\oplus 4}\\\oplus\\\mathcal{O}(2)^{\oplus 4}\\\oplus\\\mathcal{O}(2)^{\oplus 4}\\\oplus\\\mathcal{O}(0)^{\oplus 2}\end{array}&\rTo&\begin{array}{c}\mathcal{O}(9)^{\oplus 2}\\\oplus\\\mathcal{O}(8)^{\oplus 4}\\\oplus\\\mathcal{O}(7)^{\oplus 4}\\\oplus\\\mathcal{O}(6)^{\oplus 4}\\\oplus\\\mathcal{O}(5)^{\oplus 2}\end{array}&\rTo&\begin{array}{c}\mathcal{O}(14)^{\oplus 2}\\\oplus\\\mathcal{O}(13)^{\oplus 4}\\\oplus\\\mathcal{O}(12)^{\oplus 4}\\\oplus\\\mathcal{O}(11)^{\oplus 4}\\\oplus\\\mathcal{O}(10)^{\oplus 2}\end{array}&\rTo&\ldots
\end{diagram}
\end{equation}
At $n=6$ we find the antibrane of this. From table \ref{tabd8mod} we read off that the effective superpotential encoding the obstructions of the deformations of this brane is cubic.\\
To get the quasi--isomorphic finite complex we take a trivial brane in the linear sigma model, given by:
\begin{equation}
Q^{LSM}_{triv}=x_1^5\eta_1+x_1^5P\bar{\eta}_1+x_2^4\eta_2+x_2^6P\bar{\eta}_2+x_3^3\eta_3+x_3^7P\bar{\eta}_3+x_4^2\eta_4+x_4^3P\bar{\eta}_4+x_5\eta_5+x_5P\bar{\eta}_5
\end{equation}
The associated complex is:\\
{}[0]:
\begin{equation}
\begin{diagram}
\begin{array}{c}\mathcal{O}(0)\end{array}&\rTo&\begin{array}{c}\mathcal{O}(5)^{\oplus 2}\\\oplus\\\mathcal{O}(4)^{\oplus 2}\\\oplus\\\mathcal{O}(3)\end{array}&\rTo&\begin{array}{c}\mathcal{O}(10)^{\oplus 2}\\\oplus\\\mathcal{O}(9)^{\oplus 4}\\\oplus\\\mathcal{O}(8)^{\oplus 3}\\\oplus\\\mathcal{O}(7)^{\oplus 2}\end{array}&\rTo&\begin{array}{c}\mathcal{O}(15)^{\oplus 2}\\\oplus\\\mathcal{O}(14)^{\oplus 4}\\\oplus\\\mathcal{O}(13)^{\oplus 4}\\\oplus\\\mathcal{O}(12)^{\oplus 4}\\\oplus\\\mathcal{O}(11)\end{array}&\rTo&\begin{array}{c}\mathcal{O}(20)^{\oplus 2}\\\oplus\\\mathcal{O}(19)^{\oplus 4}\\\oplus\\\mathcal{O}(18)^{\oplus 4}\\\oplus\\\mathcal{O}(17)^{\oplus 4}\\\oplus\\\mathcal{O}(16)^{\oplus 2}\end{array}&\rTo&\ldots
\end{diagram}
\end{equation}
Tensoring this with $\mathcal{O}(4)$ and shifting by one position to the right we can subtract this trivial complex from (\ref{43210semininf}) to obtain the following finite complex:\\
{}[1]:
\begin{equation}
\begin{diagram}
\begin{array}{c}\mathcal{O}(4)\\\oplus\\\mathcal{O}(3)^{\oplus 4}\\\oplus\\\mathcal{O}(2)^{\oplus 4}\\\oplus\\\mathcal{O}(1)^{\oplus 4}\\\oplus\\\mathcal{O}(0)^{\oplus 2}\end{array}&\rTo&\begin{array}{c}\mathcal{O}(8)^{\oplus 2}\\\oplus\\\mathcal{O}(7)^{\oplus 3}\\\oplus\\\mathcal{O}(6)^{\oplus 4}\\\oplus\\\mathcal{O}(5)^{\oplus 2}\end{array}&\rTo&\begin{array}{c}\mathcal{O}(12)\\\oplus\\\mathcal{O}(11)^{\oplus 2}\\\oplus\\\mathcal{O}(10)^{\oplus 2}\end{array}&\rTo&\mathcal{O}(15)&\rTo&\ldots
\end{diagram}
\end{equation}
Computing the Chern character we find:
\begin{equation}
\label{43210brane}
\mathrm{ch}(E)=-8+4H+18H^2-\frac{28}{3}H^3
\end{equation}
As we can read off from table \ref{tab-d10orientifold} the tadpole cancellation condition is not satisfied if we just include the O9--plane, which has:
\begin{equation}
\label{o9d10}
\mathrm{ch}(E)^{O9}=\pm 4(8-4H-16H^2+\frac{25}{3}H^3)
\end{equation}
However, we have also found a pair of O5--planes as fixed point sets of the $\mZ_2$ action
\begin{equation}
(x_1,x_2,x_3,x_4,x_5)\longrightarrow(-x_1,x_2,x_3,x_4,-x_5).
\end{equation}
The Chern characters of this configuration have been computed to be:
\begin{equation}
\mathrm{ch}(E)=\left\{\begin{array}{c}
\pm4(3H^2-\frac{3}{2}H^3)\\
\pm4(2H^2-H^3)
\end{array}\right.
\end{equation}
Adding the second combination, $\pm4(2H^2-H^3)$, to (\ref{o9d10}) we get precisely (\ref{43210brane}). Thus we get tadpole cancellation if we take the $L=(4,3,2,1,0)$ boundary state and add an O9--plane and a particular pair of O5--planes. Note that this configuration is supersymmetric and that both the brane and the orientifolds are compatible with deformations away from the Gepner point.
\subsection{Algebraic Second Chern Class and Normal Function}
\label{sec:algebr-second-chern}
In order to get appropriate boundary conditions for the normal function we have to determine the algebraic second Chern class. It was argued in \cite{Morrison:2007bm} that this can be obtained directly from the periodic complex defined by a matrix factorization. We write a matrix factorization $Q$ as
\begin{equation}
  \label{eq:45}
  Q=
  \begin{pmatrix}
    0 & f\\ g & 0
  \end{pmatrix}
\end{equation}
If we have pairs $(Q_+,Q_-)$ of matrix factorizations such that $W\mathbbm{1}=f_{\pm}\cdot g_{\pm}$ as in (\ref{d8-mf}) and (\ref{d10-mf}), we can define $E_{\pm}=\mathrm{Ker} g_{\pm}$. Note that the complexes coming from the matrix factorizations are exact, therefore we can make use of the relation $\mathrm{Im}f_{\pm}=\mathrm{Ker}g_{\pm}$. The next step is to find a generic section $\tH^0(E_{\pm})$. In all the cases we consider we have $\mathrm{det}(f_{\pm})=W^{16}$, which implies that the bundles we are looking at have rank $16$. Grothendieck defines in~\cite{Grothendieck:1958ab} the algebraic second Chern class as the codimension two locus where $r-2$ generic sections of $E_{\pm}$ fail to be linearly independent. For a more accessible explanation see~\cite{Fulton:1998ab}. This amounts to calculating all the $14\times 14$ minors of a section in $H^0(E_{\pm})$. Out of this calculation one can extract a pair of algebraic curves $C_{\pm}$. The topological second Chern classes are then~\cite{Morrison:2007bm}:
\begin{equation}
  c_2(E_+)-c_2(E_-)=[C_+-C_-] \in \tH^4(X,\mZ) = \tH_2(X,\mZ)
\end{equation}
If $[C_+-C_-]=0\in \tH^4(X,\mZ)$ the cycle $C_+-C_-$ defines a normal function. We now perform this calculation for our two branes. 
\subsubsection{The $d=8$ brane $L=(3,3,2,1,0)$}
 From the deformed matrix factorization (\ref{d8-mf}) for $L=(3,3,2,1,0)$ we obtain the following semi--infinite complex:
\begin{equation}
\begin{diagram}
\begin{array}{c}
\mathcal{O}(3)^{\oplus 4}\\\oplus\\\mathcal{O}(2)^{\oplus 4}\\\oplus\\\mathcal{O}(1)^{\oplus 4}\\\oplus\\\mathcal{O}(0)^{\oplus 4}
\end{array}&\rTo^{f_\pm}&
\begin{array}{c}
\mathcal{O}(7)^{\oplus 4}\\\oplus\\\mathcal{O}(6)^{\oplus 4}\\\oplus\\\mathcal{O}(5)^{\oplus 4}\\\oplus\\\mathcal{O}(4)^{\oplus 4}
\end{array}&\rTo^{g_\pm}&
\begin{array}{c}
\mathcal{O}(11)^{\oplus 4}\\\oplus\\\mathcal{O}(10)^{\oplus 4}\\\oplus\\\mathcal{O}(9)^{\oplus 4}\\\oplus\\\mathcal{O}(8)^{\oplus 4}
\end{array}&\rTo^{f_\pm}&\cdots
\end{diagram}
\end{equation}
We define:
\begin{equation}
E_{\pm}=\mathrm{Ker}(\begin{diagram}
\begin{array}{c}
\mathcal{O}(7)^{\oplus 4}\\\oplus\\\mathcal{O}(6)^{\oplus 4}\\\oplus\\\mathcal{O}(5)^{\oplus 4}\\\oplus\\\mathcal{O}(4)^{\oplus 4}
\end{array}&\rTo^{g_\pm}&
\begin{array}{c}
\mathcal{O}(11)^{\oplus 4}\\\oplus\\\mathcal{O}(10)^{\oplus 4}\\\oplus\\\mathcal{O}(9)^{\oplus 4}\\\oplus\\\mathcal{O}(8)^{\oplus 4}
\end{array}
\end{diagram})
\end{equation}
Using the exactness of the complex we take a section of $\mathcal{O}(3)^{\oplus 4}\oplus\mathcal{O}(2)^{\oplus 4}\oplus\mathcal{O}(1)^{\oplus 4}\oplus\mathcal{O}(0)^{\oplus 4}$ and apply the map $f_{\pm}$ to get a section $s\in\tH^0(E_{\pm})$. Calculating all the $14\times 14$ minors of $s$ we get the following conditions:
\begin{equation}
(x_5^2\pm 4\psi x_1^2x_2^2x_3^2x_4^2)\cdot W^6\stackrel{!}{=}0\qquad (x_i^8+x_j^8)\cdot W^6\stackrel{!}{=}0\quad i\neq j,\: i,j\in\{1,2,3,4\}
\end{equation}
Up to permutations in the variables $x_1,\ldots,x_4$, we obtain the following algebraic curves:
\begin{equation}
C_{\pm}=\{x_1+\mu x_2=0,x_3+\nu x_4=0,x_5\pm2\sqrt{\psi}x_1x_2x_3x_4=0\}, \qquad \mu^8=\nu^8=-1
\end{equation}
Since $[C_+-C_-]=0\in \tH_2(X)$ the cycle $C_+-C_-$ defines a normal function for $X$.
\subsubsection{The $d=10$ brane $L=(4,3,2,1,0)$}
To obtain the geometric boundary condition from the deformed matrix factorization (\ref{d10-mf}) we define:
\begin{equation}
E_{\pm}=\mathrm{Ker}(\begin{array}{c}
\mathcal{O}(4)^{\oplus 2}\\\oplus\\\mathcal{O}(3)^{\oplus 4}\\\oplus\\\mathcal{O}(2)^{\oplus 4}\\\oplus\\\mathcal{O}(1)^{\oplus 4}\\\oplus\\\mathcal{O}(0)^{\oplus 2}\end{array}\stackrel{g_\pm}{\longrightarrow}\begin{array}{c}
\mathcal{O}(9)^{\oplus 2}\\\oplus\\\mathcal{O}(8)^{\oplus 4}\\\oplus\\\mathcal{O}(7)^{\oplus 4}\\\oplus\\\mathcal{O}(6)^{\oplus 4}\\\oplus\\\mathcal{O}(5)^{\oplus 2}\end{array} ),
\end{equation}
where $g_\pm$ refers to the $16\times16$--block of the bulk--deformed matrix factorization~(\ref{d10-mf}) associated to the $L=(4,3,2,1,0)$ boundary state. In order to obtain the algebraic second Chern class, we again calculate all the $14\times 14$ minors of a generic section $s\in \tH^0(E_{\pm})$. This yields the following conditions:
\begin{equation}
  \begin{gathered}
    (x_5^2\pm 4\psi x_1^2x_2^2x_3^2x_4^2)\cdot
    W^6\stackrel{!}{=}0\quad(x_i^{10}+x_j^{10})\cdot
    W^6\stackrel{!}{=}0\\ (x_k^{10}+x_4^5)\cdot
    W^6\stackrel{!}{=}0\qquad i\neq j\neq k, \:i,j,k\in\{1,2,3\}
  \end{gathered}
\end{equation} 
Up to permutations in $x_1,x_2,x_3$, we obtain the following algebraic curves:
\begin{equation}
C_{\pm}=\{x_1+\mu x_2=0,x_3^2+x_4=0,x_5\pm\sqrt{5\psi}x_1x_2x_3x_4=0\}\qquad\mu^{10}=-1
\end{equation}
The topological second Chern class is then:
\begin{equation}
\ch_2^{\text{top}}(E_+)-\ch_2^{\text{top}}(E_-)=[C_+-C_+]
\end{equation}
This defines a trivial class in $\tH_2(X,\mZ)$, thus defining a normal function. 
\section{Resolution of Singularities and Toric Geometry}
\label{sec:toric}
In the last section we have determined suitable geometric boundary conditions $C_\pm$ which yield a normal function on $X$. However, we are actually interested in the mirror manifold $Y$ which be get by quotienting with a suitable finite group $G_{\text{GP}}$ as prescribed by the Greene--Plesser construction \cite{Greene:1990ud}. We now have to map $C_{\pm}$ to boundary conditions on the mirror $Y$, in order to get a normal function on $Y$. Certain points on the curves $C_{\pm}$ may coincide with the fixed points of the group action. The latter induce singularities and have to be resolved. Since we are working with weighted projective spaces we can invoke standard techniques of toric geometry for resolving these singularities. This is the topic of the present section.
\subsection{$d=8$}
\begin{figure}[ht]
\begin{center}
\includegraphics{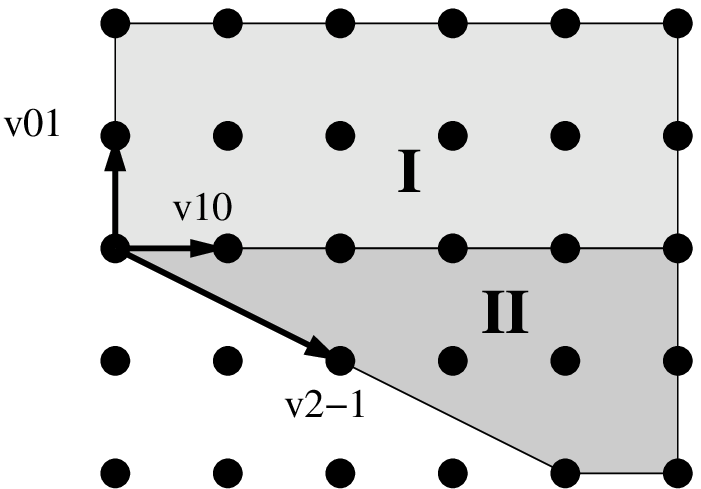}
\caption{The resolution of $S$.}\label{fig-toric1}
\end{center}
\end{figure}

\begin{figure}[ht]
\begin{center}
\includegraphics{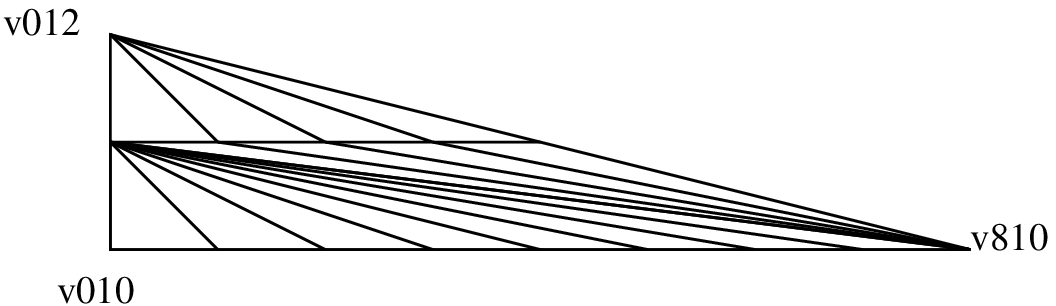}
\caption{Resolution of $\mC^3/\mZ_8\times\mZ_4$.}\label{fig-toric2}
\end{center}
\end{figure}

The mirror of the $d=8$ hypersurface $X$ is $Y=X/(\mZ_8)^3$. The generators of $(\mZ_8)^3$ can be written as:
\begin{eqnarray}
\label{z83}
g_1:&&(1,7,0,0,0)\nonumber\\
g_2:&&(1,0,7,0,0)\nonumber\\
g_3:&&(1,0,0,7,0)
\end{eqnarray}
In \cite{Klemm:1992tx} it has been observed that there is the $(\mZ_8)^3$ contains a $\mZ_4$ subgroup which acts trivially. Therefore that group we have to quotient $X$ by to get the mirror is actually $(\mZ_8)^2\times\mZ_2$. The generators of this group can be chosen to be:
\begin{eqnarray}
g'_1:&&(6,1,1,0,0)\nonumber\\
g'_2:&&(3,1,0,0,4)\nonumber\\
g'_3:&&(4,0,0,0,4),
\end{eqnarray}
where $g'_3$ generates the $\mZ_2$ factor. Let us now define a plane $P=\{x_1+\mu x_2=0,x_3+\nu x_4=0\}$ and points $p_{1}=\{x_1=-\mu x_2,x_3=x_4=x_5=0\}$ and $p_2=\{x_1=x_2=x_5=0,x_3=-\nu x_4\}$. Now we find that we can combine the generators of $(\mZ_8)^2\times\mZ_2$ in the following way (modulo $8$):
\begin{eqnarray}
\label{z82z4}
\tilde{g}_1:&& (7,7,2,0,0)\equiv5(3,1,0,0,4)+2(6,1,1,0,0)+(4,0,0,0,4)\nonumber\\
\tilde{g}_2:&& (1,7,0,0,0)\equiv-(4,0,0,0,4)-(3,1,0,0,4)\nonumber\\
\tilde{g}_3:&& (3,3,5,5,0)\equiv6(3,1,0,0,4)-3(4,0,0,0,4)+5(1,1,1,1,4)
\end{eqnarray}
We can of course also obtain these generators as combinations of the $\mZ_8$ generators (\ref{z83}).  Note that $\tilde{g}_3$ is a $\mZ_4$--generator, whereas $\tilde{g}_1$ and $\tilde{g_2}$ are $\mZ_8$--generators.\\
The reason why we have rewritten the $g_i$ in this form is that the plane $P$ is fixed under $\tilde{g}_3$. We define:  $S=P/\mZ_4$.  Furthermore $p_1$ is fixed under $\tilde{g}_1$ and $\tilde{g}_3$ and $p_2$ is fixed under $\tilde{g}_2$ and $\tilde{g}_3$. The singularities at $p_1$, $p_2$ and $S$ have to be resolved. The result will be two coordinate charts to be used around $p_1$ and $p_2$. It will turn out in Section~\ref{sec:picard-fuchs} that the relevant contributions to the integral over tubes around $C_{\pm}$ come from these points. Note that we cannot just compute the integral of a tube around $P$ because this would contain both $C_+$ and $C_-$ around the points $p_1$ and $p_2$.\\
The calculation includes the following steps:
\begin{itemize}
\item Make a proper choice of affine coordinates, suited for the points $p_1$ and $p_2$.
\item Resolve the singularity of $S$.
\item Resolve the singularity of $Y$. 
\item Choose a coordinate patch of $Y$ which reduces to the blowup of $S$ when restricting to $S$. This gives the local coordinates around $p_1$ and $p_2$.
\end{itemize}
 Let us first focus on the point $p_1$. Choosing the affine patch $x_1=1$, we have $p_1=(1,-\mu^{-1},0,0,0)$. We define the following affine coordinates:
\begin{equation}
t=\frac{x_2}{x_1}\quad u=\frac{x_3}{x_1}\quad v=\frac{x_4}{x_1}\quad w=\frac{x_5}{x_1^4}
\end{equation}
The equation defining the Calabi--Yau hypersurface then becomes:
\begin{equation}
1+t^8+u^8+v^8+w^2-4\psi t^2u^2v^2
\end{equation}
The surface $S$ is defined by $t=-\mu^{-1}$, $v=-\nu^{-1}u$.  The group action on $S$ is then:
\begin{equation}
\label{s-action1}
(u,w)\rightarrow (iu,-w)
\end{equation}
Now we can use toric geometry methods to resolve the singularity on $S$. At first we have to pick a monomial basis. An arbitrary rational monomial which is invariant under (\ref{s-action1}) can be chosen to be of the following form:
\begin{equation}
(w^2)^a(u^2w^{-1})^b=u^{2b}w^{2a-b}
\end{equation}
For this monomial to be regular we must have $b\geq0$, $2a-b\geq0$. These equations define a cone spanned by the vectors $(0,1)$ and $(2,-1)$. In order to resolve the singularity we have to subdivide the cone by adding the vector $(1,0)$. This is depicted in figure \ref{fig-toric1}.
The two subcones $I$ and $II$ are generated by the vectors $(0,1),(1,0)$ and $(1,0),(2,-1)$, respectively. We have a coordinate chart for each of these cones. For cone $I$ the coordinates are defined by:
\begin{equation}
u^{2b}w^{2a-b}=u_I^bw_I^a,
\end{equation}
which yields:
\begin{equation}
u_I=u^2w^{-1}\qquad w_I=w^2
\end{equation}
For cone $II$ we have:
\begin{equation}
u^{2b}w^{2a-b}=u_{II}^aw_{II}^{2a-b},
\end{equation}
and thus,
\begin{equation}
u_{II}=u^4\qquad w_{II}=u^{-2}w.
\end{equation}
Choosing for instance $\mu=e^{i\pi/8},\nu=e^{-i\pi/8}$, the equation for $Y$, restricted to $S$ becomes:
\begin{eqnarray}
(I):&&w_I(1-2\sqrt{\psi}u_I)(1+2\sqrt{\psi}u_I)\nonumber\\
(II):&&u_{II}(w_{II}-2\sqrt{\psi})(w_{II}+2\sqrt{\psi})
\end{eqnarray}
So, in chart $I$ we have $p_{1,\pm}=(\pm(4\psi)^{-1/2},0)$ and in chart $II$: $p_{1,\pm}=(0,\pm(4\psi)^{1/2})$.\\\\
In order to get the resolution of the singularity in $Y$ we have to consider the quotient singularity $\mC^3/\mZ_8\times\mZ_4$, where the $\mZ_8$ is generated by $\tilde{g}_1$ in (\ref{z82z4}) and the $\mZ_4$ is generated by $\tilde{g}_3$. An invariant monomial can be represented by:
\begin{equation}
(u^8)^a(uvw)^b(w^2)^c=u^{8a+b}v^bw^{2c+b}
\end{equation}
The inequalities $8a+b\geq0$, $b\geq0$ and $2c+b\geq0$ define a cone spanned by the vectors $(8,1,0)$, $(0,1,0)$ and $(0,1,2)$. We choose a particular resolution of the singularity as shown in figure \ref{fig-toric2}.
The coordinates of the triangles pointing towards $(0,1,1)$ and $(0,1,2)$ are: 
\begin{equation}
(\alpha,1,\beta)\qquad(\alpha+1,1,\beta)\qquad(0,1,\beta+1),
\end{equation}
where $\alpha$ and $\beta$ are integers whose values can be read off from figure \ref{fig-toric2}. We can then define coordinates $T=t$, $X_{\alpha\beta}$, $Y_{\alpha\beta}$, $Z_{\alpha\beta}$ via the following relation:
\begin{equation}
u^{8a+b}v^bw^{b+2c}=X_{\alpha\beta}^{a\alpha+b+c\beta}Y_{\alpha\beta}^{a(\alpha+1)+b+c\beta}Z_{\alpha\beta}^{b+c(\beta+1)}
\end{equation}
Solving this, we get:
\begin{eqnarray}
T&=&t\nonumber\\
X_{\alpha\beta}&=&u^{-7+\alpha+\beta+\alpha\beta}v^{1+\alpha+\beta+\alpha\beta}w^{(1+\alpha)(-1+\beta)}\nonumber\\
Y_{\alpha\beta}&=&u^{8-\alpha-\alpha\beta}v^{-\alpha(1+\beta)}w^{\alpha(1-\beta)}\nonumber\\
Z_{\alpha\beta}&=&u^{-\beta}v^{-\beta}w^{2-\beta}
\end{eqnarray}
The next step is to restrict to $S$ and see whether the restriction is compatible with the coordinates we have found after resolving the singularities there. Thus, if we set for instance $v=-e^{i\pi/8}u$, and scan through the values of $\alpha$ and $\beta$. We are lucky for the following two choices:
\begin{eqnarray}
\alpha=\beta=1 && X_{11}=i\quad Y_{11}=e^{-i\pi/4}u_{II}\quad Z_{11}=-e^{-i\pi/8}w_{II}\nonumber\\
\alpha=2,\beta=1 && X_{21}=e^{3i\pi/4}u_{II}\quad Y_{21}=-i\quad Z_{21}=-e^{-i\pi/8}w_{II}
\end{eqnarray}
Writing this again in terms of the $x_1$ we get the following choices of coordinates in the neighborhood of $p_{1\pm}$:
\begin{equation}
\label{p1coord}
\begin{array}{cccc}
T=\frac{x_2}{x_1}&X=\frac{x_4^4}{x_3^4}&Y=\frac{x_3^6}{x_1^4x_4^2}&Z=\frac{x_5}{x_1^2x_3x_4}\\
T'=\frac{x_2}{x_1}&X'=\frac{x_4^6}{x_1^4x_3^2}&Y'=\frac{x_3^4}{x_4^4}&Z'=\frac{x_5}{x_1^2x_3x_4}
\end{array}
\end{equation}
The defining equation of the Calabi--Yau in these patches is:
\begin{eqnarray}
&&1+T^8+XY^2+X^3Y^2+XYZ^2-4\psi T^2XY\nonumber\\
&&1+T^{'8}+X^{'2}Y^{'3}+X^{'2}Y'+X'Y'Z^{'2}-4\psi T^{'2}X'Y'
\end{eqnarray}
This concludes our discussion concerning the coordinates in the neighborhood of $p_{1\pm}$. We omit the calculation for $p_2$ because it is completely analogous. The local coordinates close to $p_{2\pm}$ can be obtained from (\ref{p1coord}) by exchanging $x_1\leftrightarrow x_3$ and $x_2\leftrightarrow x_4$.\\
For completeness we also give the coordinates on the patches defined by the triangles pointing towards $(8,1,0)$. The coordinates of these triangles are:
\begin{equation}
(\alpha,1,1)\quad (\alpha+1,1,1)\quad(\alpha,1,0)\qquad \alpha=0,\ldots,3
\end{equation}
The coordinates on the patches are defined via the following relations:
\begin{equation}
u^{8a+b}v^{b}w^{2c}=(X_{\alpha})^{a\alpha+b+c}(Y_{\alpha})^{a(\alpha+1)+b+c}(Z_{\alpha})^{8a+b}
\end{equation}
From this we get:
\begin{eqnarray}
X_{\alpha}&=&v^8w^{8+2(-7+\alpha)}\nonumber\\
Y_{\alpha}&=&v^8w^{-8+2(8-\alpha)}\nonumber\\
Z_{\alpha}&=&uvw^{-1}
\end{eqnarray}
For none of the allowed values of $\alpha$ these coordinates reduce to those on $S$.
\subsection{$d=10$}
\begin{figure}[ht]
\begin{center}
\includegraphics{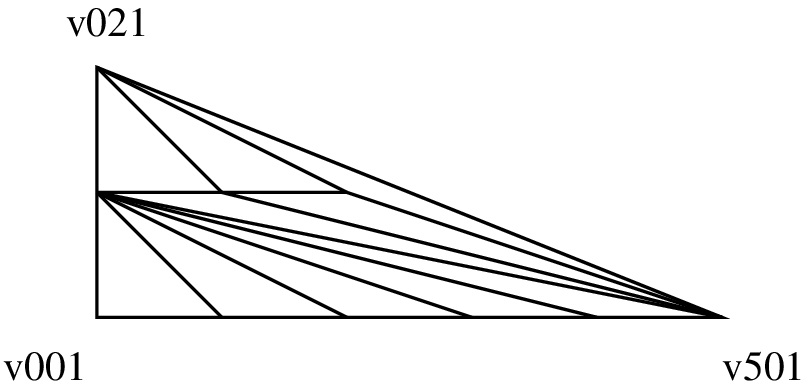}
\caption{The resolution of the singularity at $p_1$.}\label{fig-toricd10a}
\end{center}
\end{figure}

\begin{figure}[ht]
\begin{center}
\includegraphics{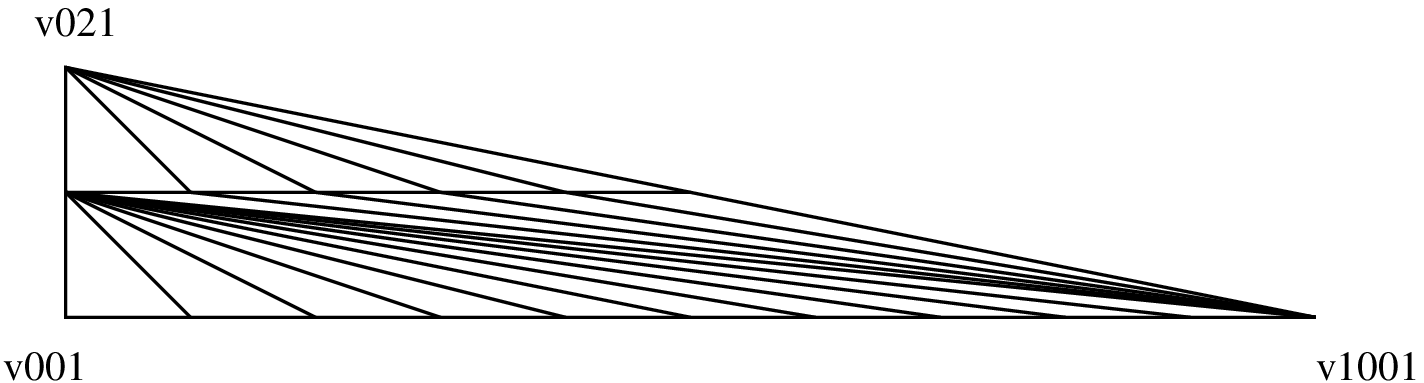}
\caption{The resolution of the singularity at $p_2$.}\label{fig-toricd10b}
\end{center}
\end{figure}

We now look at the mirror $Y=X/(\mZ_{10})^2$ of the Calabi--Yau $X$. Let us first define the points $p_1,p_2$ and the plane $P$:
\begin{eqnarray}
p_1&=&\{x_1=-\mu x_2,x_3=x_4=x_5=0\}\nonumber\\
p_2&=&\{x_1=x_2=x_5=0,x_4=-x_3^2\}\nonumber\\
P&=&\{x_1+\mu x_2=0,x_3^2+x_4=0\}
\end{eqnarray}
The generators of the $(\mZ_{10})^2$--action are:
\begin{eqnarray}
g_1:&&(1,9,0,0,0)\nonumber\\
g_2:&&(1,0,9,0,0)
\end{eqnarray}
For our purposes it is useful to reshuffle them in the following way. 
\begin{eqnarray}
\tilde{g}_1:&&(0,0,3,2,5)\equiv(1,9,0,0,0)+8(1,0,9,0,0)+(1,1,1,2,5)\nonumber\\
\tilde{g}_2:&&(1,9,0,0,0)
\end{eqnarray}
We observe that $p_1$ is fixed by $\tilde{g}_1$ and $p_2$ is fixed by $\tilde{g}_2$. In contrast to the quintic and the $d=8$ hypersurface, the plane $P$ is not fixed by any of the $\mZ_{10}$ actions but by $5\tilde{g}_1\equiv(0,0,5,0,5)$ which is a $\mZ_2$--generator. This $\mZ_2$--action is harmless in the sense that we do not have to choose new coordinates on $S = P/\mZ_2$ since it acts with an overall minus sign.  The singularities at $p_1$ and $p_2$ have to be resolved. This is what we will do next. \\\\
Let us start with the point $p_1$. We choose the affine patch $x_1=1$ and coordinates:
\begin{equation}
t=\frac{x_2}{x_1}\quad u=\frac{x_3}{x_1}\quad v=\frac{x_4}{x_1^2}\quad w=\frac{x_5}{x_1^5} 
\end{equation}
The hypersurface equation in these coordinates is:
\begin{equation}
1+t^{10}+u^{10}+v^5+w^2-5\psi t^2u^2v^2
\end{equation}
An invariant monomial under the action $\tilde{g}_1$ is:
\begin{equation}
(v^5)^a(w^2)^b(uvw)^c=u^cv^{5a+c}w^{2b+c}
\end{equation}
Regularity imposes the inequalities $c\geq0$, $5a+c\geq0$ and $2b+c\geq0$. This defines a cone spanned by the vectors $(0,0,1)$, $(5,0,1)$ and $(0,2,1)$. The toric diagram and a convenient triangulation are depicted in figure  \ref{fig-toricd10a}.
For each triangle, we get a set of local coordinates. The triangles pointing towards $(0,1,1)$ and $(0,2,1)$ have the following coordinates:
\begin{equation}
(\alpha,\beta,1)\quad(\alpha+1,\beta,1)\quad (0,\beta+1,1)
\end{equation}
The values of the integers $\alpha,\beta$ can be read off from figure \ref{fig-toricd10a}. The coordinates $T=t$, $X_{\alpha\beta}$, $Y_{\alpha\beta}$ and $Z_{\alpha\beta}$ are defined via the following relation:
\begin{equation}
u^cv^{5a+c}w^{2b+c}=X_{\alpha\beta}^{a\alpha+b\beta+c}Y_{\alpha\beta}^{a(\alpha+1)+b\beta+c}Z_{\alpha\beta}^{b(\beta+1)+c}
\end{equation}
From this, we obtain:
\begin{eqnarray}
T&=&t\nonumber\\
X_{\alpha\beta}&=&u^{1+\alpha+\beta+\alpha\beta}v^{-4+\alpha+\beta+\alpha\beta}w^{(1+\alpha)(-1+\beta)}\nonumber\\
Y_{\alpha\beta}&=&u^{-\alpha(1+\beta)}v^{5-\alpha-\alpha\beta}w^{\alpha-\alpha\beta}\nonumber\\
Z_{\alpha\beta}&=&u^{-\beta}v^{-\beta}w^{2-\beta}
\end{eqnarray}
The triangles pointing towards $(5,0,1)$ have coordinates
\begin{equation}
(\alpha,1,1)\quad (\alpha+1,1,1)\quad (5,0,1)\qquad \alpha=0,1.
\end{equation}
The coordinates $X_{\alpha}$, $Y_{\alpha}$ and $Z_{\alpha}$ are defined via:
\begin{equation}
u^cv^{5a+c}w^{2b+c}=X_{\alpha}^{a\alpha+b+c}Y_{\alpha}^{a(\alpha+1)+b+c}Z_{\alpha}^{5a+c}
\end{equation}
Solving this, we find:
\begin{eqnarray}
T&=&t\nonumber\\
X_{\alpha}&=&u^{5}w^{-3+2\alpha}\nonumber\\
Y_{\alpha}&=&u^{-5}w^{5-2\alpha}\nonumber\\
Z_{\alpha}&=&u^{}v^{}w^{-1}
\end{eqnarray}
Finally, we also have an exceptional triangle with the following coordinates:
\begin{equation}
(0,2,1)\quad (3,1,1)\quad (5,0,1)
\end{equation}
This leads to the following coordinates in this patch:
\begin{equation}
X_e=u^5w\quad Y_e=u^{-10}\quad Z_e=u^6v
\end{equation}  
Now we have to choose the patch which is most suitable for our purposes. For the quintic and the $d=8$ hypersurface we had an additional condition that the local coordinates when reduced to $S=P/G$, $G$ some discrete group, reduce to the coordinates of the resolution of $S$. Here, we do not have such a condition. It turns out that a wise choice are the coordinates $X\equiv X_{11}$, $Y\equiv Y_{11}$ and $Z\equiv Z_{11}$:
\begin{equation}
t=\frac{x_2}{x_1}\quad X=\frac{x_3^4}{x_1^2x_4}\quad Y=\frac{x_4^3}{x_1^4x_3^2}\quad Z=\frac{x_5}{x_1^2x_3x_4}
\end{equation}
If we insert the boundary condition $x_4=-x_3^2$, which amounts to the reduction to $P$, we have $Y=-X^2$. So, on $P$, the coordinates $X,Y$ behave like $x_3,x_4$. This is the analogue of the condition we had on $d=8$ and the quintic without the difficulty that we have to resolve the singularity of $S$. This only works in the patch $\alpha=\beta=1$. \\\\
Let us now turn to the point $p_2$. In contrast to the previous cases, we prefer to do the resolution of singularities all over again because the structure is not so symmetric.  We refrain from putting primes or tildes on all the coordinates. The calculations around $p_2$ will be done in the affine patch $x_3=1$, so we choose the following coordinates:
\begin{equation}
u=\frac{x_1}{x_3}\quad v=\frac{x_2}{x_3}\quad t=\frac{x_4}{x_3^2}\quad w=\frac{x_5}{x_3^5}
\end{equation} 
The hypersurface equation has the following form in these coordinates:
\begin{equation}
u^{10}+v^{10}+1+t^5+w^2-5\psi u^2v^2t^2
\end{equation}
A monomial which is invariant under the $\mZ_{10}$--action $\tilde{g}_2$ is given by:
\begin{equation}
(v^{10})^a(w^2)^b(uvw)^c=u^cv^{10a+c}w^{2b+c}
\end{equation}
This defines a cone spanned by $(0,0,1)$, $(10,0,1)$ and $(0,2,1)$. The corresponding toric diagram and a triangulation are depicted in figure \ref{fig-toricd10b}.
The coordinates of the triangles pointing towards $(0,1,1)$ and $(0,2,1)$ are:
\begin{equation}
(\alpha,\beta,1)\quad (\alpha+1,\beta,1)\quad (0,\beta+1,1)
\end{equation}
From this, we can define local coordinates $X_{\alpha\beta}$, $Y_{\alpha\beta}$, $Z_{\alpha\beta}$ through the following relation:
\begin{equation}
u^cv^{10a+c}w^{2b+c}=X_{\alpha\beta}^{a\alpha+b\beta+c}Y_{\alpha\beta}^{a(\alpha+1)+b\beta+c}Z_{\alpha\beta}^{b(\beta+1)+c}
\end{equation}
Solving for $X_{\alpha\beta}$, $Y_{\alpha\beta}$, $Z_{\alpha\beta}$, we get:
\begin{eqnarray}
T&=&t\nonumber\\
X_{\alpha\beta}&=&u^{1+\alpha+\beta+\alpha\beta}v^{-9+\alpha+\beta+\alpha\beta}w^{(1+\alpha)(-1+\beta)}\nonumber\\
Y_{\alpha\beta}&=&u^{-\alpha(1+\beta)}v^{10-\alpha-\alpha\beta}w^{\alpha-\alpha\beta}\nonumber\\
Z_{\alpha\beta}&=&u^{-\beta}v^{-\beta}w^{2-\beta}
\end{eqnarray}
The triangles pointing towards $(10,0,1)$ have the following coordinates:
\begin{equation}
(\alpha,1,1)\quad (\alpha+1,1,1)\quad (10,0,1)\qquad \alpha=0,\ldots,4
\end{equation}
For each $\alpha$ we get coordinates $X_{\alpha}$, $Y_{\alpha}$, $Z_{\alpha}$:
\begin{eqnarray}
X_{\alpha}&=&u^{10}w^{-8+2\alpha}\nonumber\\
Y_{\alpha}&=&u^{-10}w^{10-2\alpha}\nonumber\\
Z_{\alpha}&=&uvw^{-1}
\end{eqnarray}
The distinguished patch is given by $\alpha=2,\beta=1$ and we set $X\equiv X_{21}$, $Y\equiv Y_{21}$ and $Z\equiv Z_{21}$, where:
\begin{equation}
T=\frac{x_4}{x_3^2}\quad X=\frac{x_1^6}{x_2^4x_3^2}\quad Y=\frac{x_2^6}{x_1^4x_3^2}\quad Z=\frac{x_5}{x_1x_2x_3^3}
\end{equation}
Setting $x_1=-\mu x_2$ we have the simple boundary condition $Y=-X$ in the new coordinates.
\section{Picard--Fuchs equations}
\label{sec:picard-fuchs}
Having determined suitable boundary conditions and having resolved the singularities we are now ready to derive the inhomogeneous Picard--Fuchs equations. The crucial ingredient is the Griffiths--Dwork algorithm \cite{Dwork,griffiths1} (see \cite{Morrison:1991cd} for a practical description which we will follow here). We will review this method in the following subsection.
\subsection{The Griffiths--Dwork Method}
\label{sec:griffiths-dwork}
The Griffiths--Dwork method achieves the reduction of the pole order of rational differential forms on toric varieties modulo exact forms. These exact pieces will in the end be responsible for the inhomogeneous term of the Picard--Fuchs equation.\\
We denote by $\Omega_0$ the canonical holomorphic 4--form on a weighted projective space $\mP(w)=\mP(w_1,\ldots,w_{n+1})$ with weights $w_i,i=1,\dots,n+1$.
\begin{equation}
  \label{omdef}
  \Omega_0=\sum_{i=1}^{n+1}(-1)^{i+1}w_ix_i\mathrm{d}x_1\wedge\dots\wedge\widehat{\mathrm{d}x_i}\wedge\dots\mathrm{d}x_{n+1}
\end{equation}
Rational differentials of degree $n$ on toric varieties are defined as expressions of the form $\frac{P\Omega_0}{W^\ell}$ where $P$ and $W$ are weighted homogeneous polynomials of weight $\sum_{i}w_i$ with $\mathrm{deg}P+\sum_{i}w_i=\ell\,\mathrm{deg}W$. Now suppose that $W(z)=0$ defines a family of (quasi--smooth) hypersurface $Y$ in the weighted projective space $\mP(w)$, depending on some parameters $z$ (the coefficients of the polynomial $W$). The middle cohomology of such a hypersurface $Y_z$ is then described by differential forms on $\mP(w)$ with poles along $Y_z$. To each differential form $\frac{P\Omega_0}{W^{\ell}}$ one can associate a cohomology class by a residue construction: For an $(n-1)$--chain $\Gamma_z$ on $Y_z$, the tube $T(\Gamma_z)$ over $\Gamma_z$ is an $n$--chain on $\mP(w)$, disjoint from $Y_z$, analogously for $(n-1)$--cycle $\gamma_z$ on $Y_z$. The residue of $\frac{P\Omega_0}{W^{\ell}}$ is defined as follows:
\begin{equation}
  \label{eq:48}
\int_{\Gamma_z}\mathrm{Res}_{Y_z}\frac{P\Omega_0}{W^{\ell}}=\frac{1}{2\pi i}\int_{T(\Gamma_z)}\frac{P\Omega_0}{W^{\ell}}
\end{equation}
Since altering $\frac{P\Omega_0}{W^{\ell}}$ by an exact differential does not change the integrals, the cohomology of $Y_z$ is represented by equivalence classes of differential forms $\frac{P\Omega_0}{W^{\ell}}$ modulo exact forms. In particular, we obtain the holomorphic 3--form $\Omega$ on $Y_z$ in this way:
\begin{equation}
  \label{eq:Omega}
  \widehat\Omega(z) = \mathrm{Res}_{Y_z}\frac{\rho(z)\Omega_0}{W(z)}
\end{equation}
Here $\rho(z)$ is an arbitrary holomorphic function. Griffiths' reduction of pole order algorithm works as follows. For $W$ and $A_j$ weighted homogeneous polynomials with $\mathrm{deg}W=d$ and $\mathrm{deg}A_j=\ell d-k_j-w$ we define:
\begin{equation}
\label{phi-def}
\varphi=\frac{1}{W^{\ell}}\sum_{i<j}(-1)^{i+j+1}(w_ix_iA_j-w_jx_jA_i)\mathrm{d}x_1\wedge\ldots\wedge\widehat{\mathrm{d}x_i}\wedge\ldots\wedge\widehat{\mathrm{d}x_j}\wedge\ldots\wedge\mathrm{d}x_{n+1}
\end{equation}
Then one computes:
\begin{equation}
\label{pole-red}
\mathrm{d}\varphi=\frac{\left(\ell\sum A_j\frac{\partial W}{\partial x_j}- W\sum\frac{\partial A_j}{\partial x_j}\right)\Omega_0}{W^{\ell+1}}=\frac{\ell\sum A_j\frac{\partial W}{\partial x_j}\Omega_0}{W^{\ell+1}}-\frac{\sum\frac{\partial A_j}{\partial x_j}\Omega_0}{W^{\ell}}
\end{equation}
Thus, any form whose numerator lies in the Jacobian ideal of $W$ is equivalent, modulo exact forms, to a form with smaller pole order.\\
This can be used to derive Picard--Fuchs equations. Since the cycles $T(\Gamma_z)$ do not change in homology as $z$ varies locally we can exchange differentiation and integration:
\begin{equation}
\frac{\mathrm{d}^k}{\mathrm{d}z^k}\int_{\Gamma_z}\mathrm{Res}_{Y_z}\frac{P\Omega_0}{W^{\ell}}=\frac{1}{2\pi i}\int_{T(\Gamma_z)}\frac{\mathrm{d}^k}{\mathrm{d}z^k}\left(\frac{P\Omega_0}{W^{\ell}}\right)
\end{equation}
The Picard--Fuchs operator $\cL$ then has the following property:
\begin{equation}
\label{inhom-pf}
\cL\left(\frac{P\Omega_0}{W}\right) = \left(\frac{\mathrm{d}^n}{\mathrm{d}z^n}+\sum_{j=0}^{n-1}C_j(z)\frac{\mathrm{d}^j}{\mathrm{d}z^j}\right)\left(\frac{P\Omega_0}{W}\right)=\mathrm{d}\varphi
\end{equation}
If we integrate (\ref{inhom-pf}) over an $n$--cycle $T(\gamma)$ the right--hand side will be $0$. For $P=\rho(z)$, and $W$ a hypersurface in $\mP(w)$, we end up with the well--known Picard--Fuchs equation satisfied by the periods $\varpi=\int_\gamma \Omega$ in~(\ref{eq:PF_for_periods}). However, if we integrate over a chain $\Gamma$ with specific boundaries the right--hand side of (\ref{inhom-pf}) will give a non--zero contribution. To determine $C_j(z)$ and $\varphi$ we have to compute successive $z$--derivatives of $\frac{P\Omega_0}{W}$ and use reduction of pole order to determine a linear relation among these derivatives modulo exact forms.\\\\
Now, we specify $W(z)$ to be one of the one--parameter degree $d$ hypersurfaces in the weighted projective spaces $\mP(w)$ given in~(\ref{eq:the_Xs}) with the standard deformation:
\begin{equation}
  \label{eq:Wpsi}
  W(x_i,\psi)=\sum_{i=1}^{n+1}x_i^{\frac{d}{w_i}}- c \psi m(x_i)
\end{equation}
where
\begin{equation}
  \begin{aligned}
    c&=6, & m(x_i) &= x_1x_2x_3x_4x_5   &\text{for } \mP(1,1,1,1,2)[6]\\
    c&=4, & m(x_i) &= (x_1x_2x_3x_4)^2  &\text{for } \mP(1,1,1,1,4)[8]\\
    c&=5, & m(x_i) &= (x_1x_2x_3x_4)^2  &\text{for } \mP(1,1,1,2,5)[10]\\
  \end{aligned}
  \label{eq:16}
\end{equation}
Furthermore, we set 
\begin{equation}
  \label{eq:46}
  z=
  \begin{cases}
    (c\psi)^{-d} & d=6,\\
    (4c\psi)^{-d/2} & d=8,10.
  \end{cases}
\end{equation}
We define the rational differential forms:
\begin{equation}
\label{omega-def}
  \omega_{\ell}=(-1)^{\ell-1}(\ell-1)! c^{\ell-1}\frac{ m(x_i)^{\ell-1}}{W(x_i,\psi)^{\ell}}\Omega_0
\end{equation}
With this definition we have:
\begin{equation}
\label{form-derivative}
  \frac{\mathrm{d}}{\mathrm{d}\psi}\omega_{\ell}=-\omega_{\ell+1}
\end{equation}
In order to get the Picard--Fuchs operator $\cL$ one must find an expression for $\omega_{n+1}$ as a linear combination of $\omega_1,\ldots,\omega_n$ modulo exact forms. This choice of conventions makes it particularly simple to read off the Picard--Fuchs operator. The calculation is most easily done using a Gr\"obner basis algorithm which has been implemented in the computer algebra program Singular \cite{GPS05}: Given a global form $\eta_{\ell}$ with pole of order $\ell$ one uses the Gr\"obner basis of the Jacobian ideal $J$ of $W$ to reduce $\eta_{\ell}$ and $\omega_{\ell}$ to standard form. This gives a coefficient $\varepsilon_{\ell}\in\mC(\psi)$ such that the numerator of $\eta_{\ell}-\varepsilon_{\ell}\omega_{\ell}$ lies in $J$. Further application of the Gr\"obner basis reduction gives:
\begin{equation}
\eta_{\ell}-\varepsilon_{\ell}\omega_{\ell}=\sum_j A_{\ell j}\frac{\partial W}{\partial x_j}
\end{equation}
The pole order reduction formula (\ref{pole-red}) then determines forms $\varphi_{\ell}$ and $\eta_{\ell-1}$ such that:
\begin{equation}
\eta_{\ell}-\varepsilon_{\ell}\omega_{\ell}=\mathrm{d}\varphi_{\ell}+\eta_{\ell-1},
\end{equation}
where $\eta_{\ell-1}$ has a pole of order $\ell-1$. Starting with $\eta_{n+1}=\omega_{n+1}$ there is a relation:
\begin{equation}
\omega_{n+1}=\varepsilon_1\omega_1+\ldots+\varepsilon_n\omega_n+\mathrm{d}\varphi
\end{equation}
where $\varphi=\sum_{\ell=1}^n \varphi_\ell$. Using (\ref{form-derivative}) one gets the Picard--Fuchs equation~(\ref{inhom-pf}).\\
Depending on the choice of $\rho(z)$ and the $\omega_\ell$ one may still have to transform the Picard--Fuchs operator $\cL$ into its standard form $\cL_{\text{PF}}$. The fundamental period $\varpi_0(z)$, i.e. the unique holomorphic solution to $\cL_{\text{PF}}$ near the point $z=0$ of maximal unipotent monodromy is
\begin{equation}
  \label{eq:pi0}
  \varpi_0(z) = \sum_{n=0}^\infty \frac{\Gamma(1+dn)}{\prod_{i=1}^5\Gamma(1+w_in)} z^n
\end{equation}
In order to get the standard integral (symplectic) basis of periods near this point, we follow~\cite{Givental:1996ab} and define a cohomology--valued period
\begin{equation}
  \label{eq:piH}
  \varpi_0(z,H) = \sum_{n=0}^\infty \frac{\Gamma(d(n+H)+1)}{\prod_{i=1}^5\Gamma(w_i(n+H)+1)} z^{n+H} \in H^*(Y,\mZ)
\end{equation}
where $H$ is the restriction of the hyperplane class of $\mP(w)$ to $Y$. Expanding $\varpi(z,H)$ by cohomology degree yields
\begin{equation}
  \label{eq:periodexpansion}
  \varpi_0(z,H) = 
  w_0(z) + w_1(z)H + w_2(z)H^2 - w_3(z) H^3
\end{equation}
and the expansion coefficients define the period vector
\begin{equation}
  \varpi^{(L)}=(w_3,w_2,w_0,w_1)
  \label{eq:pi_L}
\end{equation}
Their behavior near $z=0$ is of the form $w_i(z) = \left(\log z\right)^i + O(z), \; n=0,1,2,3$. In terms of the deformation parameter $\psi$ in~(\ref{eq:Wpsi}), the choice of $\rho(\psi)$ which leads to the above basis of solutions is simply~\cite{Klemm:1992tx} 
\begin{equation}
  \label{eq:17}
  \rho(\psi) =
  \begin{cases}
    \psi & \text{for } \mP(1,1,1,1,2)[6]\\
    \psi^{\frac{1}{2}} &\text{for } \mP(1,1,1,1,4)[8]\\
    \psi^{\frac{1}{2}} &\text{for } \mP(1,1,1,2,5)[10]
  \end{cases}
\end{equation}
Then we also have to take into account that we are actually considering an orbifold, i.e. $Y=\{W_\psi=0 \}/G_{\text{GP}}$. Hence, when integrating over cycles or chains, we have to divide by the order of $G_{\text{GP}}$, see Appendix B of~\cite{Klemm:1992tx}. This leads us to the definition
\begin{equation}
  \label{eq:Omega_hat}
  \Omega(\psi) = \frac{|G_{\text{GP}}|}{\left(2\pi i\right)^3} \widehat\Omega(\psi)
\end{equation}
with $\widehat\Omega(\psi)$ as in~(\ref{eq:Omega}). \\

\subsection{The $d=8$ Hypersurface}
\label{sec:d=8-hypersurface}
We now discuss the $d=8$ hypersurface. One gets the following expressions for the $\varepsilon_i$:
\begin{equation}
\varepsilon_1=-\frac{1}{16(\psi^4-1)}\qquad\varepsilon_2=\frac{5\psi}{\psi^4-1}\qquad \varepsilon_3=-\frac{29\psi^2}{2(\psi^4-1)}\qquad\varepsilon_4=\frac{8\psi^3}{\psi^4-1}
\end{equation}
From this we can read off the Picard--Fuchs operator:
\begin{equation}
\label{gd-pfd8}
\mathcal{L}=\frac{d ^4}{d \psi^4}+\frac{8\psi^3}{\psi^4-1}\frac{d ^3}{d \psi^3}+\frac{29\psi^2}{2(\psi^4-1)}\frac{d ^2}{d \psi^2}+\frac{5\psi}{\psi^4-1}\frac{d }{d \psi}+\frac{1}{16(\psi^4-1)}
\end{equation}
With that we have:
\begin{equation}
  \label{eq:18}
  \cL\tfrac{\Omega_0}{W}=\mathrm{d}\varphi
\end{equation}
$\varphi$ is quite a complicated expression. Since it is not unique and its structure not very enlightening we refrain from writing it down. Now we have to integrate over the three--chain $\Gamma$. What we have to compute is therefore
\begin{equation}
  \label{eq:integral}
  \int_{T_\epsilon(\Gamma)}\mathrm{d}\varphi = \int_{T_{\epsilon}(C_+-C_-)}\varphi,
\end{equation}
where $T_{\epsilon}(C_+-C_-)$ denotes a tube of radius $\epsilon$ around $C_+$ and $C_-$. Let us focus on $p_1$. We choose the unprimed coordinates in (\ref{p1coord})\footnote{One can check that these two choices of coordinates lead to the same results.}. After inserting the boundary conditions $T=-\mu, X=\nu^4$ the coordinates for $p_{1,\pm}$ are:
\begin{equation}
Y=0\qquad Z=\pm2\mu\sqrt{\psi}
\end{equation}
We parameterize the tube $T_{\epsilon}(C_{+};p_{1+})$ as follows:
\begin{equation}
T=-\mu+\epsilon e^{i\chi}\frac{f(r)}{-8\mu^7+8Y\mu\nu^4\psi}\quad X=\nu^4\quad Y=r e^{i\phi}\quad z=2\mu\sqrt{\psi}+\epsilon(\mu\nu^4)^{-1}e^{i\chi}e^{-i\phi}
\end{equation}
where
\begin{equation}
0\leq\chi\leq 2\pi\quad 0\leq\phi\leq2\pi \quad 0\leq r\leq r_{\ast},
\end{equation}
and $f(r)$is a $C^{\infty}$--function with $f(0)=1$, and $f(r)=0$ for $r\geq r_{\ast}>0$. Inserting this into the expression for $\varphi$ the integration can be performed explicitly. For $p_{1-}$ we have to substitute $\sqrt{\psi}\rightarrow-\sqrt{\psi}$ into the result. The calculation around $p_2$ is analogous after exchanging $x_1\leftrightarrow x_3$ and $x_2\leftrightarrow x_4$ in the definitions (\ref{p1coord}).\\
Choosing $\mu=e^{-i\pi/8}$ and $\nu=e^{i\pi/8}$ we find for the expression in~(\ref{eq:integral})
\begin{equation}
  \label{eq:19}
  \int_{T_{\epsilon}(C_+-C_-)}\varphi=-\frac{1}{(\psi^4-1)}\left(\frac{3\pi^2}{8}\psi^{-\frac{5}{2}}+\frac{\pi^2}{16}\psi^{-\frac{1}{2}} \right),
\end{equation}
In contrast to the case of the quintic studied in~\cite{Morrison:2007bm}, the inhomogeneous term now consists of two contributions. Different choices of $\mu$ and $\nu$ at most give a sign change in the $\psi^{-5/2}$--term.\\\\
The next step is now to relate the Picard--Fuchs operator (\ref{gd-pfd8}) to the standard differential operator which is:
\begin{equation}
\label{d8pf}
\cL_{\text{PF}}=\theta^4-16z(8\theta+1)(8\theta+3)(8\theta+5)(8\theta+7)
\end{equation} 
In order to determine how the variable $z$ is related to $\psi$ let us go back to the deformation of the superpotential. Had we taken the standard form of the Landau--Ginzburg superpotential, i.e.
\begin{equation}
W=x_1^8+x_2^8+x_3^8+x_4^8+x_5^2-8\widetilde\psi x_1x_2x_3x_4x_5,
\end{equation}
the appropriate choice for $z$ would be $z=(8\psi)^{-8}$. To get the deformation we are using we have to use the equation of motion of $x_5$:
\begin{equation}
x_5-4\widetilde\psi x_1x_2x_3x_4=0
\end{equation} 
Inserting this back into the superpotential we get:
\begin{equation}
W=x_1^8+x_2^8+x_3^8+x_4^8+x_5^2-16\widetilde\psi^2 x_1^2x_2^2x_3^2x_4^2,
\end{equation}
However, we have used the deformation $4\psi x_1^2x_2^2x_3^2x_4^2$ because we preferred to have a deformation which is linear in the deformation parameter (for instance in the Griffiths--Dwork procedure). Taking this into account we find that in our conventions the right choice for the variable $z$ is:
\begin{equation}
\label{zd8}
z=(16\psi)^{-4}
\end{equation} 
It it easy to see that this is compatible with the choice of $z$ for the standard deformation $-8\tilde\psi x_1x_2x_3x_4x_5$:
\begin{equation}
z=(16\psi)^{-4}=(4\cdot 4\psi)^{-4} = (4\cdot 16\widetilde\psi^2)^{-4}=(8^2\widetilde\psi^2)^{-4}=(8\widetilde\psi)^{-8}
\end{equation}
It is then easy to express (\ref{d8pf}) in terms of $\psi$ since with (\ref{zd8}) we have:
\begin{equation}
z\frac{d}{dz}=-\frac{1}{4}\psi\frac{d}{d\psi}
\end{equation}
Making this change of variables in $\cL_{\text{PF}}$ in~(\ref{d8pf}) we find that the relation to $\cL$ in~(\ref{gd-pfd8}) is
\begin{eqnarray}
  \label{eq:11}
  \cL_{\text{PF}}=(\psi^4-1)\sqrt{\psi}\frac{1}{4^4}\cL\frac{1}{\sqrt{\psi}}
\end{eqnarray}
Combining~(\ref{eq:Omega_hat}) with $|G_{\text{GP}}| = 2^7$,~(\ref{eq:Omega}) with $\rho(\psi)$ as in~(\ref{eq:17}), plugging this into~(\ref{eq:48}) and applying~(\ref{eq:11}) to it we find that 
\begin{equation}
  \label{eq:15}
  \cL_{PF} \int_\Gamma \Omega = -\frac{1}{32\pi^4} (\psi^4-1)\sqrt{\psi} \int_{T_\epsilon(\Gamma)} \frac{\Omega_0}{W}
\end{equation}
Inserting the value of the integral using~(\ref{eq:integral}) and~(\ref{eq:19}) and substituting~(\ref{zd8}) we finally find that the domain wall tension $\cT_B$ in~(\ref{eq:T_B-normal}) satisfies the following inhomogeneous Picard--Fuchs equation
\begin{equation}
  \label{eq:20}
  \cL_{\text{PF}} \cT_B(z) = \cL_{\text{PF}} \int_\Gamma \Omega = \frac{1}{16\pi^2}\left(48 z^{\frac{1}{2}} + \frac{1}{32}\right)
\end{equation}
We immediately see that $\cT_B$ is also a solution to the homogeneous differential equation $\cL_B\cT_B=0$ with
\begin{equation}
  \label{eq:21}
  \cL_B  = 8 \theta ( 2 \theta - 1 )\cL_{\text{PF}}
\end{equation}
where we have introduced the factor 8 for later convenience.

\subsection{The $d=10$ Hypersurface}
\label{sec:d=10-hypersurface}
Griffiths--Dwork reduction yields the following expressions for the $\varepsilon_i$:
\begin{equation}
\varepsilon_1=-\frac{\psi}{4(4\psi^5-1)}\qquad\varepsilon_2=\frac{20\psi^2}{4\psi^5-1}\qquad\varepsilon_3=-\frac{2(29\psi^5-1)}{\psi^2(4\psi^5-1)}\qquad\varepsilon_4=\frac{2(16\psi^5+1)}{\psi(4\psi^5-1)}
\end{equation}
We also get a large expression for $\varphi$ which we do not write down.
From this we can read off the Picard--Fuchs operator:
\begin{equation}
\cL=\frac{d ^4}{d \psi^4}+\frac{2(16\psi^5+1)}{\psi(4\psi^5-1)}\frac{d ^3}{d \psi^3}+\frac{2(29\psi^5-1)}{\psi^2(4\psi^5-1)}\frac{d ^2}{d \psi^2}+\frac{20\psi^2}{4\psi^5-1}\frac{d }{d \psi}+\frac{\psi}{4(4\psi^5-1)}
\end{equation}
We have to integrate the inhomogeneous term in the Picard--Fuchs equation over a tube of radius $\epsilon$ around $C_+$ and $C_-$. Around $p_1$ we choose $T=-\mu$ and $Y=-X^2$. From the hypersurface equation we obtain the following coordinates for $p_{1,\pm}$:
\begin{equation}
X=0\qquad Z=\pm\mu\sqrt{5\psi}
\end{equation}
The tube $T_{\epsilon}(C_{+};p_{1+})$ is parameterized as follows:
\begin{equation}
T=-\mu+\epsilon e^{i\chi}\frac{f(r)}{-10\mu^9-10X^3\mu\psi}\quad X=re^{i\phi}\quad Y=-X^2\quad \quad z=\mu\sqrt{5\psi}-\epsilon\mu^{-1}e^{i\chi}e^{-3i\phi}
\end{equation}
where, as for the $d=8$ case,
\begin{equation}
0\leq\chi\leq 2\pi\quad 0\leq\phi\leq2\pi \quad 0\leq r\leq r_{\ast},
\end{equation}
and $f(r)$is a $C^{\infty}$--function with $f(0)=1$, and $f(r)=0$ for $r\geq r_{\ast}>0$. For $T_{\epsilon}(C_{-};p_{1-})$ one just has to substitute $\sqrt{\psi}\rightarrow -\sqrt{\psi}$ in the result of the integral over $T_{\epsilon}(C_{+};p_{1+})$.\\
In a similar manner we proceed for $p_2$. Here we have $T=-1$ and $Y=-X$. The coordinates for $p_{2\pm}$ are:
\begin{equation}
X=0,Z=\pm\sqrt{5\psi}
\end{equation} 
The tube  $T_{\epsilon}(C_{+};p_{2+})$ is parameterized in the following way:
 \begin{equation}
T=-1+\epsilon e^{i\chi}\frac{f(r)}{5-10X^2\psi}\quad X=re^{i\phi}\quad Y=-X\quad Z=\sqrt{5\psi}-\epsilon e^{i\chi}e^{-2i\phi}
\end{equation}
After the integration over the tube we get:
\begin{equation}
  \label{eq:22}
  \int_{T_{\epsilon}(C_+-C_-)}\varphi=\frac{1}{(4\psi^5-1)}\pi^2(5\psi)^{\frac{1}{2}}
\end{equation}
The standard differential operator is:
\begin{equation}
  \label{d10pf}
  \cL_{\text{PF}}=\theta^4-80z(10\theta+1)(10\theta+3)(10\theta+7)(10\theta+9)
\end{equation}
To find the relation between $z$ and $\psi$ we look at the $d=10$ model with the other deformation where one has $z=(10\tilde{\psi})^{-10}$:
\begin{equation}
W=x_1^{10}+x_2^{10}+x_3^{10}+x_4^5+x_5^2-10\tilde{\psi}x_1x_2x_3x_4x_5
\end{equation}
Reinserting the equations of motion for $x_5$,
\begin{equation}
x_5=5\tilde{\psi}x_1x_2x_3x_4x_5,
\end{equation}
we get:
\begin{equation}
W=x_1^{10}+x_2^{10}+x_3^{10}+x_4^5+x_5^2-25\tilde{\psi}^2x_1^2x_2^2x_3^2x_4^2
\end{equation}
As in the $d=8$ case we have chosen $25\tilde{\psi}^2\equiv 5\psi$. Taking this into account the proper choice for $z$ is:
\begin{equation}
  \label{zd10}
  z=(20\psi)^{-5}
\end{equation}
Consistency is easily checked:
\begin{equation}
z=(10\tilde{\psi})^{-10}=(10^2\tilde{\psi}^2)^{-5}=(4\cdot 25\tilde{\psi^2})^{-5}=(20\psi)^{-5}
\end{equation}
With that we find:
\begin{equation}
  \label{eq:23}
  \cL_{\text{PF}}=(4\psi^5-1)\frac{1}{4\cdot5^4}\frac{1}{\sqrt{\psi}}\cL\frac{1}{\sqrt{\psi}}
\end{equation} 
Combining~(\ref{eq:Omega_hat}) with $|G_{\text{GP}}| = 10^2$,~(\ref{eq:Omega}) with $\rho(\psi)$ as in~(\ref{eq:17}), plugging this into~(\ref{eq:48}) and applying~(\ref{eq:23}) to it we find that 
\begin{equation}
  \label{eq:24}
\cL_{PF} \int_\Gamma \Omega =\frac{5^2}{(2\pi i)^4}\frac{1}{4\cdot5^4}\frac{1}{\sqrt{\psi}}\int_{T_\epsilon(\Gamma)} \frac{\Omega_0}{W}
\end{equation}
Inserting the value of the integral using~(\ref{eq:19}) and substituting~(\ref{zd10}) we finally find that the domain wall tension $\cT_B$ in~(\ref{eq:T_B-normal}) satisfies the following inhomogeneous Picard--Fuchs equation
\begin{equation}
  \label{eq:2}
  \cL_{\text{PF}} \cT_B(z) = \cL_{\text{PF}} \int_\Gamma \Omega = \frac{1}{16\pi^2}\frac{\sqrt{5}}{100}
\end{equation}
We immediately see that $\cT_B$ is also a solution to the homogeneous differential equation $\cL_B\cT_B=0$ with
\begin{equation}
  \cL_B  = 5 \theta \cL_{\text{PF}}
\end{equation}
where we have introduced the factor 5 for later convenience.

\section{Monodromies and Instantons}
\label{sec:instantons}
This section will be concerned with the properties of solutions to differential equations of the type we have found in Section~\ref{sec:picard-fuchs}. We will study their analytic continuation to the Gepner point, their monodromies around the Gepner point, the large complex structure limit and the conifold point. Furthermore we will compute the instanton expansion and determine the BPS invariants.


\subsection{Solutions to the Picard--Fuchs equations}
\label{sec:solut-pf-equat}

We analyze the solutions to differential equations of the type
\begin{equation}
  \label{eq:26}
  \cL_B \cT_B = 0
\end{equation}
where $\cL_B$ is of the form
\begin{equation}
  \label{eq:27}
  \cL_B = (d\theta+k)\cL_{PF}
\end{equation}
and $\cL_{PF}$ is a differential operator of the generalized hypergeometric type. Here, $d$ denotes the degree of any of the hypersurfaces in~(\ref{eq:hypersurfaces}). We have seen examples in~(\ref{eq:21}) and~(\ref{eq:24}). 

A general solution of~(\ref{eq:26}) can be obtained by standard techniques of solving linear ordinary differential equations. In the context of Picard--Fuchs operators and closed string mirror symmetry this is nicely explained in~\cite{Batyrev:1993wa}. We first have a look at the two examples. For the cases $d=8$, $L=(3,3,2,1,0)$ and $d=10$, $L=(4,3,2,1,0)$ we found
\begin{align}
  \label{eq:L8}
    \cL_B^{(8)}  &= 8 \theta ( 2 \theta - 1 )\left( \theta^4-16z(8\theta+1)(8\theta+3)(8\theta+5)(8\theta+7) \right),  \quad\text{and}\\
    \cL_B^{(10)}  &= 5 \theta \left( \theta^4-80z(10\theta+1)(10\theta+3)(10\theta+7)(10\theta+9) \right)
   \label{eq:5}
\end{align}
respectively. Their indices, i.e. solutions to the indicial equations, are
\begin{equation}
  \label{eq:7}
  \begin{array}{ccc}
     \text{singular point} & \cL_B^{(8)} & \cL_B^{(10)}\\
     \hline
     z=0      & \left(0,0,0,0,0,\tfrac{1}{2}\right) & (0,0,0,0,0)\\
     z=z_c    & (0,1,1,2,3,4) & (0,1,2,3,4)\\
     z=\infty & \left(\tfrac{1}{8},\tfrac{3}{8},\tfrac{1}{2},\tfrac{5}{8},\tfrac{7}{8},1\right) & \left(\tfrac{1}{10},\tfrac{3}{10},\tfrac{7}{10},\tfrac{9}{10},1\right)
  \end{array}
\end{equation}
where $z_c=4^{-8},z_c=4^{-4}5^{-5}$ are the conifold points of the $d=8$ and $d=10$ hypersurfaces, respectively. In particular, we observe that at $z=0$ there is no solution of the form $z^{\frac{1}{2}} + O(z)$ for $\cL_B^{(10)}$. We expect to have such a solution if we want to have nontrivial instanton contributions to $\cT_A$ in~(\ref{eq:T_A}) because of the following reason~\cite{Walcher:2006rs}. The K\"ahler parameter $t$ measures the area of a holomorphic sphere. A holomorphic disc can be viewed as half a holomorphic sphere, so it should contribute to $\cT_A$ with $q^{\frac{1}{2}}$. Now, since the mirror map is of the form $q=z + O(z^2)$ and $\varpi_0 = 1 + O(z)$, we expect from~(\ref{eq:open-string-mirror}), that $\cT_B$ should look like $z^{\frac{1}{2}} + O(z^{\frac{3}{2}})$. From that we conclude that there are no instanton corrections to the mirror of the brane $L=(3,3,2,1,0)$ in the $d=10$ case. 

Hence, we will mainly focus on the differential operator $\cL_B^{(8)}$ in~(\ref{eq:L8}). The solution corresponding to the index $\frac{1}{2}$ can easily be found to be
\begin{equation}
  \label{eq:40}
  \tau(z) = \frac{192}{\pi^2}\sum_{m\geq 0} \frac{\Gamma\left(8m+5\right)}{\Gamma\left(4m+3\right)\Gamma\left(m+\frac{3}{2}\right)^4} z^{m+\frac{1}{2}}.
\end{equation}
We have chosen the normalization such that $\tau(z) = z^{\frac{1}{2}} + \frac{98560}{9} z^{\frac{3}{2}} + \dots$. \\

It would be interesting to understand the physical meaning of the constant terms in the inhomogeneous terms $f(z)$ of the Picard--Fuchs equations for the normal functions, such as~(\ref{eq:20}) and~(\ref{eq:2})\footnote{In~\cite{Walcher:2007qp} it is shown that besides the disks $\cF^{(0,1)}$ there is a second contribution $\cR^{(0,0)}$ at worldsheet Euler number $-1$, coming from the crosscaps, which also satisfies the inhomogeneous Picard--Fuchs equation. It is tempting to relate the constant term with this additional contribution.}.

\subsection{Analytic continuation}
\label{sec:analyt-cont}

For the analytic continuation of $\tau(z)$, we will first consider the slightly more general form of solutions to the differential operator $(d\theta+k)\cL$, where $k=0,\dots,d-1$. We will then set $k=\frac{d}{2}$ at the end. The Barnes integral representation for the solution to the Picard--Fuchs equation $(d\theta+k)\cL \varpi =0$ takes the form
\begin{equation}
  \label{eq:Barnes}
  \tau(z) = \frac{K}{2\pi i} \int_C \diff{}{s} \frac{\Gamma\left(ds+1\right)\Gamma\left(s+\frac{k}{d}\right)\Gamma\left(-s+\frac{d-k}{d}\right)}{\prod_{i=1}^5 \Gamma\left(w_is+1\right)} \e{\pi i \left(s-\frac{k}{d}\right)} z^s 
\end{equation}
where
\begin{equation}
  \label{eq:41}
  K = \frac{\prod_{i=1}^5 \Gamma\left(\frac{w_ik}{d}+1\right)}{\Gamma\left(k+1\right)}.
\end{equation}
and $C = \{it | t\in\mR\}$. For $|z| < z_c$ we close the contour on the positive real axis, picking up the poles at $s=m$, $m=0,1,2,\dots$ we find
\begin{equation}
  \tau(z) = K \sum_{m\geq 0} \frac{\Gamma\left(dm+k+1\right)}{\prod_{i=1}^n \Gamma\left(w_im+\frac{w_ik}{d}+1\right)} z^{m+\frac{k}{d}}
\end{equation}
Setting $d=8$ and $w=(1,1,1,1,4)$, and $k=\frac{d}{2}$ reproduces~(\ref{eq:40}). For $|z| > z_c$ we close the contour on the negative real axis picking up the poles at $s+\frac{1}{2}=-m$, $m=0,1,2,\dots$, and at $ds=-m'-1$, $m'=0,1,2,\dots$. 
Hence we get from~(\ref{eq:Barnes}) 
\begin{equation}
  \begin{aligned}
    \tau(z) &= -K\sum_{m\geq0} \frac{\Gamma\left(-dm-k+1\right)}{\prod_{i=1}^n \Gamma\left(-w_im-\frac{w_ik}{d}+1\right)} \e{-i\pi\frac{2k}{d}}z^{-m-\frac{k}{d}}\\ 
    &\phantom{=} 
    + \frac{\pi}{d}K \sum_{m\geq 1} \frac{1}{\Gamma(m)\prod_{i=1}^n \Gamma\left(1-\frac{w_im}{d}\right)} \frac{\e{-i\pi\frac{k}{d}}}{\sin\pi\left(\frac{m}{d}+\frac{k}{d}\right)} \e{i\pi\frac{ m}{d}(d-1)} z^{-\frac{m}{d}}
  \end{aligned}
  \label{eq:tau_anal}
\end{equation}
We analyze the two terms in~\eqref{eq:tau_anal} in turn for $k=\frac{d}{2}$. The first one then reads (up to the factor $K$)
\begin{equation}
  \label{eq:25}
  \sum_{m=1}^\infty \frac{\Gamma(-dm+\frac{d}{2}+1)}{\prod_{i=1}^5\Gamma(-w_im+\frac{w_i}{2}+1)} z^{-m+\frac{1}{2}}
\end{equation}
We have to be careful about possible poles in this expression. This depends on whether $d$ is even or odd. Using the facts that $d=\sum_{i=1}^5 w_i$ and that $\sum_{i=1}^5 1 - \frac{2w_i}{d} = 3$ one finds that $d$ is odd if and only if all $w_i$ are odd, and that $d$ is even if and only if at least one of the $w_i$ is even. In the former case, $-dm +\frac{d}{2}+1$ is never an integer, hence there are no poles. If $d$ is even, however, $-dm +\frac{d}{2}+1$ is a negative integer for $m\geq 1$, hence the numerator always has a first order pole. But, since at least one the $w_i$ is even, the corresponding argument $-w_im +\frac{w_i}{2}+1$ in the denominator also is a negative integer for $m\geq 1$, and hence the denominator has at least a first order pole whenever the numerator does. 

In our examples, just one of the $w_i$ is even, call it $w_5$, hence the poles cancel. Using similar techniques as in Appendix A of~\cite{Hosono:1994ax} we compute
\begin{equation}
  \label{eq:80}
  \lim_{z\to m} \frac{\Gamma\left(-d(z+\frac{1}{2})+1\right)}{\Gamma\left(-w_5(z+\frac{1}{2})+1\right)} = (-1)^{\frac{d}{2}-\frac{w_5}{2}}\frac{w_5}{d}\frac{\Gamma\left(w_5(m-\frac{1}{2})\right)}{\Gamma\left(d(m-\frac{1}{2})\right)}, \qquad m = 1,2,\dots
\end{equation}
Taking this into account, we find for the first term in~\eqref{eq:tau_anal}
\begin{equation}
  \label{eq:74}
  (-1)^{\frac{d}{2}-\frac{w_5}{2}}\frac{w_5}{d}K\sum_{m=1}^\infty \frac{\Gamma\left(w_5(m+\frac{1}{2})\right)}{\Gamma\left(d(m+\frac{1}{2})\right)\prod_{i=1}^4\Gamma\left(w_i(-m+\frac{1}{2})+1\right)} z^{-m+\frac{1}{2}}
\end{equation}
Next, we look at the second term in the analytic continuation~\eqref{eq:tau_anal} of $\tau$. From the general fact that $\tau$ is only defined up to periods, we expect that its analytic continuation will consist of a solution $\tau^{(1)}$ to the analytically continued differential equation $\cL_{\text{PF}}\tau = f(z^{-\frac{1}{d}})$ and a contribution $\tau^{(2)}$ from the Gepner point periods. We choose as a basis of periods in local coordinates near the Gepner point $\varpi^{(G)} = (\varpi_0,\varpi_1,\varpi_2,\varpi_3)$ defined by $\varpi_k(\psi) = \varpi_0(\alpha^{k}\psi)$ where $\alpha$ is a $d$th root of unity. For the one--parameter Calabi--Yau hypersurfaces under investigation, they are given in~\cite{Klemm:1992tx}
\begin{equation}
  \label{eq:1}
  \varpi_j(\psi) = -\frac{\pi}{d} \sum_{n=1}^\infty \frac{1}{\Gamma(n)\prod_{i=1}^5\Gamma(1-\frac{n}{d}w_i)}\frac{\e{i\pi\frac{n}{d}(d-1)}}{\sin\pi\frac{n}{d}}\e{2\pi i \frac{n}{d} j}\left(C\psi\right)^n, 
\end{equation}
and can be rewritten as, if $w_1=1$, 
\begin{equation}
  \label{eq:3}
  \varpi_j(\psi) = -\frac{1}{d} \sum_{n=1}^\infty \frac{\Gamma(\frac{n}{d}w_1)}{\Gamma(n)\prod_{i=2}^5\Gamma(1-\frac{n}{d}w_i)}\e{i\pi\frac{n}{d}(d-1)}\e{2\pi i \frac{n}{d} j}\left(C\psi\right)^n.
\end{equation}
On the other hand, the second term in~\eqref{eq:tau_anal} 
can be rewritten (again up to the factor $K$) as follows, if $w_1=1$, 
\begin{equation}
  \label{eq:4}
  \frac{1}{d} \sum_{n=1}^\infty \frac{\Gamma(\frac{n}{d}w_1)}{\Gamma(n)\prod_{i=2}^5\Gamma(1-\frac{n}{d}w_i)}\frac{\e{-i\pi\frac{k}{d}}\sin\pi\frac{n}{d}}{\sin \pi\left(\frac{n}{d}+\frac{k}{d}\right)}\e{i\pi \frac{n}{d}(d-1)}\left(C\psi\right)^n  
\end{equation}
We would like to express~(\ref{eq:4}) in terms of~(\ref{eq:3}). For this purpose, we first take a closer look at the periods $\varpi_j$. Suppose again that $w_5$ is the only even weight. Then we observe that the product $\prod_{i=2}^5\Gamma(1-\frac{n}{d}w_i)$ has a pole of order at least 1 for $n\in \frac{d}{w_5}\mZ$ due to the factor $\Gamma(1-\frac{w_5}{d})$, hence the corresponding coefficient vanishes. In particular, we can replace the sum in~(\ref{eq:4}) as follows
\begin{equation}
  \label{eq:6}
  \sum_{n=1}^\infty \longrightarrow \sum_{\substack{n=1\\n\not = \frac{d}{w_5} \mod d}}^\infty
\end{equation}
without changing anything. 
For the expression in~\eqref{eq:4}, however, the situation is slightly different if $k=\frac{d}{2}$. Then the factor $\sin \pi\left(\frac{n}{d}+\frac{1}{2}\right)$ has a simple zero for $n=\frac{d}{2} \mod d$, at which there is also a simple pole from the factor $\Gamma(1-\frac{w_5n}{d})$. The same way we derived~\eqref{eq:80}, we then find that 
\begin{equation}
  \label{eq:72}
  \lim_{z\to dm-\frac{d}{2}} \frac{1}{\sin \pi\left(\frac{z}{d}+\frac{1}{2}\right)\Gamma(1-\frac{w_5z}{d})} = \frac{w_5}{\pi}(-1)^{m+\frac{w_5}{2}}\Gamma(w_5m-\tfrac{w_5}{2}),\qquad m=1,2,\dots
\end{equation}
So we can decompose the sum into the terms for which $n\not=\frac{w_5}{d} \mod d$ (this also excludes $n=\frac{d}{2} \mod d$) and those for which $n=\frac{d}{2} \mod d$, the remaining terms vanish. We find
\begin{equation}
  \label{eq:73}
  \begin{aligned}
    &  \frac{1}{d} \sum_{\substack{n=1\\n\not = \frac{d}{w_5} \mod d}}^\infty \frac{\Gamma(\frac{n}{d}w_1)}{\Gamma(n)\prod_{i=2}^5\Gamma(1-\frac{n}{d}w_i)}\frac{\e{-i\pi\frac{1}{2}}\sin\pi\frac{n}{d}}{\sin \pi\left(\frac{n}{d}+\frac{1}{2}\right)}\e{i\pi \frac{n}{d}(d-1)}\left(C\psi\right)^n  \\
    & +(-1)^{\frac{w_5}{2}-\frac{d}{2}}\frac{w_5}{d\pi} \sum_{m=1}^\infty
    \frac{\Gamma(w_1m-\frac{w_1}{2})\Gamma(w_5m-\frac{w_5}{2})}{\Gamma(dm-\frac{d}{2})\prod_{i=2}^4\Gamma(-w_im+\frac{w_i}{2}+1)}\sin\pi\left(m-\tfrac{1}{2}\right)\left(C\psi\right)^{dm-\frac{d}{2}}
  \end{aligned}
\end{equation}
Using $\sin (\pi z) \Gamma(z) = \frac{\pi}{\Gamma(1-z)}$ we can rewrite the second term of~(\ref{eq:73}) and obtain
\begin{equation}
  \label{eq:77}
  \begin{aligned}
    &\frac{1}{d} \sum_{\substack{n=1\\n\not = \frac{d}{w_5} \mod d}}^\infty \frac{\Gamma(\frac{n}{d}w_1)}{\Gamma(n)\prod_{i=2}^5\Gamma(1-\frac{n}{d}w_i)}\frac{\e{-i\pi\frac{k}{d}}\sin\pi\frac{n}{d}}{\sin \pi\left(\frac{n}{d}+\frac{k}{d}\right)}\e{i\pi \frac{n}{d}(d-1)}\left(C\psi\right)^n  \\
    &+(-1)^{\frac{w_5}{2}-\frac{d}{2}}\frac{w_5}{d} \sum_{m=1}^\infty
    \frac{\Gamma(w_5m-\frac{w_5}{2})}{\Gamma(dm-\frac{d}{2})\prod_{i=1}^4\Gamma(-w_im+\frac{w_i}{2}+1)}\left(C\psi\right)^{dm-\frac{d}{2}}
  \end{aligned}  
\end{equation}
Including the contribution from~(\ref{eq:74}) and observing that $(-1)^{\frac{w_5}{2}-\frac{d}{2}}=1$ we find that the analytic continuation of $\tau$ becomes:
\begin{equation}
  \label{eq:75}
  \begin{aligned}
    \tau &= \tau^{(1)} + \tau^{(2)}\\
    &=\frac{2w_5}{d} K \sum_{m=1}^\infty
    \frac{\Gamma(w_5m-\frac{w_5}{2})}{\Gamma(dm-\frac{d}{2})\prod_{i=1}^4\Gamma(-w_im+\frac{w_i}{2}+1)}\left(C\psi\right)^{dm-\frac{d}{2}}\\
    &\phantom{=} + \frac{1}{d} K \sum_{\substack{n=1\\n\not = \frac{d}{w_5} \mod d}}^\infty \frac{\Gamma(\frac{n}{d}w_1)}{\Gamma(n)\prod_{i=2}^5\Gamma(1-\frac{n}{d}w_i)}\frac{\e{-i\pi\frac{1}{2}}\sin\pi\frac{n}{d}}{\sin \pi\left(\frac{n}{d}+\frac{1}{2}\right)}\e{i\pi \frac{n}{d}(d-1)}\left(C\psi\right)^n  
  \end{aligned}
\end{equation}
For the remainder of this subsection we restrict ourselves to the case $d=8$.
In order to express $\tau^{(2)}$ in terms of the periods $\varpi_j$ we have to rewrite the quotient involving the sines. 
We set $g=\e{-\pi i \frac{2m-1}{d}}$ and $\alpha=\e{2\pi i \frac{k}{d}}$. Note that $g^d=-1$. Then we find
\begin{equation}
  \label{eq:79}
  \frac{\e{-\frac{i\pi}{d}}\sin\pi(\frac{2m-1}{d})}{\sin\pi(\frac{2m-1}{d}+\frac{1}{2})} = -\sum_{l=1}^{\frac{d}{2}-1} \left(-g^2\right)^l  
\end{equation}
Hence, we find for $d$ even (and $k=\frac{d}{2}$) that
\begin{equation}
  \label{eq:8}
  \begin{aligned}
    \tau^{(2)} &= \frac{K}{d} \sum_{n=1}^\infty
    \frac{\Gamma(\frac{n}{d})}{\Gamma(n)\prod_{i=1}^4\Gamma(1-\frac{n}{d}w_i)}
    \sum_{l=1}^{\frac{d}{2}-1} (-1)^{l+1} g^{2l}
    \e{i\pi \frac{n}{d}(d-1)}\left(C\psi\right)^n\\
    & = -\sum_{l=1}^{\frac{d}{2}-1} (-1)^l \varpi_{l}
  \end{aligned}
\end{equation}
where we take the index of $\varpi_l$ modulo $d$. 
For the case $d=8$ we obtain therefore 
\begin{equation}
  \label{eq:28}
  \tau^{(1)} = \frac{192}{\pi^2}\sum_{m=1}^\infty \frac{\Gamma(-8m+5)}{\Gamma(-4m+3)\Gamma(-m+\frac{3}{2})^4} z^{-m+\frac{1}{2}}  
\end{equation}
and
\begin{equation}
  \label{eq:12}
  \tau^{(2)} = \frac{192}{\pi^2}\left(-\varpi_1 +\varpi_2 - \varpi_3\right)
\end{equation}


\subsection{Monodromies}
\label{sec:monodromies}

Next, we want to study the behavior of $\tau$ around the Gepner point.  There is a $\mZ_d$ monodromy $A$ around this point, sending $\varpi_i(\alpha\psi)$ to $A\varpi_i(\alpha\psi)=\varpi_i(\alpha\psi) = \varpi_{i+1}(\psi)$ in the Gepner point basis of periods~(\ref{eq:3}). Here we need to express $\varpi_4$ in terms of $\varpi_0,\dots,\varpi_3$ depending on the model. This basis is related to the one near the large complex structure limit~(\ref{eq:pi_L}) by
\begin{equation}
  \label{eq:basis transformation}
  \varpi^{(L)} = M\varpi^{(G)}.
\end{equation}
This allows us to express the monodromy $A$ in terms of the basis $\varpi^{(L)}=(w_3,w_2,w_0,w_1)$. We consider the case $d=8$, $k=4$. First, we need the monodromy matrices and the change of basis $M$, which we can take from~\cite{Klemm:1992tx} or~\cite{Scheidegger:1999ed} up to permutation of the rows and columns. We reproduce them here in our conventions. The monodromy matrix $A^{(G)}$ in the basis $\varpi^{(G)}$ reads
\begin{equation}
  \label{eq:29}
  A^{(G)} = \left( \begin {array}{cccc} 0&1&0&0\\\noalign{\medskip}0&0&1&0
\\\noalign{\medskip}0&0&0&1\\\noalign{\medskip}-1&0&0&0\end {array}
 \right)
\end{equation}
The basis transformation is
\begin{equation}
  \label{eq:30}
  M =   \left( \begin {array}{cccc} -1&1&0&0\\\noalign{\medskip}\frac{3}{2}&\frac{3}{2}&\frac{1}{2}&-\frac{1}{2}\\\noalign{\medskip}1&0&0&0\\\noalign{\medskip}-\frac{1}{2}&\frac{1}{2}&\frac{1}{2}&\frac{1}{2}
\end {array} \right)
\end{equation}
This yields the monodromy matrix $A^{(L)} = M A^{(G)} M^{-1}$. We will also need the monodromy matrix $T^{(L)}$ around the conifold point and the monodromy matrix $T_\infty^{(L)}$ around the large complex structure limit in the large volume basis $\varpi^{(L)}$ 
\begin{equation}
  \label{eq:31}
  \begin{aligned}
    A^{(L)} &= \left( \begin {array}{cccc}
        -3&1&-4&1\\\noalign{\medskip}-1&1&-1&2
        \\\noalign{\medskip}1&0&1&0\\\noalign{\medskip}-1&0&-1&1\end
      {array} \right)
    & 
  T^{(L)} &= \left( \begin {array}{cccc} 1&0&0&0\\\noalign{\medskip}0&1&0&0
\\\noalign{\medskip}-1&0&1&0\\\noalign{\medskip}0&0&0&1\end {array}
 \right)  
    &
  T_\infty^{(L)} &= \left( \begin {array}{cccc} 1&-1&4&1\\\noalign{\medskip}0&1&-1&-2
\\\noalign{\medskip}0&0&1&0\\\noalign{\medskip}0&0&1&1\end {array}
 \right)
  \end{aligned}
\end{equation}
Following the argument of~\cite{Walcher:2006rs}, we can assume from the general form of the A--model domain wall tension $\cT_A$ in~(\ref{eq:T_A}) that the $\cT_{B,\pm}$ takes the form 
\begin{equation}
  \label{eq:32}
  \cT_{B,\pm}(z) = \frac{w_1(z)}{2} \pm b w_0(z) \pm a\tau(z) 
\end{equation}
where we write $\tau = \tau^{(1)} + \tau^{(2)}$ as before, and $a$ and $b$ are yet to be determined. Applying $M^{-1}$ to the expression in~(\ref{eq:12}), we can write it in terms of the large complex structure basis
\begin{equation}
  \label{eq:33}
  \tau^{(2)} = \frac{\pi^2}{192}\left(7w_0 + 4w_3 - 2 w_2\right)  
\end{equation}
With this information we can now determine $A\cT_{\pm}$. The monodromy sends $\psi \to \e{\frac{2\pi i}{8}}\psi$, hence $z^{-\frac{1}{8}} \to \e{\frac{2\pi i}{8}} z^{-\frac{1}{8}}$. A look at~(\ref{eq:28}) yields for $\tau^{(1)}$ 
\begin{equation}
  \label{eq:34}
  A^{(G)}\tau^{(1)} = -\tau^{(1)}
\end{equation}
Using~(\ref{eq:12}) and~(\ref{eq:29}) we obtain
\begin{equation}
  \label{eq:35}
  A^{(G)}\tau^{(2)} = \frac{\pi^2}{192}\left(-\varpi_0 +\varpi_2 - \varpi_3  \right)
\end{equation}
where we have used the relations~\cite{Klemm:1992tx}
\begin{equation}
  \label{eq:81}
  \varpi_j + \varpi_{4+j} = 0,\qquad j=0,\dots,3.
\end{equation}
Next, we express this in terms of the basis $\varpi^{(L)}$, by applying $M^{-1}$ to it which yields
\begin{equation}
  \label{eq:36}
  A^{(L)}\tau^{(2)} = \frac{\pi^2}{192}\left(-7w_0 + 2w_2 - 3w_3\right)
\end{equation}
Finally, we need the transformation of $w_0$ and $w_1$ under $A^{(L)}$ which we can read off directly from~(\ref{eq:31}). Plugging this and (\ref{eq:34}), (\ref{eq:36}), into (\ref{eq:32}) we find
\begin{equation}
  \label{eq:39}
    A^{(L)}\cT_{+} = \frac{w_1}{2} + \left(-\frac{1}{2}+b\right)w_0 - a\tau + \left(-\frac{1}{2}+b+\frac{a\pi^2}{192}\right)w_3
\end{equation}
In order to obtain $\cT_{-}$ the $w_3$ has to vanish, and together with the condition for the $w_0$ term we find $a=\frac{48}{\pi^2}$ and $b=\frac{1}{4}$. We could equally well take an integer multiple of $a$, however plugging this result into the ansatz~(\ref{eq:32}) shows that
\begin{equation}
  \label{eq:13}
    \cT_{B,\pm} = \frac{w_1}{2} \pm \frac{w_0}{4} \pm  \frac{48}{\pi^2}\sum_{m\geq 0} \frac{\Gamma\left(8m+5\right)}{\Gamma\left(4m+3\right)\Gamma\left(m+\frac{3}{2}\right)^4} z^{m+\frac{1}{2}}.
\end{equation}
is precisely a solution to the inhomogeneous Picard--Fuchs equation~(\ref{eq:20}). 

Next, we consider the monodromy around the conifold point. Using an extension of an argument of~\cite{Candelas:1990rm}, the following was shown in~\cite{Walcher:2006rs} for the quintic. According to~(\ref{eq:31}), $w_0 \to w_0-w_3$ when going around $z=z_c$. Therefore, $w_0(z) = w_3(z)(z-z_c) + g(z)$ with $g(z)$ a holomorphic function. Furthermore, $w_3$ is the vanishing period at the conifold point, i.e. it has a expansion of the form $w_3 = A(z-z_c) + B(z-z_c)^2+O\left((z-z_c)^3\right)$. Suppose $\tau(z)$ has an expansion $C w_3(z)(z-z_c) + h(z)$ with $h(z)$ a holomorphic function. Then upon taking second derivatives and taking the limit $z\to z_c$ one finds that $C=1$ and $g=h$. The same is true here. Therefore:
\begin{equation}
  \label{eq:38}
  T^{(L)} \cT_{B,\pm} = \frac{w_1}{2} \pm \frac{w_0}{4} \mp \frac{w_3}{4} \pm  \frac{48}{\pi^2}\tau \pm \frac{48}{\pi^2}\frac{\pi^2}{192}w_3 = \cT_{B,\pm}
\end{equation}
To conclude we have the following behavior of $\cT_{B,\pm}$ under the monodromies of $\cL_{\text{PF}}$:
\begin{itemize}
  \item invariance under the conifold monodromy $T$
  \item the behavior under the $B$-field monodromy $T_\infty$ as described in (2.5) and (2.6) of~\cite{Walcher:2006rs}, i.e.
    \begin{equation}
      \label{eq:56}
      \begin{aligned}
        \cT_{B,-}(w_1+w_0)&=\cT_{B,+}(w_1), & \cT_{B,+} + \cT_{B,-} &= w_1
      \end{aligned}
    \end{equation}
  \item the behavior under the Gepner monodromy 
    \begin{equation}
      \begin{aligned}
        A^{(L)}\cT_{+} &= \cT_{-} \\
        A^{(L)}\cT_{-} &= \cT_{+} 
      \end{aligned}
      \label{eq:57}
    \end{equation}
  \item and finally
    \begin{equation}
      \label{eq:58}
      A^{(L)}T^{(L)}T_\infty^{(L)}=\bb1
    \end{equation}
\end{itemize}
These conditions are all consistent, and the extended monodromy matrices take the form
\begin{equation}
  \label{eq:60}
    \begin{aligned}
  A^{(L)} &=  \left( \begin {array}{ccccc} -1&0&0&0&1\\\noalign{\medskip}0&-3&1&-4&1\\\noalign{\medskip}0&-1&1&-1&2
\\\noalign{\medskip}0&1&0&1&0\\\noalign{\medskip}0&-1&0&-1&1\end {array}
 \right)
  &    
  T^{(L)} &= \left( \begin {array}{ccccc} 1&0&0&0&0\\\noalign{\medskip}0&1&0&0&0\\\noalign{\medskip}0&0&1&0&0
\\\noalign{\medskip}0&-1&0&1&0\\\noalign{\medskip}0&0&0&0&1\end {array}
 \right)  
  &
  T_\infty^{(L)} &= \left( \begin {array}{ccccc} -1&0&0&1&1\\\noalign{\medskip}0&1&-1&4&1\\\noalign{\medskip}0&0&1&-1&-2
\\\noalign{\medskip}0&0&0&1&0\\\noalign{\medskip}0&0&0&1&1\end {array}
 \right)
  \end{aligned}  
\end{equation}
We note that $\left(A^{(L)}\right)^8=\bb1$. We observe that the extension takes the same form as the one for the quintic in~\cite{Walcher:2006rs}. This can probably be argued to be true in general on the basis of the behavior of the original monodromies and the conditions imposed above.

\subsection{Real BPS invariants}
\label{sec:instanton-expansion}

Now we are ready to compute the instanton expansion in~(\ref{eq:T_A}). We collect the ingredients for performing the open string mirror computation
\begin{equation}
  \label{eq:open-string-mirror2}
  \cT_A(t) = \varpi_0(z(t))^{-1} \cT_B(z(t)).
\end{equation}
We get the fundamental period of the Picard--Fuchs operator in~(\ref{d8pf}) from~(\ref{eq:pi0}):
\begin{eqnarray}
  \label{eq:w0}
  \varpi_0(z) = w_0(z) &=&\sum_{m=0}^{\infty}\frac{(8m)!}{(m!)^4(4m)!}z^m \nonumber \\
&=& 1+1680\,z+32432400\,{z}^{2} + O(z^3)
\end{eqnarray}
The normal function part of the domain wall tension $\cT_B$, which satisfies the inhomogeneous Picard--Fuchs equation (\ref{eq:20}), is:
\begin{eqnarray}
  \label{eq:TB}
  \cT_B(z) &=&\frac{48}{\pi^2}\tau(z)=\frac{1}{4}\sum_{m=0}^{\infty}\frac{\Gamma(8m+5)}{\Gamma(4m+3)\Gamma(m+\frac{3}{2})^4}z^{m+\frac{1}{2}}\nonumber\\
&=&\frac{1}{\pi^2}\left( 48\,z^{\frac{1}{2}}+{\frac {1576960}{3}}\,{z}^{\frac{3}{2}}+{\frac {339028738048}{25}}\,{z}^{\frac{5}{2}} + O(z^{\frac{7}{2}})\right)
\end{eqnarray}
Furthermore we need the logarithmic solution $w_1$ from~(\ref{eq:periodexpansion})
\begin{eqnarray}
  \label{eq:w1}
  w_1(z) &=&w_0(z) \log z+4\sum_{m=0}^{\infty}\frac{(8m)!}{(m!)^4(4m)!}z^m[2\Psi(1+8m)-\Psi(1+4m)-\Psi(1+m)] \nonumber\\
&=& w_0(z) \log z + 15808\,z+329980320\,{z}^{2}+ O(z^3),
\end{eqnarray}
where $\Psi$ is the Polygamma function. This yields for the inverse $z(q)$ mirror map $q(z)=\e{2 \pi i t(z)}$ with $t(z) = \frac{w_1(z)}{w_0(z)}$
\begin{equation}
  \label{eq:inversemirrormap}
  z(q) = q-15808\,{q}^{2}+71416416\,{q}^{3} + O(q^4)
\end{equation}
Inserting all this into~(\ref{eq:open-string-mirror2}), we get:
\begin{equation}
  \label{eq:F01B}
  \cT_A(q) = 48\,{q}^{\frac{1}{2}}+{\frac {196864}{3}}\,{q}^{\frac{3}{2}} + O({q}^{\frac{5}{2}})
\end{equation}
By definition this is the (quantum part of the) generation function $\cF^{(0,1)}$ of maps of holomorphic discs whose expansion is~\cite{Ooguri:1999bv}
\begin{equation}
  \label{eq:F01A}
  \cT_A(q) = \cF^{(0,1)}(q) = \sum_{\substack{d \geq 0\\ d \in 2\mZ+1}} \sum_{k|d} \frac{1}{k^2} n^{(0,1)}_d q^{\frac{dk}{2}}.
\end{equation}
From this we can read off the BPS invariants $n^{(0,1)}_d$. The result is displayed in Table~\ref{tab:d8instantons}.\\\\

We can now try to compute the BPS invariants for open worldsheets with Euler character $\chi=0$. As mentioned in Section~\ref{sec:general-remarks}, we have not only to consider holomorphic maps of annuli but also to include unoriented worldsheets like the Klein bottle. In principle, we can determine the annulus invariants using the extended holomorphic anomaly equations~\cite{Walcher:2007tp,Walcher:2007qp}. The central ingredient here is the Griffiths infinitesimal invariant~\cite{Griffiths:1983ab} $\Delta_{zz}$ which, in the holomorphic limit, is related to the domain wall tension $\cT_B$ in the following way\footnote{In the following we will restrict to the one--parameter case.}:
\begin{equation}
  \label{eq:42}
  \Delta_{zz} = D_zD_z\cT_B(z).
\end{equation}
Furthermore we will need the terminator $\Delta^z$ which has the following simple form:
\begin{equation}
  \label{eq:44}
  \Delta^z = -\frac{\Delta_{zz}}{C_{zzz}}
\end{equation}
The holomorphic anomaly equation for the generating function $\cF^{(0,2)}$ of holomorphic maps of Riemann surfaces with genus $g=0$ and $h=2$ boundaries~(\ref{eq:HAE_Annuli}) reads
\begin{equation}
  \label{eq:HAE_F02}
  \partial_{\ibar} \partial_j\cF^{(0,2)}_B = -\Delta_{jk} \Delta^{k}_{\ibar} + \frac{N}{2}g_{\ibar j} 
\end{equation}
Since there is exactly one B--brane before the orientifold projection we set $N=0$. Then,~(\ref{eq:HAE_F02}) can be integrated to yield (for $h^{1,1}(X)=1$)
\begin{equation}
  \label{eq:43}
  \partial_z \cF^{(0,2)}_B = -\Delta_{zz} \Delta^{z} + f^{(0,2)}_z(z)
\end{equation}
where $f^{(0,2)}_z(z)$ is the holomorphic ambiguity. In~\cite{Walcher:2007tp} it was observed in three examples by comparison to a localization computation in the A--model that this ambiguity can be set to zero. The localization computation is not available for our brane, since we do not know the explicit form of the A--brane $L$. Therefore, we naively assume that we can set the ambiguity to zero in our example as well. Defining the expansion
\begin{equation}
  \label{eq:F02A}
  4\cA_A = \cF^{(0,2)}_A = \sum_{\substack{d \geq 0\\d \in 2\mZ}} \sum_{\substack{k|d\\k \in 2\mZ+1}} \frac{1}{k} n^{(0,2)}_d q^{\frac{dk}{2}},
\end{equation}
we find from integrating~(\ref{eq:43}) using~(\ref{eq:44}) with 
\begin{equation}
  \label{eq:Yukawa}
  C_{zzz} = \frac{2}{1-z_c}
\end{equation}
the following result
\begin{equation}
  \label{eq:annulus_exp}
  \cA_A = 144\,q+709632\,{q}^{2}+21513266688\,{q}^{3} + O(q^4).
\end{equation}
There is also a holomorphic anomaly equation for the Klein bottle contribution $\cK_B$~\eqref{eq:HAE_Klein} which depends on the choice of the orientifold projection. We pointed out in Section~\ref{sec:LG-CY} that the brane $L=(3,3,2,1,0)$ satisfies the tadpole cancellation with the trivial orientifold projection. The relevant holomorphic anomaly equation is then
\begin{equation}
  \label{eq:HAE_Klein2}
    \partial_{\ibar} \partial_j\cK_B = \frac{1}{2} C_{jkl}C_{\ibar}^{kl} - G_{\ibar j} 
\end{equation}
This can be integrated using the special geometry relation to 
\begin{equation}
  \label{eq:Klein}
  \cK_B = \frac{1}{2} \log\left( \frac{q}{z} \pd{z}{q} \left|f^{(1,0)_k}\right|^2\right) 
\end{equation}
In~\cite{Walcher:2007tp} the holomorphic limit of $\cK_B$ was again compared with the localization computation in the A--model with the result that the holomorphic ambiguity $f^{(1,0)_k}$ seems to have the universal property that
\begin{equation}
  \label{eq:54}
  f^{(1,0)_k} = \delta^{-\frac{1}{4}}
\end{equation}
where $\delta=1-z_c$ is the discriminant at the conifold point. In the same way as before, due to the lack of a localization computation in the A--model, we assume that this behavior persists in our example as well. We find
\begin{equation}
  \label{eq:Klein_exp}
  \cK_A = -288\,q-22933088\,{q}^{2}-867789979648\,{q}^{3} + O(q^4)
\end{equation}
Finally, according to~\cite{Walcher:2007tp}, we sum annulus and Klein bottle contributions and expand in the holomorphic limit 
\begin{equation}
  \label{eq:50}
  \cA_A+\cK_A = 2 \sum_{\substack{d\geq 0\\d\in 2\mZ}}\sum_{\substack{k|d\\k\in 2\mZ+1}} \frac{1}{k} n^{(1,\text{real})}_d q^{\frac{kd}{2}} 
\end{equation}
to extract the real BPS invariants $n^{(1,\text{real})}$. They are listed in Table~\ref{tab:d8instantons}.
\begin{table}[ht]
  \begin{center}
    $
    \begin{array}{|r|r|r|}
      \hline
      d & n_d^{(0,\text{real})} & n_d^{(1,\text{real})}\\
      \hline
       1 & 48 & \\
       2 & & -72 \\
       3 & 65616 & \\
       4 & & -11111728 \\
       5 & 919252560& \\
       6 & & -423138356456\\
       7 & 17535541876944& \\
       8 & & -15627318184690224\\
       9 & 410874634758297216& \\
      10 & & -580819145044133296088 \\
      11 & 10854343378339853472336& \\
      12 & & -21851106460968509703283952 \\
      13 & 310521865321872322311676752& \\
      14 & & -31963310253709759062935592792857136\\
      15 & 9401030537961826351061423123760& \\
      \hline
    \end{array}
    $
  \end{center}
  \caption{BPS invariants $n_d^{(m,\text{real})}$, $m=0,1$ for the A--brane on $\mP(1,1,1,1,4)[8]$ mirror to the B--brane $L=(3,3,2,1,0)$.}
  \label{tab:d8instantons}
\end{table}
We do not know a reason for the $n_d^{(1,\text{real})}$ all being negative. The obvious guess is that this is due to a wrong choice of the holomorphic ambiguities. The integrality of the $n_d^{(1,\text{real})}$ is not a particularly strong consistency check on this choice since it seems to be easy to adjust them and still get integers. It could also, however, hint at a geometric property of the special Lagrangian submanifold $L$. Similar effects are known from closed string BPS invariants. 

One could now go on to worldsheets with larger Euler number for which would have to solve in general a recursive system of holomorphic anomaly equations. Here, the canonical generators for these recursions found in~\cite{Alim:2007qj} should turn out to be very useful.

\subsection{Semi--Periods}
\label{sec:semiperiods}

This section is more of speculative nature.  Choose a local coordinate patch $x_l=1$, $l\not=1$  of the weighted projective space $\mP(w)$ with homogeneous coordinates $(x_1:\ldots:x_5)$ and inhomogeneous coordinates $\xi^{(l)}_i$, $i=1,2,3,4$. Then one can define~\cite{Berglund:1993yn} particular V--shaped 3--chains $V_k$ on $X=\mP(w)[d]$ through 
\begin{equation}
  \label{eq:V}
  \begin{aligned}
    V^{(l)}_k = \{(\xi^{(l)}_i) | &\xi^{(l)}_l=1 , \xi^{(l)}_{i_1},\xi^{(l)}_{i_2},\xi^{(l)}_{i_3}, \text{ real and positive, for } i_1,i_2,i_3 \not=1, l;\\ & x_1^{(l)} \text{ is a solution to $W=0$ on the branch arg$(\xi_1^{(l)}) \to \pi + \tfrac{2\pi
        k}{d}$ as $z^{-\frac{1}{d}} \to 0$} \}.
  \end{aligned}
\end{equation}
The 3--chain on the Calabi-Yau threefold $X$ is then the union over all patches $V_k = \bigcup_{l=1}^5 V^{(l)}_k$. The monodromy matrix $A$ around the Gepner point (cf.~\ref{sec:monodromies}) acts naturally on $V$ by multiplying the coordinates $x_i$ by phases and shifting arg$(\xi_1^{(l)})$ by the angle $\frac{2\pi i}{d}$. Hence one can build linear combinations of 3--chains $\Gamma_{m,k} = A^mV_k$ such that certain boundaries $B=\partial  \Gamma_{m,k}$ get identified. In particular, the 3--chain
\begin{equation}
  \label{eq:51}
  \gamma = (1-A^{w_2})(1-A^{w_3})(1-A^{w_4})V_j
\end{equation}
is a cycle. In the same way, one can build 3--chains $\Gamma$ such that $\partial\Gamma = C_+ - C_-$, as required for a normal function. We would like to do this in such a way that 
\begin{equation}
  \label{eq:49}
  \cL_{\text{PF}} \int_\Gamma \Omega(z) = f(z)
\end{equation}
for some $f(z)$, say $f(z) = z^{\frac{1}{2}}$. Here, the semi--periods come into the story. By definition~\cite{Avram:1995wt} a semi--period is a solution $\sigma$ to the GKZ hypergeometric system $\cL_{\text{GKZ}}\sigma = 0$ associated to a Calabi--Yau threefold, which is not a solution of the corresponding Picard--Fuchs operator, i.e. $\cL_{\text{PF}}\sigma \not = 0$. This means that $\sigma$ is necessarily an integral of $\Omega$ over a 3--chain with nontrivial boundary. Let us therefore briefly recall the relation between the GKZ hypergeometric system and the PF system~\cite{Batyrev:1994hm,Hosono:1993qy,Stienstra:2005nr}. A weighted projective space is a toric variety, and toric varieties can be encoded in terms of fans of cones in a lattice polytopes (for a concise review in the context of Calabi--Yau threefolds see~\cite{Kreuzer:2006ax}). We will not go into the details of toric varieties here except for the fact that there are linear relations $l^{(a)}, a=1,\ldots,h$ among the lattice points of such a polytope. For example, the lattice polytope, traditionally called $\Delta^*$, of the weighted projective space $\mP(w)$ with $w_1=1$ is given by the vertices
\begin{equation}
  \label{eq:vertices}
  \begin{aligned}
    \rho_i &= e_i, i=1,\dots,4, &\quad \rho_5 &= -w_2e_1-w_3e_2-w_4e_3-w_5e_4
  \end{aligned}
\end{equation}
where $e_i$ is the standard basis for the lattice $\mZ^4$. It is straightforward to see that there is one relation $\sum_{i=1}^5 l^{(1)}_i\rho_i = 0$ among these vertices with $l_i^{(1)}= w_i$. Now, define $l_0^{(a)} = -\sum_{i=1}^5 l^{(a)}_i$. In our example, $l^{(1)}_0=-d$. Given a basis $l^{(a)}$ of such linear relations, the GKZ system of differential operators $\cL_a$ is~\cite{Hosono:1994ax}
\begin{equation}
  \label{eq:gkz}
  \cL_a=\prod_{l_i^{(a)}>0}\prod_{j=0}^{l_i^{(a)}-1}\left(\sum_{b=1}^hl_i^{(b)}\theta_{b}-j\right)-\prod_{l_i^{(a)}<0}\prod_{j=0}^{-l_i^{(a)}-1}\left(\sum_{b=1}^hl_i^{(b)}\theta_{b}-j \right)z_a,
\end{equation}
where $\theta_a=z_a\frac{d }{d  z_a}$ and we represent a linear relation $l^{(a)}$ by a vector 
\begin{equation}
l^{(a)}=(l^{(a)}_{0};l^{(a)}_1,\ldots,l^{(a)}_s).
\end{equation} 
For one--parameter hypersurfaces this simplifies a lot. Here we have just 
\begin{equation}
l=(-d;w_1,w_2,w_3,w_4,w_5)
\end{equation}
The fundamental period is:
\begin{equation}
\label{fperiod}
\varpi_0=\sum_{n_1,\ldots,n_h}\left[\frac{\left(-\sum_{a=1}^hn_al_{0}^{(a)}\right)!}{\prod_{j=1}^s\left(\sum_{a=1}^hn_al_{j}^{(a)}\right)!}\prod_{a=1}^h\left((-1)^{l_{0}^{(a)}}z_a\right)^{n_a} \right],
\end{equation}
and reduces for $\mP(w)$ to the one given in~(\ref{eq:pi0}). Considering the case $\mP(1,1,1,1,4)[8]$, we obtain from~(\ref{eq:gkz})
\begin{equation}
  \label{eq:gkzd8}
  \cL_{\text{GKZ}}^{(8)} = \left(4\theta\right)\left(4\theta-1\right)\left(4\theta-2\right)\left(4\theta-3\right)\left(\theta^4 - 16\left(8\theta-1\right)\left(8\theta-3\right)\left(8\theta-5\right)\left(8\theta-7\right)z\right)
\end{equation}
Returning to the semi--periods $\sigma$, it was argued, though not proven, in~\cite{Avram:1995wt} that there are 3--chains $\Gamma$ that are appropriate linear combinations of the $\Gamma_{m,k}$ such that $\sigma = \int_\Gamma \Omega$ is a solution to $\cL_{\text{GKZ}}\sigma=0$. We observe that the last factor in~(\ref{eq:gkzd8}) is precisely the Picard--Fuchs operator $\cL_{\text{PF}}$ in~(\ref{d8pf}) and moreover that
\begin{equation}
  \label{eq:52}
  \cL_{\text{GKZ}}^{(8)} = \left(4\theta-1\right)\left(4\theta-3\right)\cL_B^{(8)}
\end{equation}
where $\cL_B^{(8)}$ is precisely the differential operator annihilating the normal function $\cT_B$ in~(\ref{eq:21}). Hence, we conclude that the normal function is a semi--period. This was already observed in the case of the cubic elliptic curve in~\cite{Brunner:2004mt}. This discussion reveals that we have an alternative representation of the 3--chain $\Gamma$ in Section~\ref{sec:d=8-hypersurface} that led to $\cT_B$, namely in terms of the building blocks $\Gamma_{m,k} = A^mV_k$. It would be interesting to relate the two ways of obtaining such 3--chains. In particular, we could turn the argument around and use the procedure presented in this section to construct normal functions with desired properties such as inhomogeneous terms $f(z)$ of a particular form like e.g. $z^{\frac{1}{2}}$. If this should turn out to be a practical way to get such normal functions, it would be interesting to know which branes, i.e. which complexes $E$ or matrix factorizations $Q$ they could be associated to. Note also that in the context of Landau--Ginzburg models in one variable such chains were directly related to domain walls of the Landau--Ginzburg superpotential in~\cite{Hori:2000ck}. In the case of B--branes on noncompact Calabi--Yau threefolds, the domain wall tension is annihilated by an extended GKZ system, first discussed in~\cite{Mayr:2001xk,Lerche:2001cw}, the extension being relevant for the dependence on continuous open string moduli, which are absent here.

\section{Conclusions and Outlook}
\label{sec:conclusions}
In this article we have discussed B--branes for the one--parameter Calabi--Yau hypersurfaces in weighted $\mP^4$. In particular we focused on their normal function and derived its associated inhomogeneous Picard--Fuchs system.  Our starting point was a discussion of tensor product branes and their moduli using the language of matrix factorizations. We found that for the degree $d=6$ hypersurface none of tensor product branes has moduli. For the $d=8$ and the $d=10$ hypersurface we then picked a certain D--brane and derived the inhomogeneous Picard--Fuchs equations. \\
In our choice of branes we used the following criteria: The brane moduli should be obstructed by the bulk deformation in such a way that there are two brane vacua separated by a single domain wall. To find such a brane we have calculated the effective superpotential which encodes the obstructions to deformations of the brane and the bulk. A cubic effective superpotential indicates the desired structure. Our second criterion was tadpole cancellation. In order to check the tadpole cancellation condition we transported the matrix factorization to the large volume limit using the methods of~\cite{Herbst:2008jq}. We chose branes where the tadpole cancellation condition was satisfied by adding as few O--planes as possible.\\
Having picked a certain brane we derived geometric boundary conditions from the matrix factorizations and showed the existence of a normal function by verifying the vanishing of the algebraic second Chern class in cohomology. These boundary conditions could then, after resolving singularities coming from the Greene--Plesser orbifolds, be inserted into the inhomogeneous Picard--Fuchs equations which were derived using the Griffiths--Dwork algorithm. For the $d=8$ model we were able to calculate the domain wall tension and compute the BPS invariants for maps of holomorphic disks. As a consistency check we verified that the domain wall tension can be analytically continued to the Gepner point and is well--behaved under large radius monodromies. Moreover, we gave a prediction for the BPS invariants for maps of annuli and Klein bottles. Finally, we speculated on a connection between the solution to the inhomogeneous Picard--Fuchs equations and the GKZ system of differential equations.\\\\
Given the technical and conceptual complexity of the subject we have not achieved a complete discussion of open string mirror symmetry for this class of models. Our calculation has been done in the B--model, and of course our results should be verified by comparing with the A--model. Due to the limited knowledge of A--branes this may turn out to be very tedious since, at present, the only Lagrangians that are known are those defined by the real locus of the hypersurface equations. \\
In contrast to the quintic there are much more tensor product branes with moduli\footnote{In the quintic there was no need for a discussion which brane to choose because there was only one which has a modulus.}, and one should repeat the calculation for all of them. In particular, it would be interesting to go through the whole program for a brane which has a bicubic effective superpotential. In general, it is desirable to find models with branes whose moduli are obstructed at higher order or even not obstructed at all. \\
Of course the tensor product branes, corresponding to the Recknagel--Schomerus boundary states, are not the only ones which appear in these models and one might miss interesting phenomena by just focusing on those. It may therefore be instructive to study more general branes. \\
Another obvious direction to continue research on this subject is to study models with more than one bulk parameter. In analogy to closed string mirror symmetry, these models are expected to be more complicated but may reveal a deeper understanding of the concepts presented here.\\ 
In the inhomogeneous Picard--Fuchs equations we have derived here, we noticed the appearance of an additional constant term that was not present on the quintic. We believe that this term deserves further understanding.\\
As for the real BPS invariants, we have only computed them for the sets of worldsheets with Euler numbers $-1$ and $0$. For completeness one should also compute the BPS invariants for worldsheets with greater than zero. This would involve solving the extended holomorphic anomaly equations and this might turn to be hard since we do not know the corresponding A--brane explicitly.\\
Finally, it would be interesting to check whether the suggestion to find normal functions from the GKZ system can be used to simplify the whole procedure we have gone through in this paper.\\\\
{\bf Acknowledgements:} We would like to thank Johannes Walcher for informing us about \cite{walcher1par}. We thank Murad Alim, Manfred Herbst, Calin Lazaroiu and Duco van Straten for instructive discussions. Thanks also to Maximilian Kreuzer for support and providing computing resources. J. K. would like to thank Stephan Stieberger for interest in her work. \\
Furthermore, we would like to thank the Mathematics Research Center at the University of Warwick for hospitality during the school and conference on ``The Geometry and Integrability of Topological Field and String Theories'' where part of this work has been carried out. E. S. thanks the Fifth Simons Workshop in Mathematics and Physics at Stony Brook, which stimulated the present work, for hospitality.

\appendix
\section{Orientifolds}
\label{app-orientifolds}
In this appendix we discuss orientifolds for the one--parameter hypersurfaces. Apart from the fact that tadpole cancellation arguments lead us to a particular choice of branes to work with this discussion is independent of the rest of the paper. \\
Let us start with some generalities. Given a space--time manifold ${\bf X}$ with an involution $\tau$ and a D--brane supporting the complex vector bundle $E$. The tadpole cancellation condition for the $\tau$--orientifold of this system is:
\begin{equation}
\label{tadpole}
\mathrm{ch}(E)e^{-B}\sqrt{\hat{A}({\bf X})}=2^{2\mathrm{dim}_c({\bf X}^{\tau})-\mathrm{dim}_c{\bf X}}\epsilon\:[{\bf X}^{\tau}]\sqrt{\frac{L(\frac{1}{4}T{\bf X}^{\tau})}{L(\frac{1}{4}N{\bf X}^{\tau})}}
\end{equation}
In the above formula $B$ is the B--field, ${\bf X}^{\tau}$ is the O--plane, $\mathrm{dim}_c$ is the complex dimension not only of the internal components but includes also the space--time part which contributes two complex dimensions. Furthermore, $[{\bf X}^{\tau}]$ denotes the Poincar\'e dual of the O--plane and $\epsilon$ is a sign determined by the orientation of the orientifold plane. $\hat{A}$ is the Dirac genus.\\
We also review some facts about parity actions following \cite{Brunner:2003zm,Brunner:2004zd}.  The total parity action consists of an orientation reversal on the worldsheet, usually denoted by $\Omega$ and an action on the target space variables, denoted by $\tau$. There are two kinds of parity actions, A--parity and B--parity. A--parity is compatible with the topological A--twist and it is an antiholomorphic, isometric involution. B--parity is compatible with the topological B--twist; it is a holomorphic involution. Orientifold planes are the fixed point loci of the parity actions. For A--type parities these are O6--planes, wrapping Lagrangian submanifolds in the Calabi--Yau. Orientifold planes fixed under B--type parity are O3--, O5--, O7-- and O9--planes. Under mirror symmetry, A--parity translates to B--parity on the mirror. In order for supersymmetry to be preserved, we have to pick O$d_i$--planes such that $d_i=0\:\mathrm{mod}\:4$. Furthermore the orientifold action $\tau$ must be compatible with the $U(1)$ action on the Calabi--Yau.\\
Note that for O9--planes the tadpole cancellation condition (\ref{tadpole}) reduces to \cite{Brunner:2004zd}:
\begin{equation}
\label{o9tad}
\mathrm{ch}(E)e^{-B}=32+2\mathrm{ch}_2(X)
\end{equation}
Let us now start with the description of orientifolds in the linear sigma model \cite{Brunner:2004zd} with superpotential $W=PG(x_i,a_i)$, where $G(x_i,a_i)$ is assumed to be a tensor product of minimal models of type $A_k$ with superpotential $x_i^{k_i+2}$ and possible complex structure deformations parameterized by the $a_i$. A--parities combine a worldsheet parity action $\Omega_A$ with:
\begin{equation}
\tau^A_{{\bf m},\sigma}:\quad\begin{array}{rcl}
P&\rightarrow&\bar{P}\\
x_i&\rightarrow&e^{\frac{2\pi i m_i}{k_i+2}}\overline{x_{\sigma(i)}}\\
\end{array}
\end{equation}
The vector ${\bf m}$ labels elements of the global symmetry $\left(\prod_{i=1}^N\mZ_{k_i+2}\right)/\mZ_H$, where $H=\mathrm{lcm}(k_i+2)$. $\sigma(i)$ is an order 2 permutation with $k_{\sigma(i)}=k_i$. The above transformation is involutive if and only if:
\begin{equation}
m_i=m_{\sigma(i)}\quad (\mathrm{mod}\:k_i+2)
\end{equation}
We also have to take into account that a change of variables of the form $x_i'=e^{\frac{2\pi i n_i}{k_i+2}}x_i$. Therefore there is an equivalence relation ${\bf m}\equiv {\bf m'}$ iff $m_i'=m_i+n_i+n_{\sigma(i)}\quad (\mathrm{mod}\:k_i+2)$. From the above transformations we also deduce: $G(e^{\frac{2\pi i m_i}{k_i+2}}\overline{x_{\sigma(i)}},a_i)=\overline{G(x_i,a_i)}$.\\
B--parity combines a worldsheet action $\Omega_B$ with:
\begin{equation}
\tau^B_{{\bf m},\sigma}:\quad\begin{array}{rcl}
P&\rightarrow&-P\\
x_i&\rightarrow&e^{\frac{2\pi i m_i}{k_i+2}}x_{\sigma(i)}
\end{array}
\end{equation}
Here, $\sigma$ is again an order two permutation, as above. The condition to be involutive is now:
\begin{equation}
m_i+m_{\sigma(i)}=0\quad (\mathrm{mod}\:k_i+2)
\end{equation}
Due to possible reparameterizations, two vectors ${\bf m}$ and ${\bf m'}$ are equivalent iff $m_i'=m_i+n_i-n_{\sigma(i)}\quad (\mathrm{mod}\:k_i+2)$. This implies $G(e^{\frac{2\pi i m_i}{k_i+2}}x_{\sigma(i)},a_i)=G(x_i,a_i)$.\\
If we go to the Gepner point, the field $P$ gets an expectation value. The Gepner model is the IR limit of a Landau--Ginzburg orbifold of $G(x_i)$. The Landau--Ginzburg fields then transform as follows under A-- and B--parity:
\begin{eqnarray}
\tau^A_{{\bf m},\sigma}:&&x_i\rightarrow e^{\frac{2\pi i m_i}{k_i+2}}\overline{x_{\sigma(i)}}\nonumber \\
\tau^B_{{\bf m},\sigma}:&&x_i\rightarrow e^{\frac{2\pi i m_i}{k_i+2}}e^{\frac{\pi i}{k_i+1}}x_{\sigma(i)}
\end{eqnarray}
Note that we have to combine B--parity with a gauge transformation such that the Landau--Ginzburg superpotential changes its sign. This can be understood in terms of the Landau--Ginzburg action \cite{Brunner:2003zm}. The action contains a term with the superpotential and two fermions. Worldsheet parity flips the positions of the two fermions. If we demand parity invariance of the action we get an minus sign from putting the fermions into their original order. This sign is compensated by demanding that the Landau--Ginzburg superpotential changes its sign under B--parity.\\
At the Gepner point there are extra symmetries, the quantum symmetries, which form the group $\hat{\Gamma}\cong\mZ_H$. We can modify the parity action using this quantum symmetry. The most general parity transformations at the Gepner point are then the following:
\begin{eqnarray}
P^A_{\omega;{\bf m},\sigma}&=&g_{\omega}\tau^A_{{\bf m},\sigma}\Omega_A\nonumber\\
P^B_{\omega;{\bf m},\sigma}&=&g_{\omega}\tau^B_{{\bf m},\sigma}\Omega_B,
\end{eqnarray}
where the quantum symmetry $g_{\omega}$ associated to the $H$--th root of unity $\omega$ multiplies the $\ell$--twisted states by a phase $\omega^{\ell}$. \\\\
In the following we will discuss $B$--type orientifolds for the one--parameter hypersurfaces, restricting ourselves to parity actions where the permutation $\sigma$ is trivial.\\
Note that in all cases we will insert the value $B=-\frac{H}{2}$ for the B--field. If an O--plane has different components we can add or subtract them with relative sign factors. In the tables below we have chosen a particular combination of signs in most cases.
\subsection{$d=6$}
Let us start by collecting the relevant data of this model. The hypersurface equation is:
\begin{equation}
x_1^6+x_2^6+x_3^6+x_4^6+x_5^3
\end{equation}
In the geometric regime the Calabi--Yau threefold $X$ is defined by the hypersurface equations $W=0$. We have the following $U(1)$ action:
\begin{equation}
(x_1,x_2,x_3,x_4,x_5)\longrightarrow(\lambda x_1,\lambda x_2,\lambda x_3,\lambda x_2,\lambda^2 x_5) \qquad \lambda^6=1
\end{equation}
We will need the following topological data:
\begin{eqnarray}
c(X)&=&1+14H^2-68H^3\nonumber\\
\hat{A}(X)&=&1+\frac{7}{6}H^2
\end{eqnarray}
We collect the data about the orientifolds in table \ref{tab-d6orientifold}.
\begin{table}[ht]
\begin{center}
\begin{tabular}{|c|c|c|}
\hline
Parity Action&O--plane type&Tadpole cancellation\\
\hline
$(x_1,x_2,x_3,x_4,x_5)$&O9&$\mathrm{ch}(E)=4(8-4H-6H^2+\frac{10}{3}H^3)$\\
\hline
$(x_1,x_2,x_3,x_4,-x_5)$&O7&$\mathrm{ch}(E)=\pm4(4H-2H^2-\frac{10}{3}H^3)$\\
\hline
$(-x_1,x_2,x_3,x_4,x_5)$&O3/O7&$\mathrm{ch}(E)=\pm4(2H-H^2-\frac{5}{3}H^3)$\\
\hline
$(-x_1,-x_2,x_3,x_4,x_5)$&O5&$\mathrm{ch}(E)=\pm4(H^2-\frac{H^3}{2})$\\
\hline
\end{tabular}\caption{O--planes for the $d=6$ hypersurface.}\label{tab-d6orientifold}
\end{center}
\end{table}
\subsection{$d=8$}
The hypersurface equation is:
\begin{equation}
x_1^8+x_2^8+x_3^8+x_4^8+x_5^2
\end{equation}
We have following the $U(1)$-action:
\begin{equation}
(x_1,x_2,x_3,x_4,x_5)\longrightarrow(\lambda x_1,\lambda x_2,\lambda x_3,\lambda x_4,\lambda^4 x_5)
\end{equation}
The Chern class and A--roof genus of the Calabi--Yau hypersurface are:
\begin{eqnarray}
c(X)&=&1+22H^2-148H^3\nonumber\\
\hat{A}(X)&=&1+\frac{11}{6}H^2
\end{eqnarray}
We collect the data about the orientifolds in table \ref{tab-d8orientifold}.
\begin{table}[ht]
\begin{center}
\begin{tabular}{|c|c|c|}
\hline
Parity Action&O--plane type&Tadpole cancellation\\
\hline
$(x_1,x_2,x_3,x_4,x_5)$&O9&$\mathrm{ch}(E)=4(8-4H-10H^2+\frac{16}{3}H^3)$\\
\hline
$(x_1,x_2,x_3,x_4,-x_5)$&O7&$\mathrm{ch}(E)=\pm 4(8H-4H^2-\frac{38}{3}H^3)$\\
\hline
$(-x_1,x_2,x_3,x_4,x_5)$&O3/O7&$\mathrm{ch}(E)=\left\{\begin{array}{c}
\pm4(2H-H^2-\frac{29}{12}H^3)\\
\pm4(2H-H^2-\frac{8}{3}H^3)
\end{array}\right.$\\
\hline
$(-x_1,-x_2,x_3,x_4,x_5)$&O5&$\mathrm{ch}(E)e^{-B}=\pm2kH^2\quad k=0,\ldots,2$\\
\hline
\end{tabular}\caption{O--planes for the $d=8$ hypersurface.}\label{tab-d8orientifold}
\end{center}
\end{table}
\subsection{$d=10$}
The hypersurface equation is:
\begin{equation}
x_1^{10}+x_2^{10}+x_3^{10}+x_4^5+x_5^2
\end{equation}
The $U(1)$--action on the variables is:
\begin{equation}
(x_1,x_2,x_3,x_4,x_5)\longrightarrow(\lambda x_1,\lambda x_2,\lambda x_3,\lambda^2 x_4,\lambda^5 x_5)
\end{equation}
The Chern class and the A--roof genus can be shown to be:
\begin{eqnarray}
c(X)&=&1+34H^2-288H^3\nonumber\\
\hat{A}(X)&=&1+\frac{17}{6}H^2
\end{eqnarray}
The O--plane data can be found in table \ref{tab-d10orientifold}.
\begin{table}[ht]
\begin{center}
\begin{tabular}{|c|c|c|}
\hline
Parity Action&O--plane type&Tadpole cancellation\\
\hline
$(x_1,x_2,x_3,x_4,x_5)$&O9&$\mathrm{ch}(E)=4(8-4H-16H^2+\frac{25}{3}H^3)$\\
\hline
$(x_1,x_2,x_3,x_4,-x_5)$&O3/O7&$\mathrm{ch}(E)=\left\{\begin{array}{c}
\pm4(10H-5H^2-\frac{301}{12}H^3)\\
\pm4(10H-5H^2-\frac{76}{3}H^3)
\end{array}\right.$\\
\hline
$(x_1,x_2,x_3,-x_4,x_5)$&O7&$\mathrm{ch}(E)=\pm4(4H-2H^2-\frac{25}{3}H^3)$\\
\hline
$(-x_1,x_2,x_3,x_4,x_5)$&O3/O7&$\mathrm{ch}(E)=\left\{\begin{array}{c}
\pm4(2H-H^2-\frac{41}{12}H^3)\\
\pm4(2H-H^2-\frac{14}{3}H^3)
\end{array}\right.$\\
\hline
$(-x_1,x_2,x_3,x_4,-x_5)$&O5&$\mathrm{ch}(E)=\left\{\begin{array}{c}
\pm4(3H^2-\frac{3}{2}H^3)\\
\pm4(2H^2-H^3)
\end{array}\right.$\\
\hline
\end{tabular}\caption{O--planes for the $d=10$ hypersurface.}\label{tab-d10orientifold}
\end{center}
\end{table}
\bibliographystyle{fullsort}
\bibliography{bibliography}

\providecommand{\href}[2]{#2}\begingroup\raggedright\begin{thebibliography}{10}

\bibitem{Candelas:1990rm}
P.~Candelas, X.~C. De~La~Ossa, P.~S. Green, and L.~Parkes, ``{A pair of
  Calabi-Yau manifolds as an exactly soluble superconformal theory},'' {\em
  Nucl. Phys.} {\bf B359} (1991)
21--74.

\bibitem{Witten:1991zz}
E.~Witten, ``{Mirror manifolds and topological field theory},''
\href{http://www.arXiv.org/abs/hep-th/9112056}{{\tt hep-th/9112056}}.

\bibitem{Bershadsky:1993ta}
M.~Bershadsky, S.~Cecotti, H.~Ooguri, and C.~Vafa, ``{Holomorphic anomalies in
  topological field theories},'' {\em Nucl. Phys.} {\bf B405} (1993) 279--304,
\href{http://www.arXiv.org/abs/hep-th/9302103}{{\tt hep-th/9302103}}.

\bibitem{Bershadsky:1993cx}
M.~Bershadsky, S.~Cecotti, H.~Ooguri, and C.~Vafa, ``{Kodaira-Spencer theory of
  gravity and exact results for quantum string amplitudes},'' {\em Commun.
  Math. Phys.} {\bf 165} (1994) 311--428,
\href{http://www.arXiv.org/abs/hep-th/9309140}{{\tt hep-th/9309140}}.

\bibitem{Ooguri:1999bv}
H.~Ooguri and C.~Vafa, ``{Knot invariants and topological strings},'' {\em
  Nucl. Phys.} {\bf B577} (2000) 419--438,
\href{http://www.arXiv.org/abs/hep-th/9912123}{{\tt hep-th/9912123}}.

\bibitem{Aganagic:2000gs}
M.~Aganagic and C.~Vafa, ``{Mirror symmetry, D-branes and counting holomorphic
  discs},''
\href{http://www.arXiv.org/abs/hep-th/0012041}{{\tt hep-th/0012041}}.

\bibitem{Bouchard:2007ys}
V.~Bouchard, A.~Klemm, M.~Marino, and S.~Pasquetti, ``{Remodeling the
  B-model},''
\href{http://www.arXiv.org/abs/arXiv:0709.1453 [hep-th]}{{\tt arXiv:0709.1453
  [hep-th]}}.

\bibitem{Witten:1992fb}
E.~Witten, ``{Chern-Simons gauge theory as a string theory},'' {\em Prog.
  Math.} {\bf 133} (1995) 637--678,
\href{http://www.arXiv.org/abs/hep-th/9207094}{{\tt hep-th/9207094}}.

\bibitem{Kontsevich}
M.~Kontsevich, ``{Homological Algebra of Mirror Symmetry},'' {\em Proc.
  Internat. Congress Math.} {\bf 1} (1995) 120--139,
  \href{http://www.arXiv.org/abs/alg-geom/9411018}{{\tt alg-geom/9411018}}.

\bibitem{Walcher:2006rs}
J.~Walcher, ``{O}pening {M}irror {S}ymmetry on the {Q}uintic,'' {\em Commun.
  Math. Phys.} {\bf 276} (2007) 671--689,
\href{http://www.arXiv.org/abs/hep-th/0605162}{{\tt hep-th/0605162}}.

\bibitem{Pandharipande:2006ab}
R.~Pandharipande, J.~Solomon, and J.~Walcher, ``Disk enumeration on the quintic
  3--fold,'' (2006) \href{http://www.arXiv.org/abs/math.SG/0610901}{{\tt
  math.SG/0610901}}.

\bibitem{Walcher:2007tp}
J.~Walcher, ``{Extended Holomorphic Anomaly and Loop Amplitudes in Open
  Topological String},''
\href{http://www.arXiv.org/abs/arXiv:0705.4098 [hep-th]}{{\tt arXiv:0705.4098
  [hep-th]}}.

\bibitem{Neitzke:2007yw}
A.~Neitzke and J.~Walcher, ``{Background Independence and the Open Topological
  String Wavefunction},''
\href{http://www.arXiv.org/abs/arXiv:0709.2390 [hep-th]}{{\tt arXiv:0709.2390
  [hep-th]}}.

\bibitem{Morrison:2007bm}
D.~R. Morrison and J.~Walcher, ``{D}-branes and {N}ormal {F}unctions,''
\href{http://www.arXiv.org/abs/arXiv:0709.4028 [hep-th]}{{\tt arXiv:0709.4028
  [hep-th]}}.

\bibitem{Walcher:2007qp}
J.~Walcher, ``{Evidence for Tadpole Cancellation in the Topological String},''
\href{http://www.arXiv.org/abs/arXiv:0712.2775 [hep-th]}{{\tt arXiv:0712.2775
  [hep-th]}}.

\bibitem{Cook:2007dj}
P.~L.~H. Cook, H.~Ooguri, and J.~Yang, ``{Comments on the Holomorphic Anomaly
  in Open Topological String Theory},'' {\em Phys. Lett.} {\bf B653} (2007)
  335--337,
\href{http://www.arXiv.org/abs/arXiv:0706.0511 [hep-th]}{{\tt arXiv:0706.0511
  [hep-th]}}.

\bibitem{Bonelli:2007gv}
G.~Bonelli and A.~Tanzini, ``{The holomorphic anomaly for open string
  moduli},'' {\em JHEP} {\bf 10} (2007) 060,
\href{http://www.arXiv.org/abs/arXiv:0708.2627 [hep-th]}{{\tt arXiv:0708.2627
  [hep-th]}}.

\bibitem{Alim:2007qj}
M.~Alim and J.~D. Lange, ``{Polynomial Structure of the (Open) Topological
  String Partition Function},'' {\em JHEP} {\bf 10} (2007) 045,
\href{http://www.arXiv.org/abs/arXiv:0708.2886 [hep-th]}{{\tt arXiv:0708.2886
  [hep-th]}}.

\bibitem{Konishi:2007qx}
Y.~Konishi and S.~Minabe, ``{On Solutions to Walcher's Extended Holomorphic
  Anomaly Equation},''
\href{http://www.arXiv.org/abs/arXiv:0708.2898 [hep-th]}{{\tt arXiv:0708.2898
  [hep-th]}}.

\bibitem{Cook:2008eu}
P.~L.~H. Cook, H.~Ooguri, and J.~Yang, ``{New Anomalies in Topological String
  Theory},''
\href{http://www.arXiv.org/abs/arXiv:0804.1120 [hep-th]}{{\tt arXiv:0804.1120
  [hep-th]}}.

\bibitem{walcher1par}
D.~Krefl and J.~Walcher, ``{to appear.},''.

\bibitem{Hori:2006ic}
K.~Hori and J.~Walcher, ``{D-brane Categories for Orientifolds: The
  Landau-Ginzburg Case},'' {\em JHEP} {\bf 04} (2008) 030,
\href{http://www.arXiv.org/abs/hep-th/0606179}{{\tt hep-th/0606179}}.

\bibitem{Diaconescu:2006id}
D.-E. Diaconescu, A.~Garcia-Raboso, R.~L. Karp, and K.~Sinha, ``{D-brane
  Superpotentials in Calabi-Yau Orientifolds (projection)},''
\href{http://www.arXiv.org/abs/hep-th/0606180}{{\tt hep-th/0606180}}.

\bibitem{Donaldson:1998ab}
S.~K. Donaldson and R.~P. Thomas, ``Gauge theory in higher dimensions,'' in
  {\em The Geometric Universe}, S.~A. Huggett, ed., pp.~31 -- 47.
\newblock Oxford University Press, 1998.

\bibitem{griffiths1}
P.~Griffiths, ``{On the periods of certain rational integrals I},'' {\em Ann.
  of Math.} {\bf 90} (1969) 460--495.

\bibitem{griffiths2}
P.~Griffiths, ``{On the periods of certain rational integrals II},'' {\em Ann.
  of Math.} {\bf 90} (1969) 495--541.

\bibitem{Griffiths:1979ab}
P.~A. Griffiths, ``A theorem concerning the differential equations satisfied by
  normal functions associated to algebraic cycles,'' {\em Amer. J. Math.} {\bf
  101} (1979) 94--131.

\bibitem{Witten:1997ep}
E.~Witten, ``{Branes and the dynamics of {QCD}},'' {\em Nucl. Phys.} {\bf B507}
  (1997) 658--690,
\href{http://www.arXiv.org/abs/hep-th/9706109}{{\tt hep-th/9706109}}.

\bibitem{Kachru:2000ih}
S.~Kachru, S.~H. Katz, A.~E. Lawrence, and J.~McGreevy, ``{Open string
  instantons and superpotentials},'' {\em Phys. Rev.} {\bf D62} (2000) 026001,
\href{http://www.arXiv.org/abs/hep-th/9912151}{{\tt hep-th/9912151}}.

\bibitem{Dijkgraaf:2002fc}
R.~Dijkgraaf and C.~Vafa, ``{Matrix models, topological strings, and
  supersymmetric gauge theories},'' {\em Nucl. Phys.} {\bf B644} (2002) 3--20,
\href{http://www.arXiv.org/abs/hep-th/0206255}{{\tt hep-th/0206255}}.

\bibitem{GreenM}
M.~L. Green, {\em {Infinitesimal Methods in Hodge Theory}}, vol.~1594 of {\em
  Lecture Notes in Mathematics}.
\newblock Springer, 1994.

\bibitem{Voisin1}
C.~Voisin, {\em {Hodge Theory and Complex Algebraic Geometry I}}, vol.~76 of
  {\em Cambridge Studies in Advanced Mathematics}.
\newblock Cambridge University Press, 2002.

\bibitem{Voisin2}
C.~Voisin, {\em {Hodge Theory and Complex Algebraic Geometry II}}, vol.~77 of
  {\em Cambridge Studies in Advanced Mathematics}.
\newblock Cambridge University Press, 2003.

\bibitem{Griffiths:1983ab}
P.~Griffiths, ``{I}nfinitesimal {V}ariations of {H}odge {S}tructure. {III}
  {D}eterminantal {V}arieties and the {I}nfinitesimal {I}nvariant of {N}ormal
  {F}unctions,'' {\em Compositio Math.} {\bf 50} (1983), no.~2-3, 267--324.

\bibitem{Greene:1990ud}
B.~R. Greene and M.~R. Plesser, ``{Duality in Calabi-Yau Moduli Space},'' {\em
  Nucl. Phys.} {\bf B338} (1990)
15--37.

\bibitem{Font:1992uk}
A.~Font, ``{P}eriods and {D}uality {S}ymmetries in {C}alabi-{Y}au
  {C}ompactifications,'' {\em Nucl. Phys.} {\bf B391} (1993) 358--388,
\href{http://www.arXiv.org/abs/hep-th/9203084}{{\tt hep-th/9203084}}.

\bibitem{Klemm:1992tx}
A.~Klemm and S.~Theisen, ``{Considerations of one modulus Calabi-Yau
  compactifications: Picard-Fuchs equations, Kahler potentials and mirror
  maps},'' {\em Nucl. Phys.} {\bf B389} (1993) 153--180,
\href{http://www.arXiv.org/abs/hep-th/9205041}{{\tt hep-th/9205041}}.

\bibitem{Roiban:2002iv}
R.~Roiban, C.~Romelsberger, and J.~Walcher, ``{Discrete torsion in singular
  G(2)-manifolds and real LG},'' {\em Adv. Theor. Math. Phys.} {\bf 6} (2003)
  207--278,
\href{http://www.arXiv.org/abs/hep-th/0203272}{{\tt hep-th/0203272}}.

\bibitem{Kapustin:2002bi}
A.~Kapustin and Y.~Li, ``{D-branes in Landau-Ginzburg models and algebraic
  geometry},'' {\em JHEP} {\bf 12} (2003) 005,
\href{http://www.arXiv.org/abs/hep-th/0210296}{{\tt hep-th/0210296}}.

\bibitem{Brunner:2003dc}
I.~Brunner, M.~Herbst, W.~Lerche, and B.~Scheuner, ``{L}andau-{G}inzburg
  {R}ealization of {O}pen {S}tring {TFT},'' {\em JHEP} {\bf 11} (2006) 043,
\href{http://www.arXiv.org/abs/hep-th/0305133}{{\tt hep-th/0305133}}.

\bibitem{Jockers:2007ng}
H.~Jockers and W.~Lerche, ``{Matrix Factorizations, D-Branes and their
  Deformations},'' {\em Nucl. Phys. Proc. Suppl.} {\bf 171} (2007) 196--214,
\href{http://www.arXiv.org/abs/arXiv:0708.0157 [hep-th]}{{\tt arXiv:0708.0157
  [hep-th]}}.

\bibitem{Knapp:2007vc}
J.~Knapp, ``{D-Branes in Topological String Theory},''
\href{http://www.arXiv.org/abs/arXiv:0709.2045 [hep-th]}{{\tt arXiv:0709.2045
  [hep-th]}}.

\bibitem{orlov1}
D.~Orlov, ``{D}erived {C}ategories of {C}oherent {S}heaves and {T}riangulated
  {C}ategories of {S}ingularities,''
  \href{http://www.arXiv.org/abs/math.AG/0503632}{{\tt math.AG/0503632}}.

\bibitem{Herbst:2008jq}
M.~Herbst, K.~Hori, and D.~Page, ``{Phases Of N=2 Theories In 1+1 Dimensions
  With Boundary},''
\href{http://www.arXiv.org/abs/arXiv:0803.2045 [hep-th]}{{\tt arXiv:0803.2045
  [hep-th]}}.

\bibitem{Witten:1993yc}
E.~Witten, ``{Phases of N = 2 theories in two dimensions},'' {\em Nucl. Phys.}
  {\bf B403} (1993) 159--222,
\href{http://www.arXiv.org/abs/hep-th/9301042}{{\tt hep-th/9301042}}.

\bibitem{Brunner:2005fv}
I.~Brunner and M.~R. Gaberdiel, ``{M}atrix {F}actorisations and {P}ermutation
  {B}ranes,'' {\em JHEP} {\bf 07} (2005) 012,
\href{http://www.arXiv.org/abs/hep-th/0503207}{{\tt hep-th/0503207}}.

\bibitem{Recknagel:1997sb}
A.~Recknagel and V.~Schomerus, ``{D-branes in Gepner models},'' {\em Nucl.
  Phys.} {\bf B531} (1998) 185--225,
\href{http://www.arXiv.org/abs/hep-th/9712186}{{\tt hep-th/9712186}}.

\bibitem{Brunner:1999jq}
I.~Brunner, M.~R. Douglas, A.~E. Lawrence, and C.~Romelsberger, ``{D-branes on
  the quintic},'' {\em JHEP} {\bf 08} (2000) 015,
\href{http://www.arXiv.org/abs/hep-th/9906200}{{\tt hep-th/9906200}}.

\bibitem{Scheidegger:1999ed}
E.~Scheidegger, ``{D-branes on some one- and two-parameter Calabi-Yau
  hypersurfaces},'' {\em JHEP} {\bf 04} (2000) 003,
\href{http://www.arXiv.org/abs/hep-th/9912188}{{\tt hep-th/9912188}}.

\bibitem{Naka:2000he}
M.~Naka and M.~Nozaki, ``{Boundary states in Gepner models},'' {\em JHEP} {\bf
  05} (2000) 027,
\href{http://www.arXiv.org/abs/hep-th/0001037}{{\tt hep-th/0001037}}.

\bibitem{Govindarajan:2000vi}
S.~Govindarajan and T.~Jayaraman, ``{D-branes, exceptional sheaves and quivers
  on Calabi-Yau manifolds: From Mukai to McKay},'' {\em Nucl. Phys.} {\bf B600}
  (2001) 457--486,
\href{http://www.arXiv.org/abs/hep-th/0010196}{{\tt hep-th/0010196}}.

\bibitem{Brunner:2004zd}
I.~Brunner, K.~Hori, K.~Hosomichi, and J.~Walcher, ``{O}rientifolds of {G}epner
  {M}odels,'' {\em JHEP} {\bf 02} (2007) 001,
\href{http://www.arXiv.org/abs/hep-th/0401137}{{\tt hep-th/0401137}}.

\bibitem{Morrison:1991cd}
D.~R. Morrison, ``{Picard-Fuchs equations and mirror maps for hypersurfaces},''
\href{http://www.arXiv.org/abs/hep-th/9111025}{{\tt hep-th/9111025}}.

\bibitem{Dwork}
B.~Dwork, ``{On the Zeta Function of a Hypersurface II},'' {\em Ann. of Math.}
  {\bf 80} (1964) 227--299.

\bibitem{Stienstra:2005nr}
J.~Stienstra, ``{GKZ Hypergeometric Structures},''
\href{http://www.arXiv.org/abs/math/0511351}{{\tt math/0511351}}.

\bibitem{Batyrev:1994hm}
V.~V. Batyrev, ``{Dual polyhedra and mirror symmetry for Calabi-Yau
  hypersurfaces in toric varieties},'' {\em J. Alg. Geom.} {\bf 3} (1994)
493--545.

\bibitem{Hosono:1993qy}
S.~Hosono, A.~Klemm, S.~Theisen, and S.-T. Yau, ``{Mirror symmetry, mirror map
  and applications to Calabi-Yau hypersurfaces},'' {\em Commun. Math. Phys.}
  {\bf 167} (1995) 301--350,
\href{http://www.arXiv.org/abs/hep-th/9308122}{{\tt hep-th/9308122}}.

\bibitem{Hori:2004ja}
K.~Hori and J.~Walcher, ``{F-term equations near Gepner points},'' {\em JHEP}
  {\bf 01} (2005) 008,
\href{http://www.arXiv.org/abs/hep-th/0404196}{{\tt hep-th/0404196}}.

\bibitem{siqveland}
A.~Siqveland, ``{T}he {M}ethod of {C}omputing {F}ormal {M}oduli,'' {\em J.
  Alg.} {\bf 241} (2001) 292--237.

\bibitem{Knapp:2006rd}
J.~Knapp and H.~Omer, ``{M}atrix {F}actorizations, {M}inimal {M}odels and
  {M}assey {P}roducts,'' {\em JHEP} {\bf 05} (2006) 064,
\href{http://www.arXiv.org/abs/hep-th/0604189}{{\tt hep-th/0604189}}.

\bibitem{Hori:2004zd}
K.~Hori and J.~Walcher, ``{D}-branes from {M}atrix {F}actorizations,'' {\em
  Comptes Rendus Physique} {\bf 5} (2004) 1061--1070,
\href{http://www.arXiv.org/abs/hep-th/0409204}{{\tt hep-th/0409204}}.

\bibitem{Herbst:2004jp}
M.~Herbst, C.-I. Lazaroiu, and W.~Lerche, ``{S}uperpotentials, {A}(infinity)
  {R}elations and {WDVV} {E}quations for open topological {S}trings,'' {\em
  JHEP} {\bf 02} (2005) 071,
\href{http://www.arXiv.org/abs/hep-th/0402110}{{\tt hep-th/0402110}}.

\bibitem{Ashok:2004xq}
S.~K. Ashok, E.~Dell'Aquila, D.-E. Diaconescu, and B.~Florea, ``{O}bstructed
  {D}-branes in {L}andau-{G}inzburg {O}rbifolds,'' {\em Adv. Theor. Math.
  Phys.} {\bf 8} (2004) 427--472,
\href{http://www.arXiv.org/abs/hep-th/0404167}{{\tt hep-th/0404167}}.

\bibitem{Kapustin:2003ga}
A.~Kapustin and Y.~Li, ``{T}opological {C}orrelators in {L}andau-{G}inzburg
  {M}odels with {B}oundaries,'' {\em Adv. Theor. Math. Phys.} {\bf 7} (2004)
  727--749,
\href{http://www.arXiv.org/abs/hep-th/0305136}{{\tt hep-th/0305136}}.

\bibitem{Walcher:2004tx}
J.~Walcher, ``{S}tability of {L}andau-{G}inzburg {B}ranes,'' {\em J. Math.
  Phys.} {\bf 46} (2005) 082305,
\href{http://www.arXiv.org/abs/hep-th/0412274}{{\tt hep-th/0412274}}.

\bibitem{Grothendieck:1958ab}
A.~Grothendieck, ``La th\'eorie des classes de {C}hern,'' {\em Bull. Soc. Math.
  France} {\bf 86} (1958) 137--154.

\bibitem{Fulton:1998ab}
W.~Fulton, {\em Intersection theory}, vol.~2 of {\em Ergebnisse der Mathematik
  und ihrer Grenzgebiete. 3. Folge}.
\newblock Springer, Berlin, 2nd ed.~ed., 1998.

\bibitem{GPS05}
G.-M. Greuel, G.~Pfister, and H.~Sch\"onemann, ``{\sc Singular} 3.0,'' {A
  Computer Algebra System for Polynomial Computations}, Centre for Computer
  Algebra, University of Kaiserslautern, 2005.
\newblock {\tt http://www.singular.uni-kl.de}.

\bibitem{Givental:1996ab}
A.~B. Givental, ``Equivariant {G}romov-{W}itten invariants,'' {\em Internat.
  Math. Res. Notices} (1996), no.~13, 613--663,
  \href{http://www.arXiv.org/abs/alg-geom/9603021}{{\tt alg-geom/9603021}}.

\bibitem{Batyrev:1993wa}
V.~V. Batyrev and D.~van Straten, ``{Generalized hypergeometric functions and
  rational curves on Calabi-Yau complete intersections in toric varieties},''
  {\em Commun. Math. Phys.} {\bf 168} (1995) 493--534,
\href{http://www.arXiv.org/abs/alg-geom/9307010}{{\tt alg-geom/9307010}}.

\bibitem{Hosono:1994ax}
S.~Hosono, A.~Klemm, S.~Theisen, and S.-T. Yau, ``{Mirror symmetry, mirror map
  and applications to complete intersection Calabi-Yau spaces},'' {\em Nucl.
  Phys.} {\bf B433} (1995) 501--554,
\href{http://www.arXiv.org/abs/hep-th/9406055}{{\tt hep-th/9406055}}.

\bibitem{Berglund:1993yn}
P.~Berglund, E.~Derrick, T.~Hubsch, and D.~Jancic, ``{On periods for string
  compactifications},'' {\em Nucl. Phys.} {\bf B420} (1994) 268--287,
\href{http://www.arXiv.org/abs/hep-th/9311143}{{\tt hep-th/9311143}}.

\bibitem{Avram:1995wt}
A.~C. Avram, E.~Derrick, and D.~Jancic, ``{On Semi-Periods},'' {\em Nucl.
  Phys.} {\bf B471} (1996) 293--308,
\href{http://www.arXiv.org/abs/hep-th/9511152}{{\tt hep-th/9511152}}.

\bibitem{Kreuzer:2006ax}
M.~Kreuzer, ``{Toric Geometry and Calabi-Yau Compactifications},''
\href{http://www.arXiv.org/abs/hep-th/0612307}{{\tt hep-th/0612307}}.

\bibitem{Brunner:2004mt}
I.~Brunner, M.~Herbst, W.~Lerche, and J.~Walcher, ``{Matrix Factorizations and
  Mirror Symmetry: The Cubic Curve},'' {\em JHEP} {\bf 11} (2006) 006,
\href{http://www.arXiv.org/abs/hep-th/0408243}{{\tt hep-th/0408243}}.

\bibitem{Hori:2000ck}
K.~Hori, A.~Iqbal, and C.~Vafa, ``{D-branes and mirror symmetry},''
\href{http://www.arXiv.org/abs/hep-th/0005247}{{\tt hep-th/0005247}}.

\bibitem{Mayr:2001xk}
P.~Mayr, ``{N = 1 Mirror Symmetry and open/closed String Duality},'' {\em Adv.
  Theor. Math. Phys.} {\bf 5} (2002) 213--242,
\href{http://www.arXiv.org/abs/hep-th/0108229}{{\tt hep-th/0108229}}.

\bibitem{Lerche:2001cw}
W.~Lerche and P.~Mayr, ``{On N = 1 Mirror Symmetry for Open Type II Strings},''
\href{http://www.arXiv.org/abs/hep-th/0111113}{{\tt hep-th/0111113}}.

\bibitem{Brunner:2003zm}
I.~Brunner and K.~Hori, ``{Orientifolds and mirror symmetry},'' {\em JHEP} {\bf
  11} (2004) 005,
\href{http://www.arXiv.org/abs/hep-th/0303135}{{\tt hep-th/0303135}}.

\end{thebibliography}\endgroup

\end{document}